\documentclass[useAMS,usenatbib]{mn2e}

\usepackage{graphicx}
\usepackage{natbib}
\title[A survey of Low Luminosity Compact sources]{A survey of Low Luminosity 
Compact sources and its implication for evolution of radio-loud AGNs. I. Radio data}

\author[M. Kunert-Bajraszewska et al.]{M. Kunert-Bajraszewska$^{1}$\thanks
{E-mail: magda@astro.uni.torun.pl}, M. P. Gawro{\'n}ski$^{1}$, A.
Labiano$^{2}$ and A. Siemiginowska$^{3}$\\
$^{1}$Toru{\'n} Centre for Astronomy, N. Copernicus University,
Gagarina 11, 87-100 Toru{\'n}, Poland\\
$^{2}$European Space Agency (ESA), European Space Astronomy Centre (ESAC),
28691 Villanueva de la Canada, Madrid, Spain\\
$^{3}$Harvard-Smithsonian Centre for Astrophysics, 60 Garden St., Cambridge,
MA 02138, USA}

\begin{document}

\date{Accepted 1988 December 15. Received 1988 December 14; in original form 1988 October 11}

\pagerange{\pageref{firstpage}--\pageref{lastpage}} \pubyear{2002}

\maketitle

\label{firstpage}

\begin{abstract}
We present a new sample of Compact Steep
Spectrum (CSS) sources with radio luminosity below $10^{26}\, {\rm
W~Hz^{-1}}$ at 1.4\,GHz called the low luminosity compact (LLC) objects.
The sources have been selected from FIRST survey and observed with MERLIN at 
L-band and C-band. The main criterion used for selection was luminosity of the 
objects and approximately one third of the CSS sources from the
new sample have a value of radio luminosity comparable to FR\,Is.
About 80$\%$ of the sources have been resolved and about 30$\%$ 
of them have weak extended emission and disturbed structures when
compared with the observations of higher luminosity CSS sources.
We studied correlation between radio power and linear size, and
redshift with a larger sample that included also published samples of 
compact objects and large scale FR\,IIs and FR\,Is. 
The low luminosity compact objects occupy the space in radio power versus
linear size diagram below the main evolutionary path of radio objects. We
suggest that many of them might be short-lived objects, 
and their radio emission may be disrupted several times before becoming
FR\,IIs.
We conclude that there exists a large population of short-lived low luminosity
compact objects unexplored so far and part of them can be precursors of
large scale FR\,Is.  

\end{abstract}


\section{Introduction}
Radio sources are divided into two distinct morphological groups of objects:
FR\,Is and FR\,IIs \citep{fr74}. There is a relatively sharp luminosity
boundary between them at low frequency. The nature of the FR-division is still an open
issue, as are the details of the evolutionary process in which 
younger and smaller Gigahertz-Peaked Spectrum (GPS) and Compact Steep
Spectrum (CSS) sources become large scale radio structures. It is unclear whether
FR\,II objects evolve to become FR\,Is, or whether a division has
already occurred amongst CSS sources and some of these then become FR\,Is
and some FR\,IIs. A majority of CSS sources known to date
have high radio luminosities and, if unbeamed, have FR\,II structures.
It seems reasonable
to suspect that some of the CSSs with lower radio luminosity could be
the progenitors of less luminous FR\,I objects.

The GPS and CSS sources form a well-defined class of compact radio objects 
and are considered to be entirely contained within the host galaxy. During
their evolution the radio jets start to cross the ISM and try to leave
the host galaxy. The interaction with the ISM can be very strong in GPS/CSS 
sources
\citep{holt, labiano} 
and it seems to be a crucial point in the evolution of radio sources.
Recently developed models \citep{kb07} show that all
sources start out with a FR\,II morphology. 
As the radio source expands the interaction with the dense environment of
the host galaxy can disrupt jets and change their morphology to FR\,Is or 
hybrid objects \citep{gopal00,gawron06}. Some sources with disrupted jets
can fade away \citep{alex00}. Detection of several candidates for
dying compact sources supports this view \citep{gir05,kun05,kun06,mar06}. 
What's more, the activity of compact radio sources can be an episodic event
\citep{sn99,mar03b,gug05,kun06}, like in the case of large scale radio
objects.

As recently suggested by \citet{czerny} there could be a connection
between the existence of short-lived compact radio sources and the intermittent 
activity of the central engine caused by a radiation pressure
instability within an accretion disc. According to the authors a radio
source powered by a short-lived outburst of the central activity is 
not able to escape from the host galaxy unless the active phase lasts
longer than $\sim$$10^4$~years. It is then likely that among the low
luminosity compact (LLC) sources we can find objects affected during their
evolution either by a strong
interaction with the ISM, which changes their morphology and luminosity, or by
the instability of the accretion disc. 

In this paper we present the L-band and C-band MERLIN observations of 44 low luminosity CSS
sources selected from the FIRST survey. This is a new approach since the
main criterion used for selection was not flux density but luminosity of the
objects.
The selection criteria caused that 
approximately one third of the CSS sources from the
new sample have a value of radio luminosity lower than the
luminosity
bo\-un\-da\-ry found for FR sources \citep{fr74}, which means that they are
compact young sources with luminosities comparable to FR\,Is
(Fig.~\ref{sample_plot}).
The goal of this project is to study the properties of low luminosity
compact (LLC)
objects and the evolution of the compact object population. 
Our previous multifrequency observations of CSS sources have shown
that some of the small-scale objects can be strong candidates for compact
faders \citep{kun05, kun06}. This finding supports the idea that there exist a 
group of short-lived 
radio objects that has been largely neglected to date.

\section{Selection and observations of a new sample}
\label{obs}

Using the final release of FIRST, combined with the GB6 survey at 4.85\,GHz we
looked for unresolved, isolated sources i.e. more compact than FIRST
beam (5\farcs4) and surrounded by an empty field (we adopted 1\arcmin~as
a radius of that field). We required that the redshifts of the objects
identified with radio sources are known and we extracted them from NED and
Sloan Digital Sky Survey (SDSS). Consequently, we were able to impose the low power
criterion. The limit on flux density was chosen in order to produce a sample
of manageable size, but also to exclude objects with flux densities too low 
to be detected in snapshot observations.
Eventually, the selection criteria were as follows:

\begin{itemize}
\item {Low luminosity
criterion: $ L_{1.4{\rm GHz}} < 10^{26}$\,${\rm W~Hz^{-1}}$.} (for ${\rm
H_0}$=100${\rm\,km\,s^{-1}\,Mpc^{-1}}$, $q_{0}$=0.5; in this paper we use another
cosmology);
\item{Flux density criterion:
70\,mJy $\leq S_{1.4{\rm GHz}} \leq$ 1\,Jy;}
\item {Spectral index criterion: $\alpha_{1.4 {\rm GHz}}^{4.85
{\rm GHz}} >0.7$
($S\propto\nu^{-\alpha}$).}
\end{itemize}

\begin{table*}
\begin{center}
\caption[]{Basic parameters of 44 target sources}
\begin{tabular}{@{}c c c c c c c c c c c c@{}}
\hline
~~~Source & RA & Dec & ID& 
\multicolumn{1}{c}{\it z}&
\multicolumn{1}{c}{${\rm S_{1.4}}$}&
\multicolumn{1}{c}{log${\rm L_{1.4}}$}&
\multicolumn{1}{c}{${\rm S_{4.85}}$}&
\multicolumn{1}{c}{log${\rm L_{4.85}}$}&
\multicolumn{1}{c}{$\alpha_{1.4}^{4.85}$}&
\multicolumn{1}{c}{LLS}&
\multicolumn{1}{c}{Type}\\
~~~Name   & h~m~s & $\degr$~$\arcmin$~$\arcsec$ &  & &
\multicolumn{1}{c}{Jy}&
\multicolumn{1}{c}{$\rm W~Hz^{-1}$} &
\multicolumn{1}{c}{Jy}&
\multicolumn{1}{c}{$\rm W~Hz^{-1}$} & &
\multicolumn{1}{c}{$h^{-1}~{\rm kpc}$}&~~~ \\
~~~(1)& (2)& (3) &(4)&
\multicolumn{1}{c}{(5)}&
\multicolumn{1}{c}{(6)}&
\multicolumn{1}{c}{(7)}&
\multicolumn{1}{c}{(8)}&
\multicolumn{1}{c}{(9)}&
\multicolumn{1}{c}{(10)}&
\multicolumn{1}{c}{(11)}&
\multicolumn{1}{c}{(12)}\\
\hline
0025+006  & 00:28:33.42 & 00:55:11.00 & G & 0.104 &0.219&24.77&0.081&24.34&
            0.80&3.13&Cj? \\
0754+401  & 07:57:56.69 & 39:59:36.00 & G & 0.066 &0.098 &24.00&0.030&23.49&
            0.95&0.25&S\\
0801+437  & 08:04:54.91 & 43:35:37.20 & G & 0.123*&0.352&25.13&0.135&24.72&
            0.77&0.44&D?\\
0810+077  & 08:13:23.76 & 07:34:05.80 & Q & 0.112 &0.435&25.14&0.158&24.70&
            0.82&2.78&D?\\
0821+321  & 08:25:04.55 & 31:59:57.30 & G & 0.265 &0.076&25.21&0.030&24.81&
            0.75&7.69&D\\
0835+373  & 08:38:25.00 & 37:10:36.90 & Q & 0.396 &0.383&26.32&0.148&25.91&
            0.76&1.01&S\\
0846+017  & 08:48:56.56 & 01:36:47.40 & G & 0.349 &0.085&25.54&0.034& 25.14&
            0.74&6.12&Cj\\
0850+024  & 08:53:14.23 & 02:14:53.70 & G & 0.460 &0.112&25.94&0.034&25.42&
            0.96&5.98&D\\
0851+024  & 08:54:08.45 & 02:13:15.80 & G & 0.399 &0.118&25.82&0.039&25.33&
            0.89&3.41&D\\
0854+210  & 08:57:20.98 & 20:48:53.80 & G & 0.032 &0.082&23.27&0.033&22.87&
            0.76&1.76&O\\
0907+049  & 09:09:51.13 & 04:44:22.13 & G & 0.640p&0.178&26.49&0.049&25.93
            &1.04&8.32&D?\\
0914+114  & 09:17:16.38 & 11:13:36.90 & G & 0.178*&0.742&25.81&0.124&25.03&
            1.44&0.68&O\\
0914+504  & 09:17:34.82 & 50:16:38.20 & Q & 0.633 &0.104&26.24&0.039&25.82&
            0.79&4.86&D\\
0921+143  & 09:24:05.29 & 14:10:21.60 & Q & 0.135 &0.105&24.69&0.031&24.17&
            0.98&0.73&Cj\\
0923+079  & 09:26:07.99 & 07:45:26.60 & Q & 0.442 &0.158&26.05&0.056&25.60&
            0.83&7.94&O\\
0931+033  & 09:34:30.74 & 03:05:45.50 & G & 0.225 &0.280&25.61&0.119&25.24&
            0.69&1.61&Cj\\
0942+355  & 09:45:25.89 & 35:21:03.50 & G & 0.208 &0.140&25.24&0.042&24.71&
            0.97&4.41&D\\
1007+142  & 10:09:55.51 & 14:01:54.30 & Q & 0.213 &0.995&26.11&0.390&25.70&
            0.75&3.26&D\\
1009+053  & 10:12:04.73 & 05:06:13.20 & G & 0.460p&0.206&26.21&0.055&25.63
            &1.06&8.71&D?\\
1037+302  & 10:40:29.96 & 29:57:58.00 & G & 0.091 &0.364&24.87&0.107&24.33&
            0.98&3.63&D\\
1053+505  & 10:56:28.21 & 50:19:52.20 & Q & 0.820 &0.079&26.40&0.032&26.01&
            0.73&8.19&D\\
1140+058  & 11:43:11.03 & 05:35:15.90 & Q & 0.497 &0.194&26.26&0.061&25.76&
            0.93&16.97&Cj\\
1154+435  & 11:57:27.60 & 43:18:06.80 & Q & 0.230 &0.247&25.58&0.106&25.21&
            0.68&4.55&Cj\\
1156+470  & 11:59:19.99 & 46:45:44.80 & G & 0.467 &0.081&25.82&0.031&25.40&
            0.77&4.98&O\\
1308+451  & 13:10:57.00 & 44:51:46.60 & G & 0.391 &0.097&25.71&0.034&25.25&
            0.84&3.85&D?\\
1321+045  & 13:24:19.70 & 04:19:07.20 & G & 0.263 &0.128&25.43&0.046&24.98&
            0.82&16.90&D\\
1359+525  & 14:00:51.62 & 52:16:06.60 & G & 0.118 &0.170&24.78&0.064&24.35&
            0.78&0.42&S\\
1402+415  & 14:04:16.37 & 41:17:48.80 & G & 0.361 &0.210&25.96&0.073&25.50&
            0.85&1.00&S\\
1407+363  & 14:09:42.46 & 36:04:16.00 & G & 0.148 &0.141&24.91&0.045&24.41&
            0.92&0.07&Cj?\\
1411+553  & 14:13:27.21 & 55:05:29.30 & G & 0.282 &0.126&25.49&0.033&24.91&
            1.08&0.80&S\\
1418+053  & 14:21:04.25 & 05:08:44.90 & G & 0.455 &0.283&26.33&0.108&25.91&
            0.77&1.67&S\\
1506+345  & 15:08:05.68 & 34:23:23.30 & Gp & 0.045 &0.121&23.75&0.044&23.31&
            0.81&0.15&O\\
1521+324  & 15:23:49.35 & 32:13:50.10 & G & 0.110 &0.168&24.71&0.051&24.19&
            0.96&0.40&S\\
1532+303  & 15:34:09.90 & 30:12:04.00 & q & 0.001? &0.071&20.17&0.030&19.80&
            0.70&0.004&S\\
1542+390  & 15:43:49.49 & 38:56:01.40 & G & 0.553 &0.189&26.36&0.049&25.77&
            1.09&8.08&D\\
1543+465  & 15:45:25.46 & 46:22:44.70 & G & 0.400 &0.459&26.41&0.110&25.79&
            1.15&6.57&D?\\
1550+444  & 15:52:35.37 & 44:19:06.10 & G & 0.452 &0.139&26.02&0.049&25.56&
            0.84&6.90&O\\
1558+536  & 15:59:27.66 & 53:30:54.70 & G & 0.179 &0.170&25.17&0.058&24.71&
            0.86&5.08&D?\\
1601+528  & 16:02:46.39 & 52:43:58.70 & G & 0.106 &0.557&25.19&0.208&24.76&
            0.79&0.38&D\\
1610+407  & 16:11:48.55 & 40:40:20.90 & G & 0.151 &0.553&25.52&0.166&25.00&
            0.97&2.60&Cj?\\
1624+049  & 16:26:50.30 & 04:48:50.50 & G & 0.040p&0.162&23.77&0.056&23.31
            &0.85&1.94&D\\
1641+320  & 16:43:11.35 & 31:56:18.00 & Qb & 0.586 &0.113&26.20&0.041&25.76&
            0.82&10.56&O$\dagger$\\
1715+499  & 17:16:46.34 & 49:56:44.30 & G & 0.628 &0.095 &26.20&0.035&25.76&
            0.80&2.38&D\\
1717+547  & 17:18:54.40 & 54:41:48.20 & G & 0.147 &0.323&25.26&0.110&24.80&
            0.87&0.18&D\\
\hline
\end{tabular}
\end{center}

\begin{minipage}{165mm}
Description of the columns:
(1) source name;
(2) and (3) source coordinates (J2000) extracted from FIRST;
(4) optical identification: G - galaxy, Q - quasar, q - star-like object,
Gp - galaxy pair, Qb - binary quasar;
(5) redshift; 'p' means photometric redshift taken from literature or
SDSS; '?' means the spectroscopic redshift is very uncertain because of a poor
quality of the spectrum; $\ast$ means uncertain redshift probably belongs to
a random foreground galaxy;
(6) total flux density at 1.4\,GHz extracted from FIRST;
(7) log of the radio luminosity at 1.4\,GHz;
(8) total flux density at 4.85\,GHz extracted from GB6;
(9) log of the radio luminosity at 4.85\,GHz;
(10) spectral index between
1.4 and 4.85\,GHz calculated using flux densities in columns (6) and (8);
(11) largest linear size (LLS) calculated based on the largest angular size
measurements in the 1.4\,GHz MERLIN image -- in most cases, as a separation 
between the outermost component peaks, otherwise measured in the image contour
plot; in the case of single sources the linear size means the upper limit
estimated using deconvolved component major axis angular size from L-band
MERLIN image;
(12) radio morphology based on the available radio images, this is a simple
classification in agreement with the \citet{kun09}, more detailed description is in
section~\ref{obs}; S - single means unresolved or slightly resolved in available
radio image, Cj - core-jet, D - double-lobed, O - other, more complex
structures; '?' means the most uncertain classification.\\
$\dagger$ 1641+320 will be described in a separate paper.\\

\end{minipage}
\label{table1}
\end{table*}

\begin{figure*}
\centering
\includegraphics[width=7cm,height=7cm]{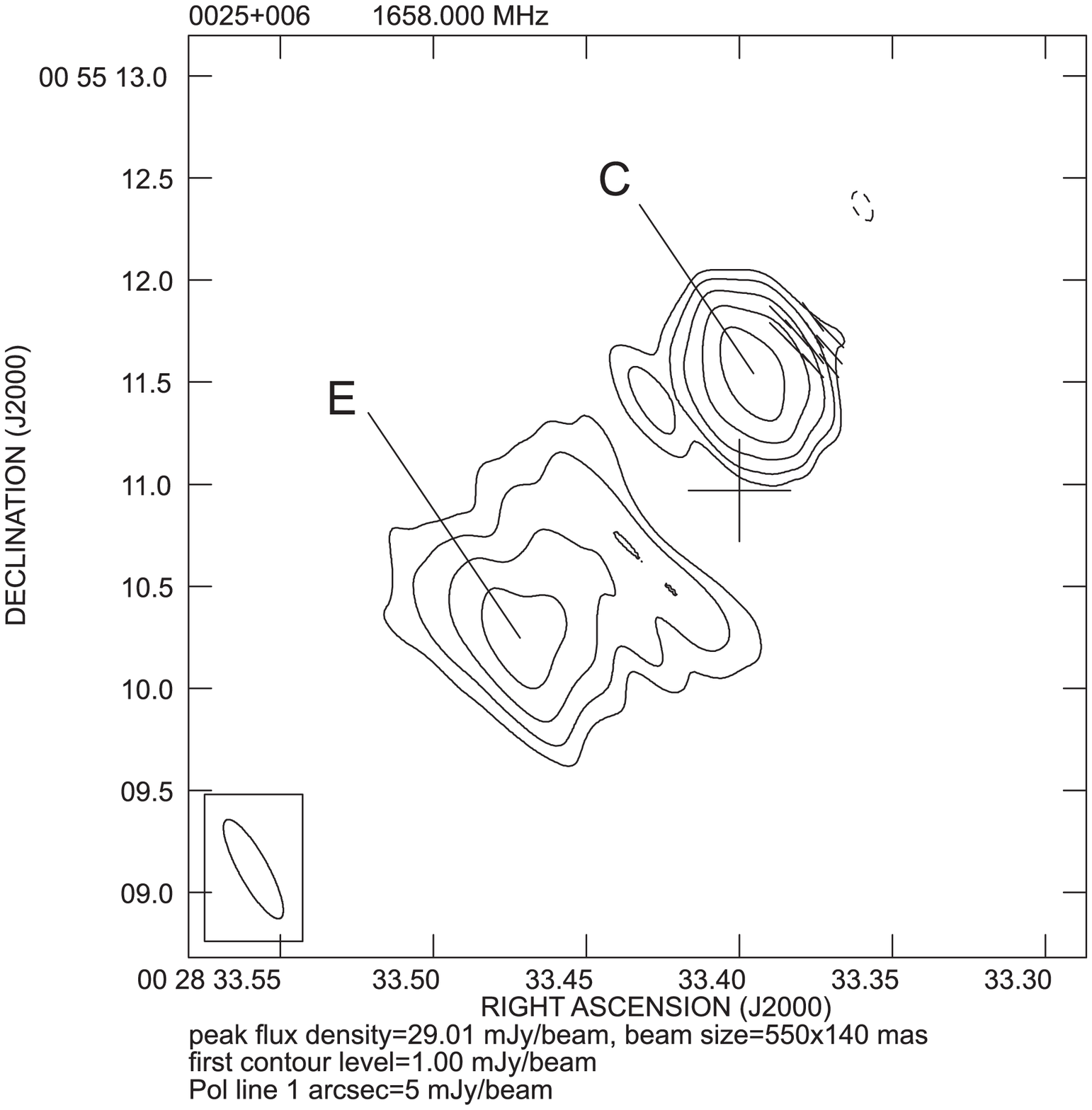}
\includegraphics[width=7cm,height=7cm]{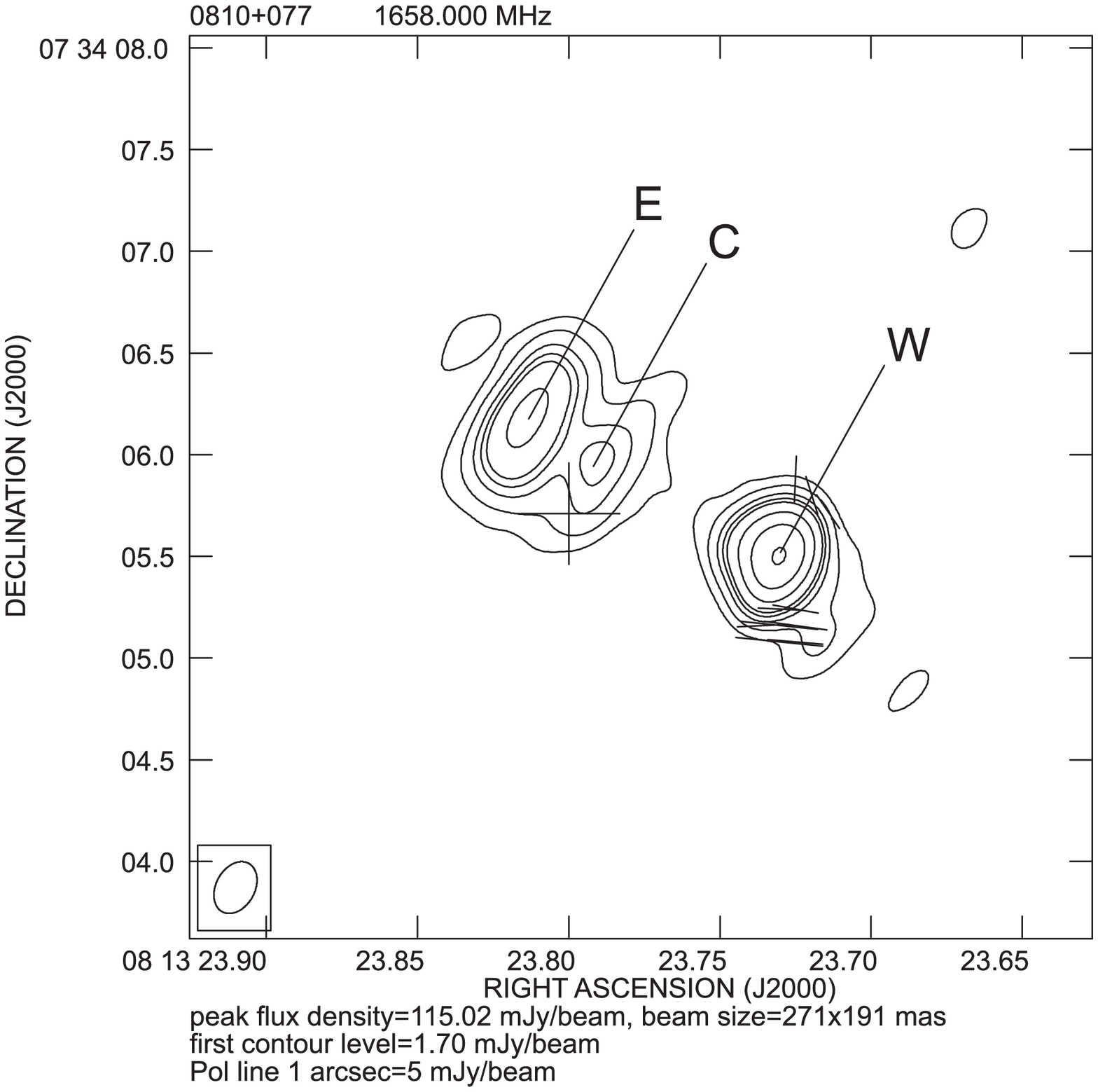}
\includegraphics[width=7cm,height=7cm]{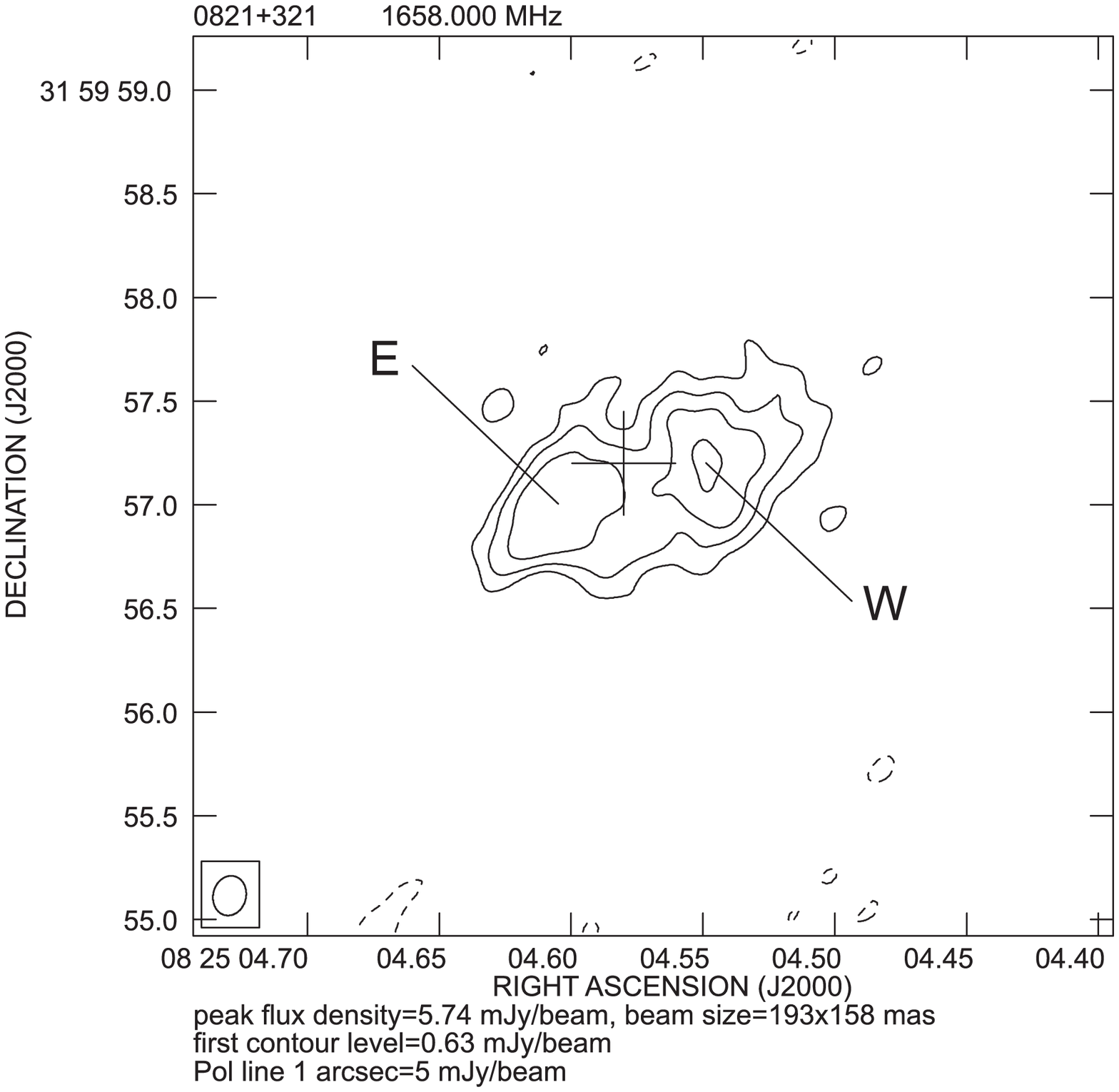}
\includegraphics[width=7cm,height=7cm]{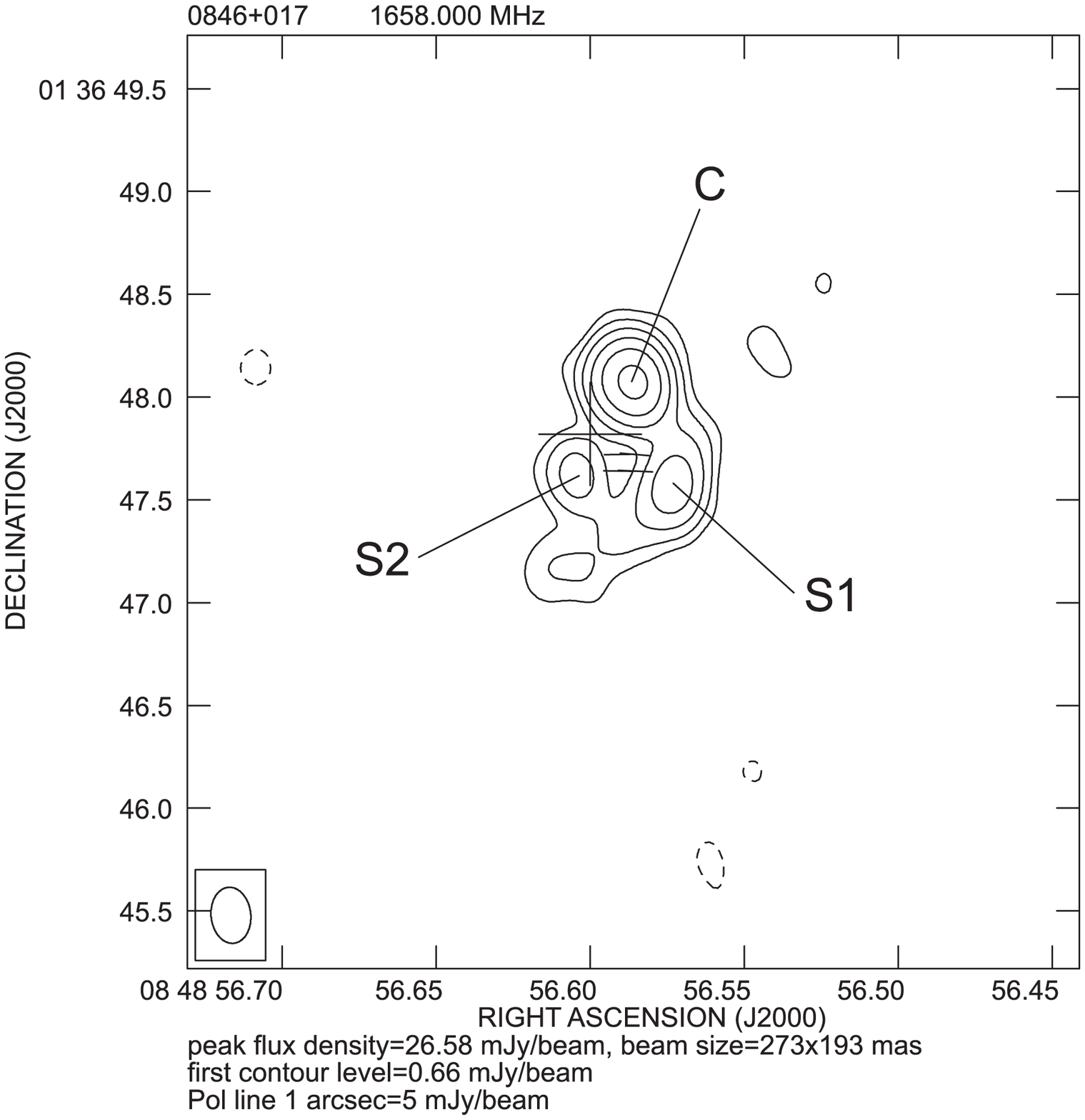}
\includegraphics[width=7cm,height=7cm]{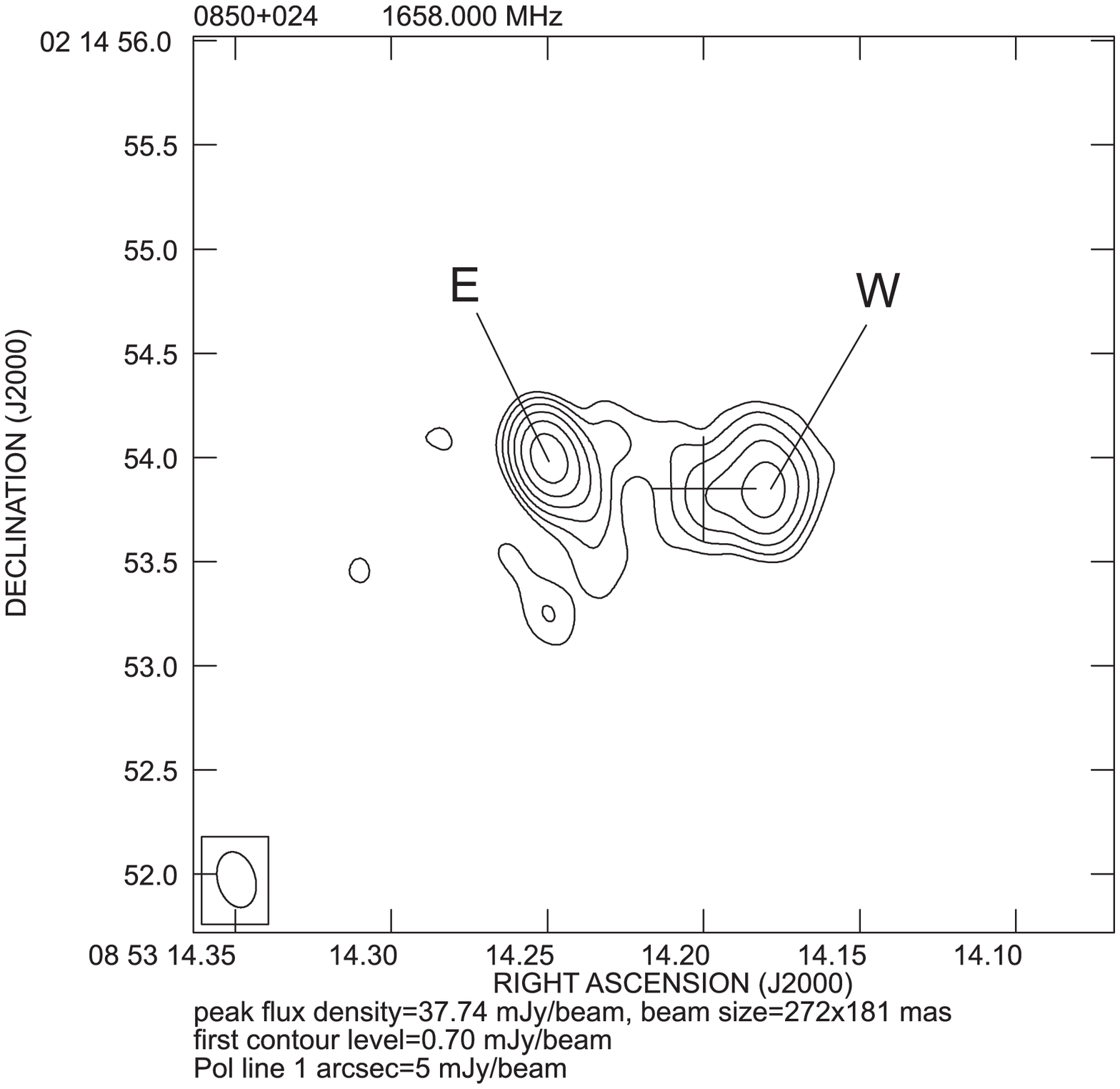}
\includegraphics[width=7cm,height=7cm]{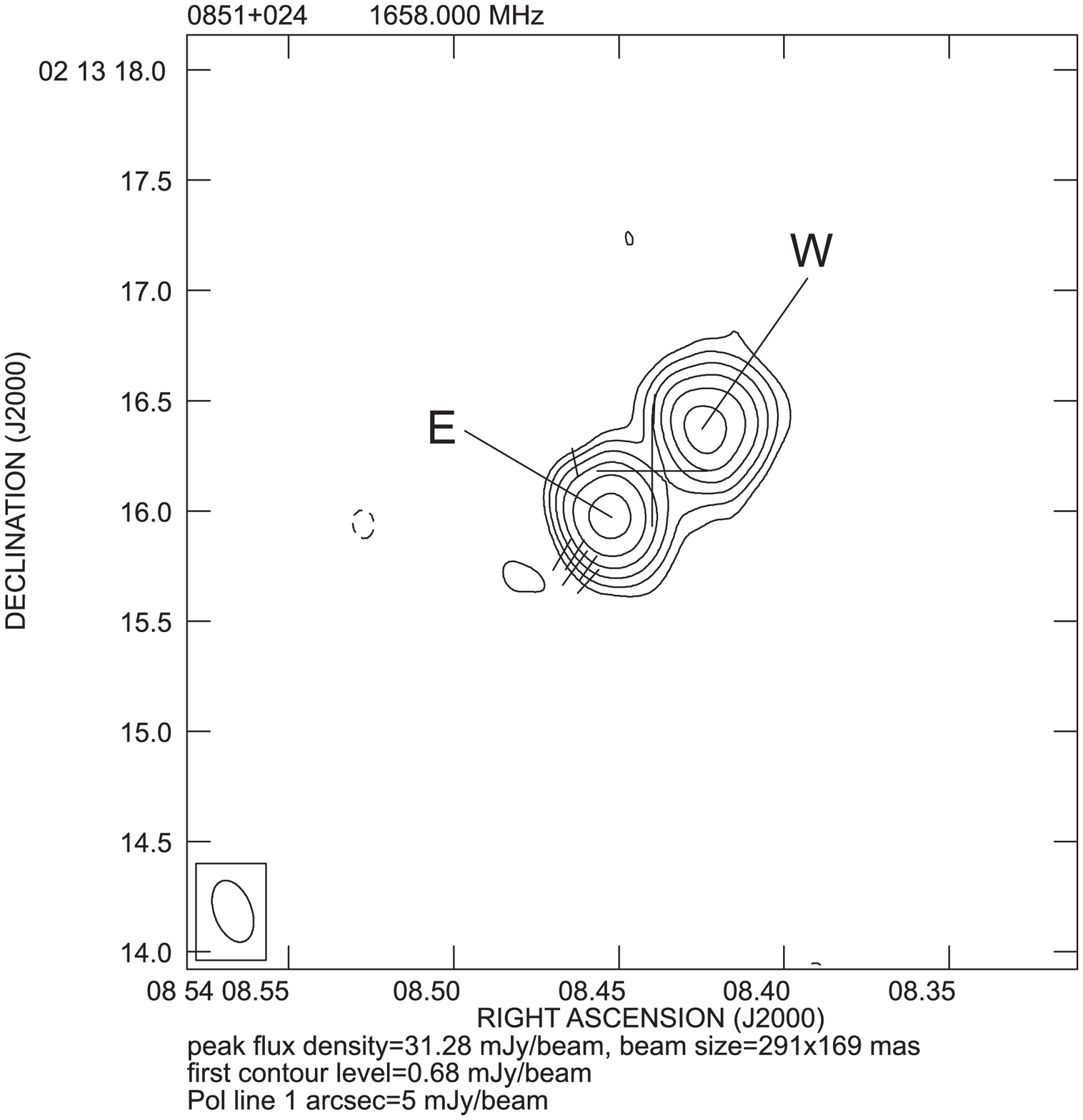}
\caption{MERLIN L-band images. Contours increase
by a factor 2, and the first contour level corresponds to $\approx
3\sigma$, vectors represent the polarized flux density.
A cross indicates the position of an optical object found using the
most actual version of SDSS/DR7.}
\label{18cm_images}
\end{figure*}

\setcounter{figure}{0}
\begin{figure*}
\centering
\includegraphics[width=7cm,height=7cm]{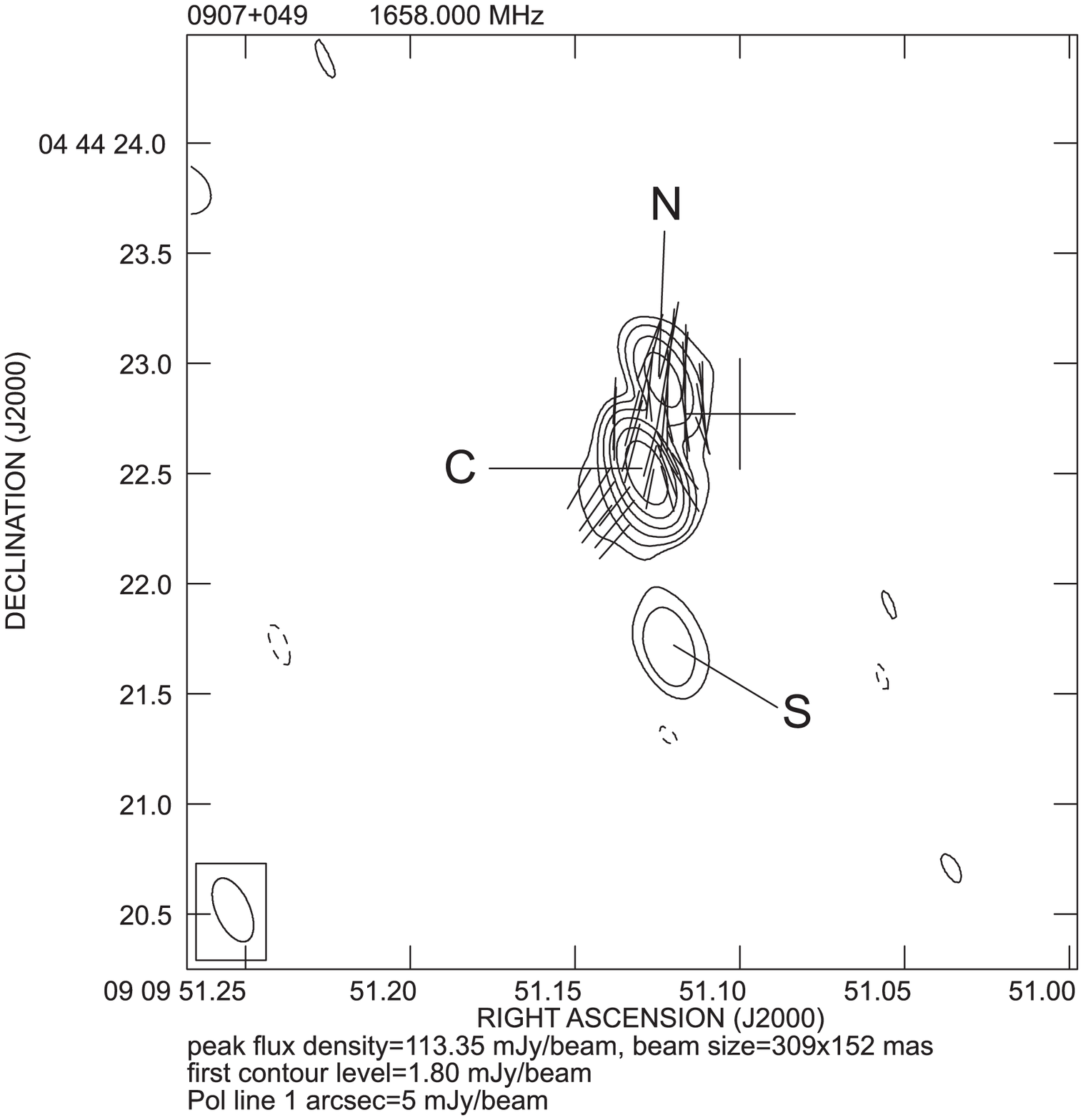}
\includegraphics[width=7cm,height=7cm]{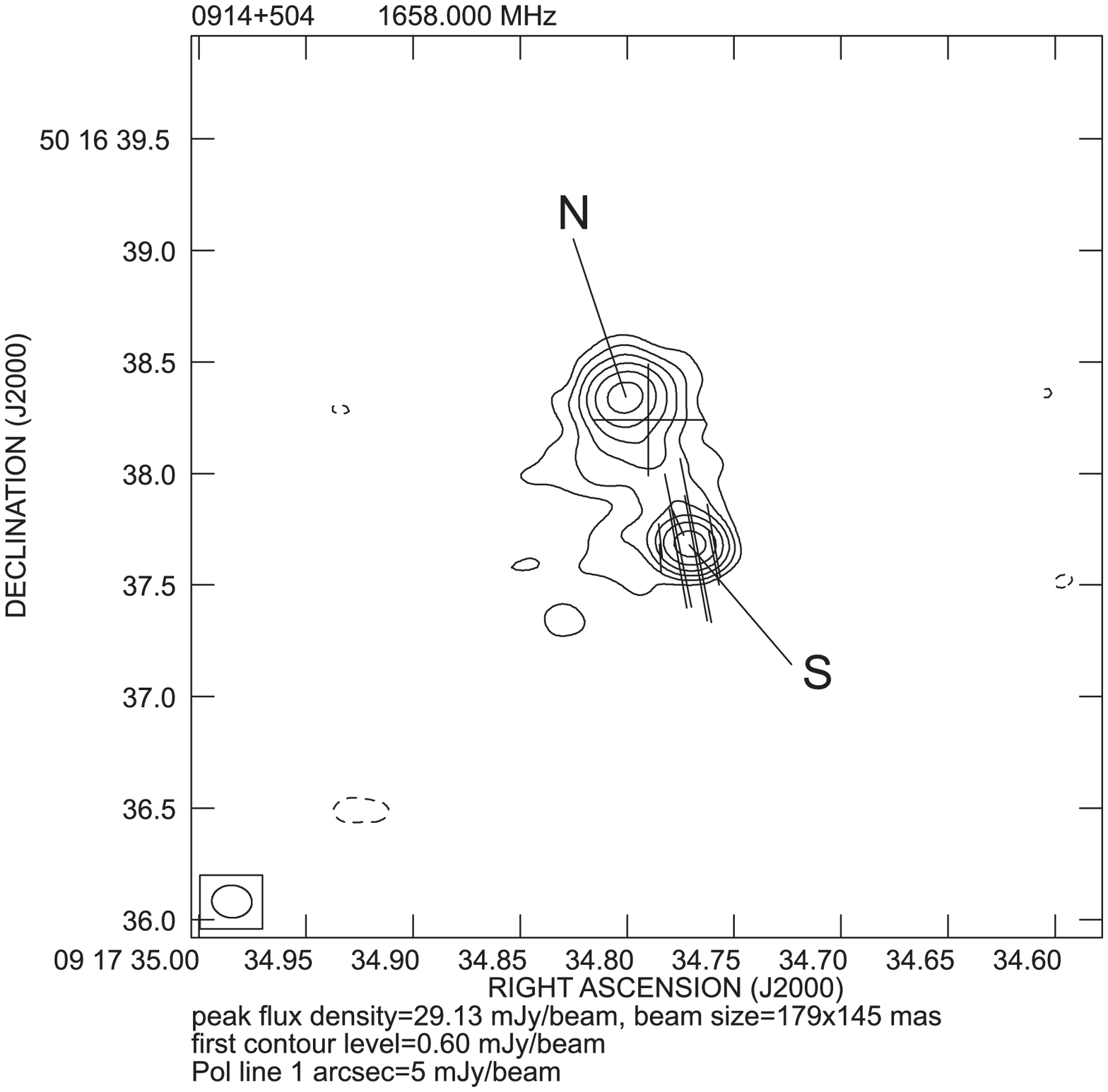}
\includegraphics[width=7cm,height=7cm]{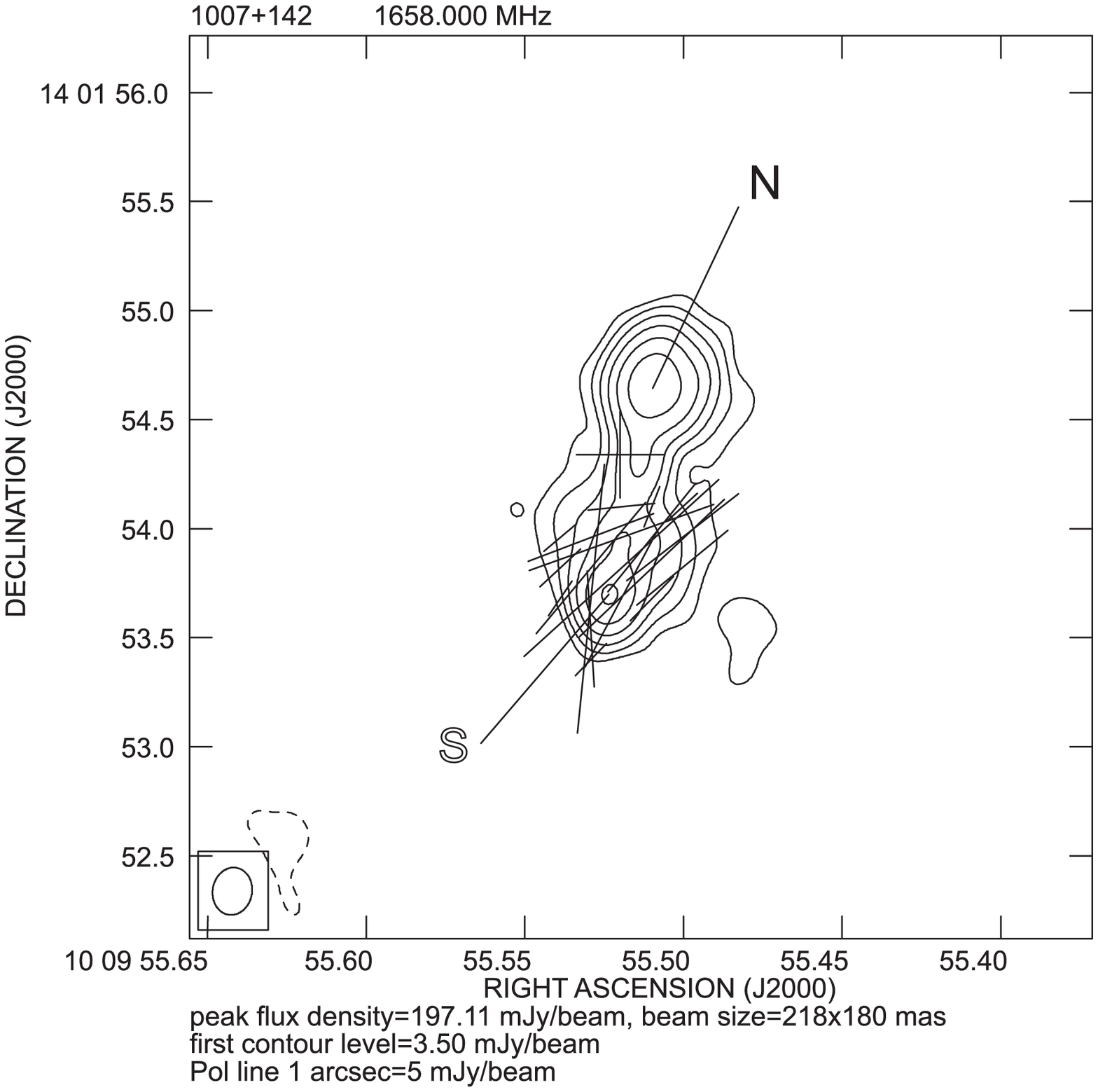}
\includegraphics[width=7cm,height=7cm]{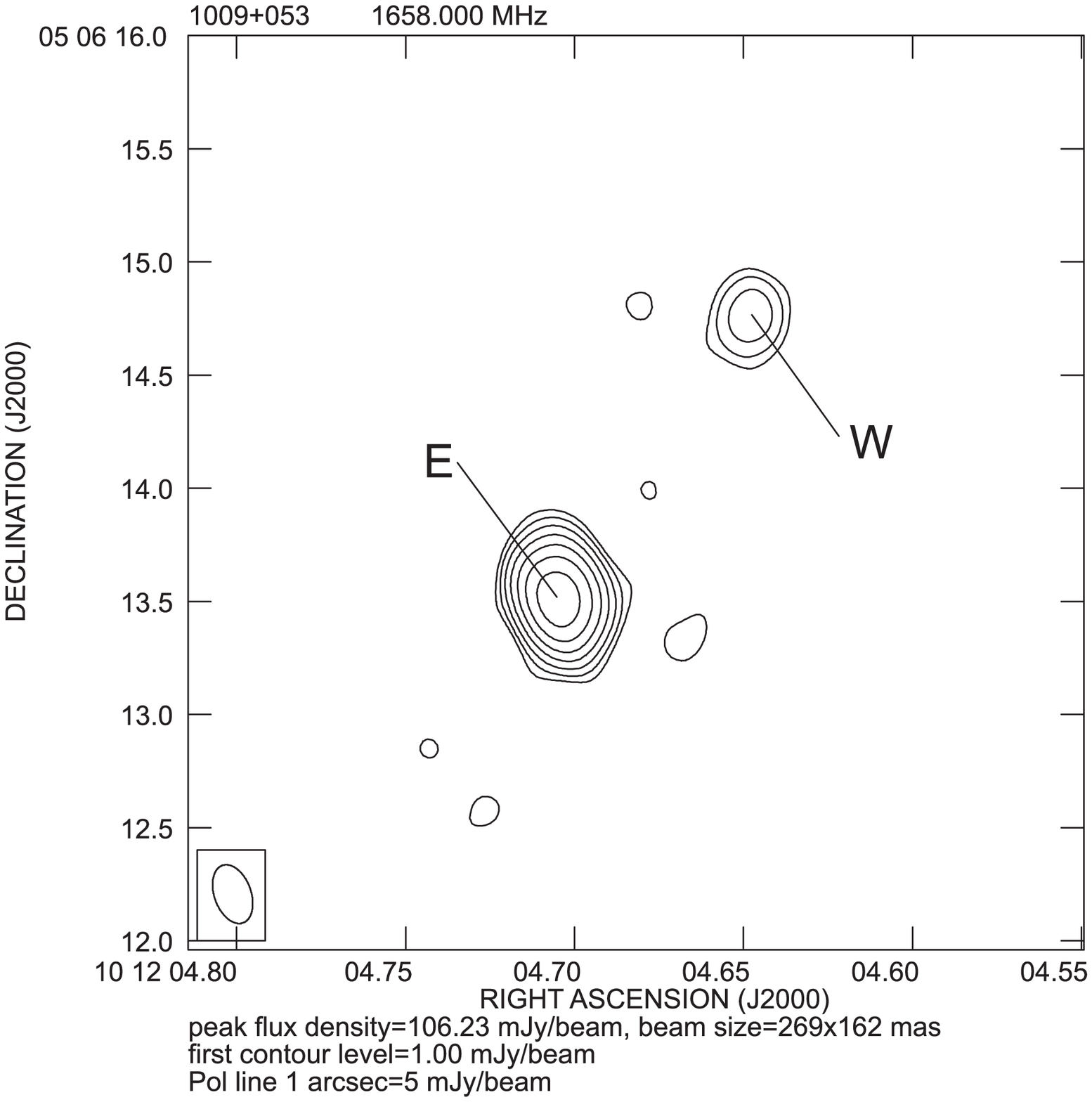}
\includegraphics[width=7cm,height=7cm]{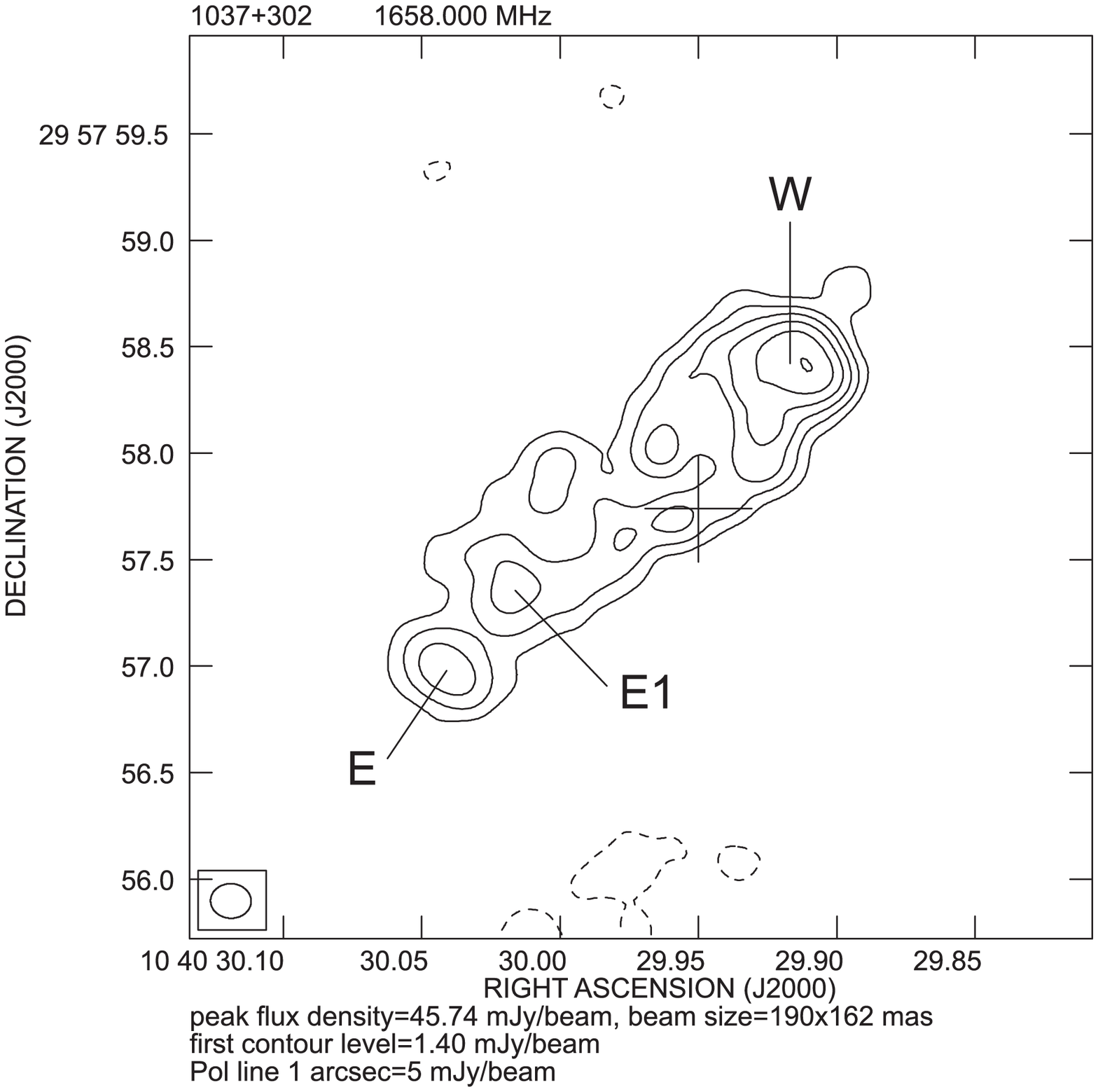}
\includegraphics[width=7cm,height=7cm]{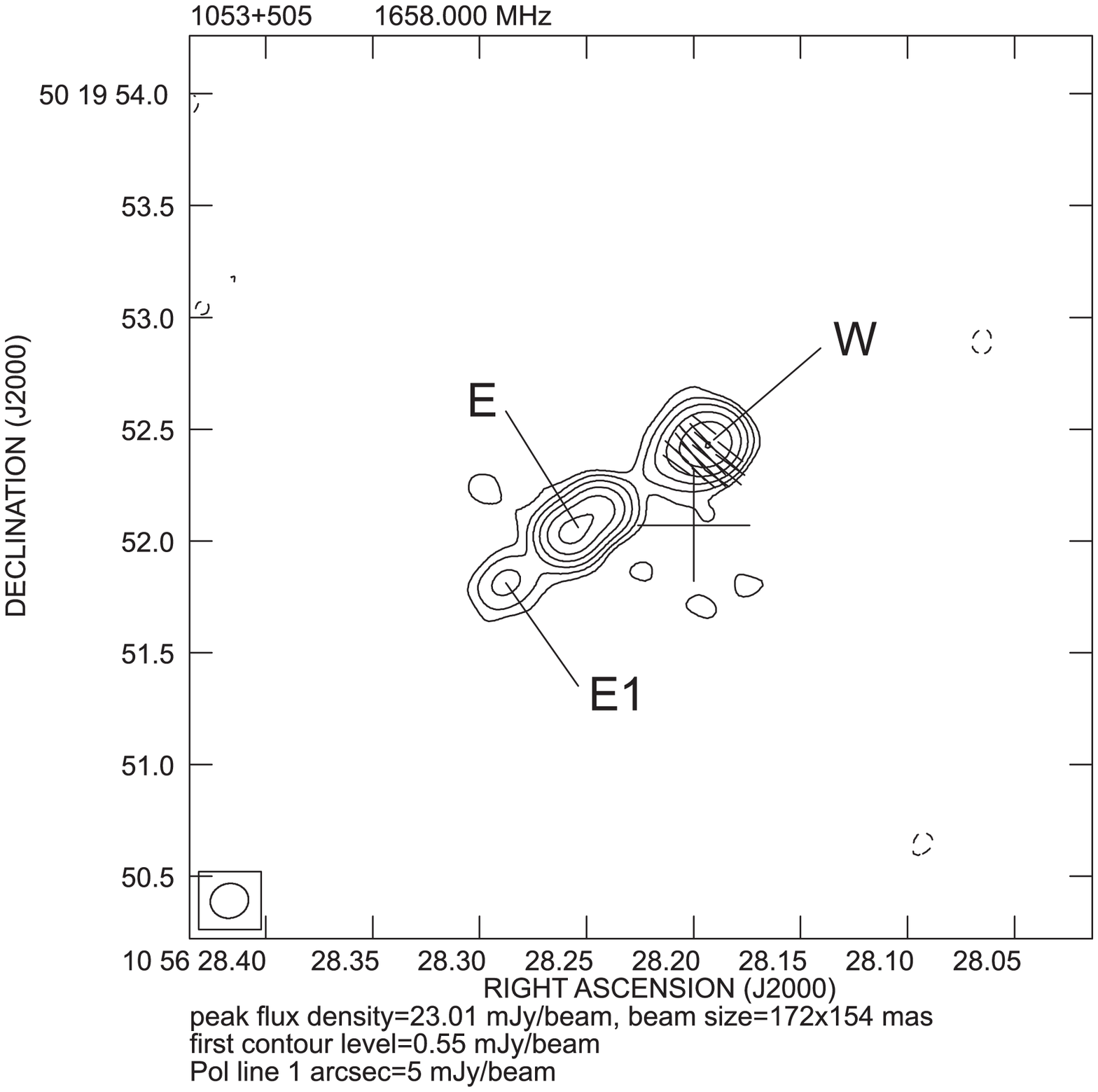}
\caption{MERLIN L-band (cont.)}
\end{figure*}

\setcounter{figure}{0}
\begin{figure*}
\centering
\includegraphics[width=7cm,height=7cm]{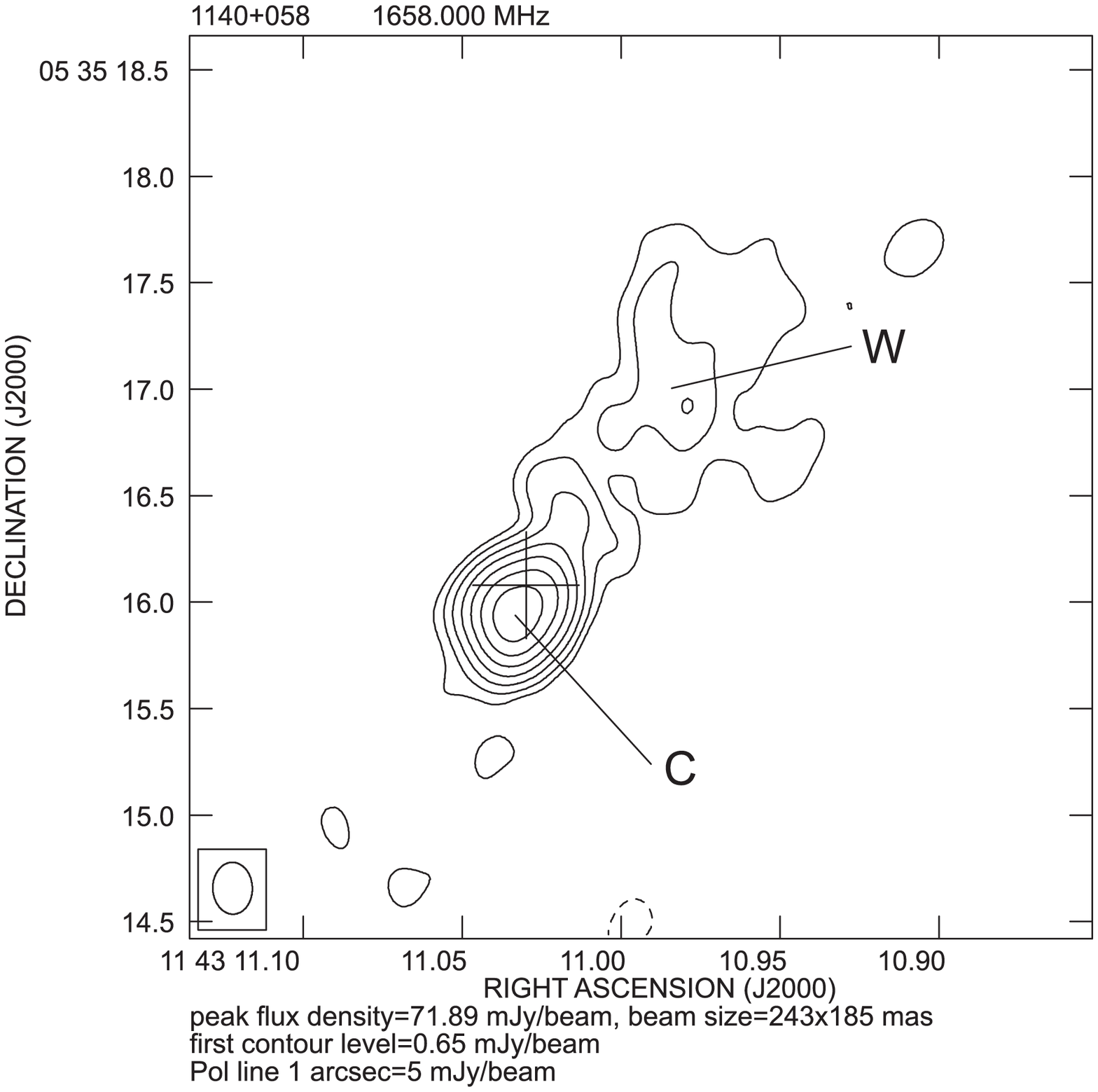}
\includegraphics[width=7cm,height=7cm]{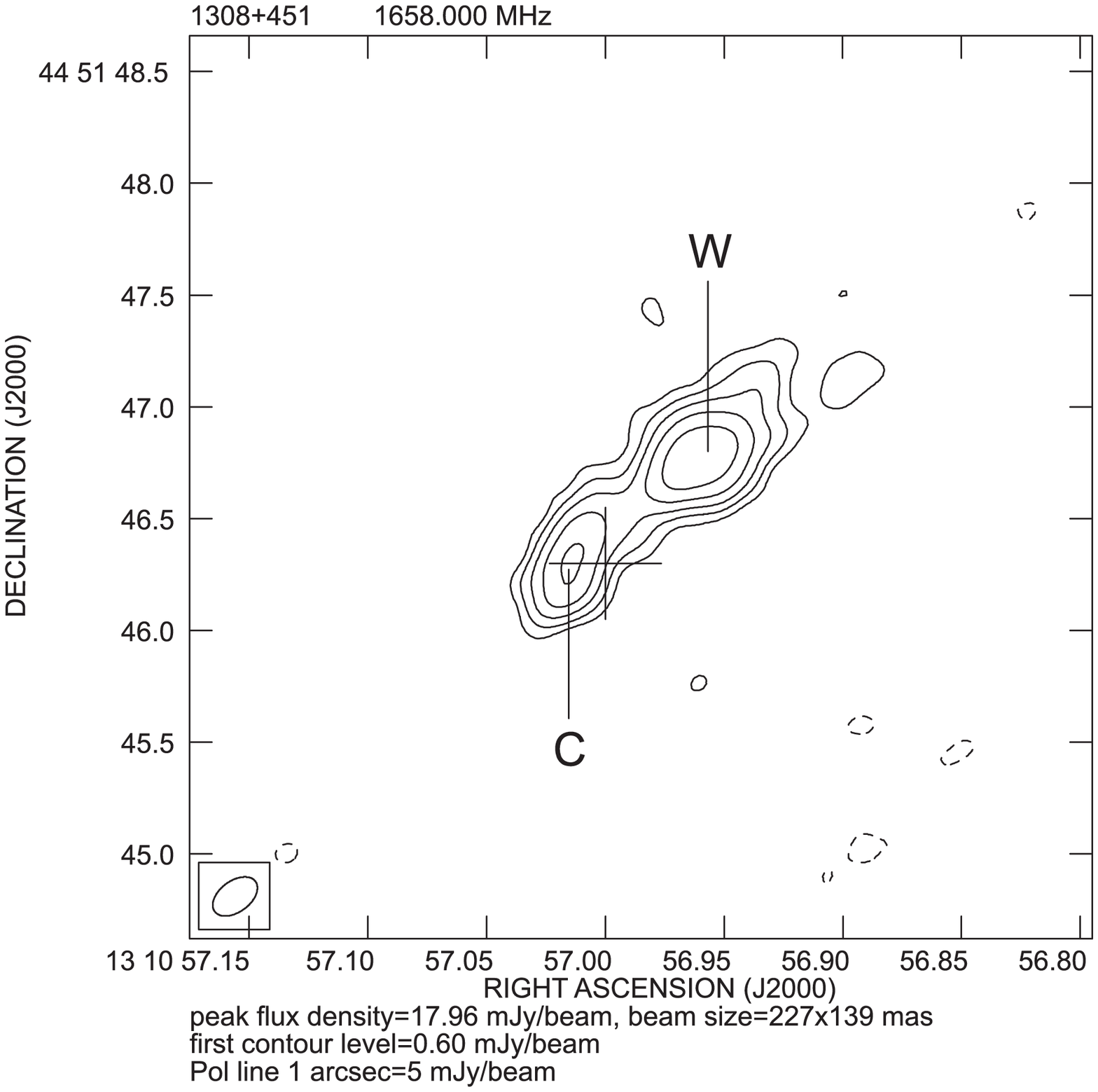}
\includegraphics[width=7.5cm,height=6.5cm]{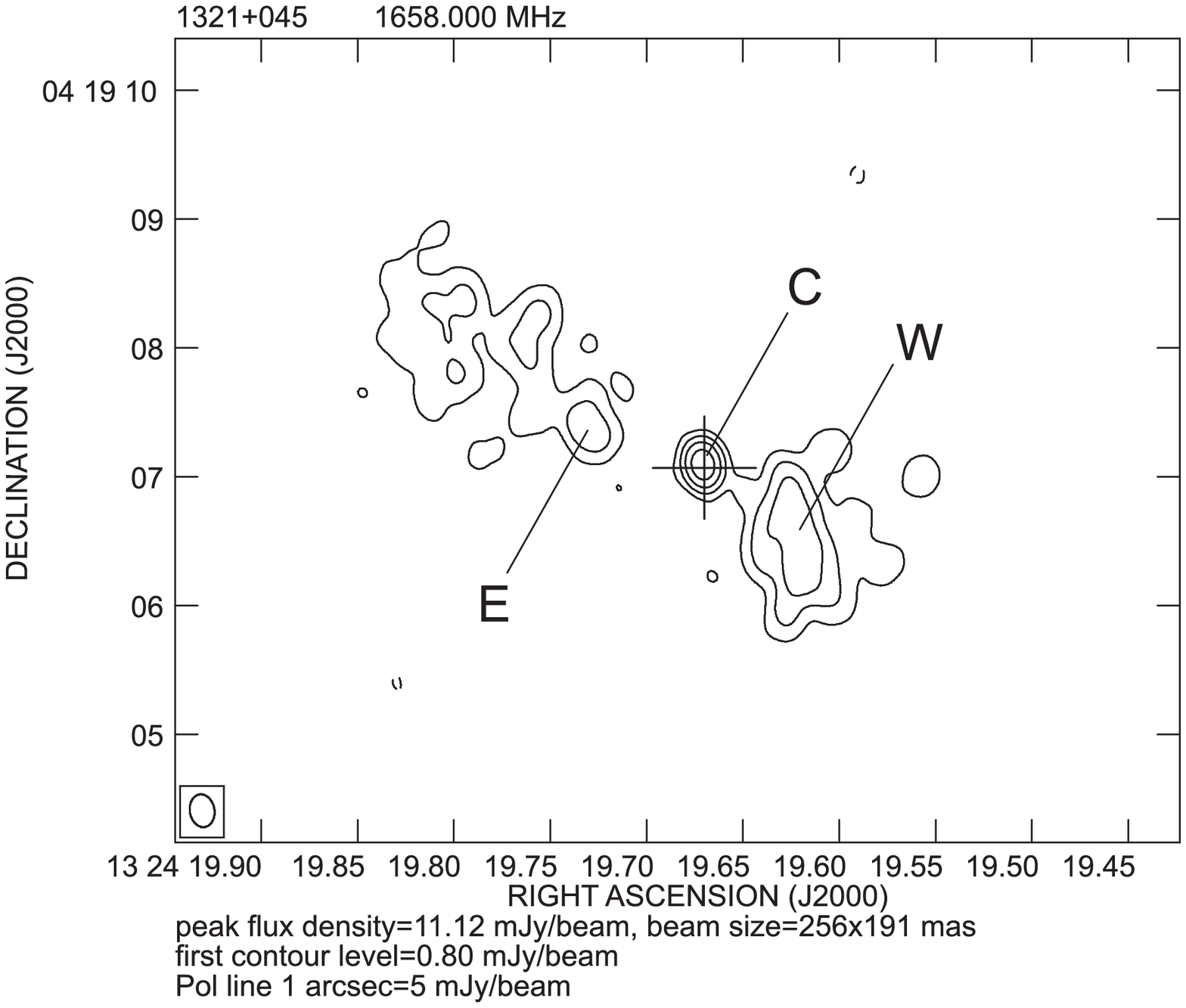}
\includegraphics[width=7cm,height=7cm]{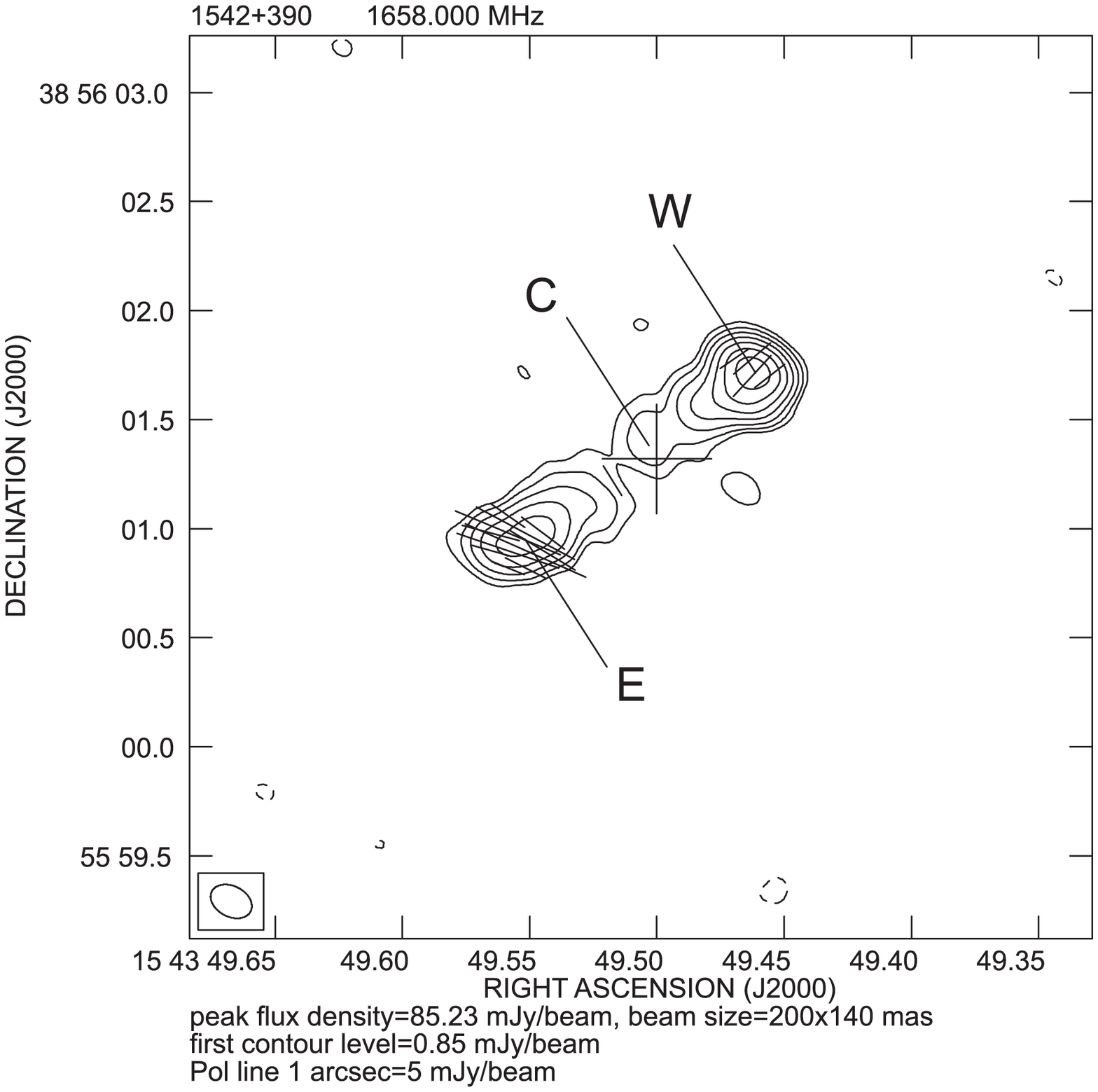}
\includegraphics[width=7cm,height=7cm]{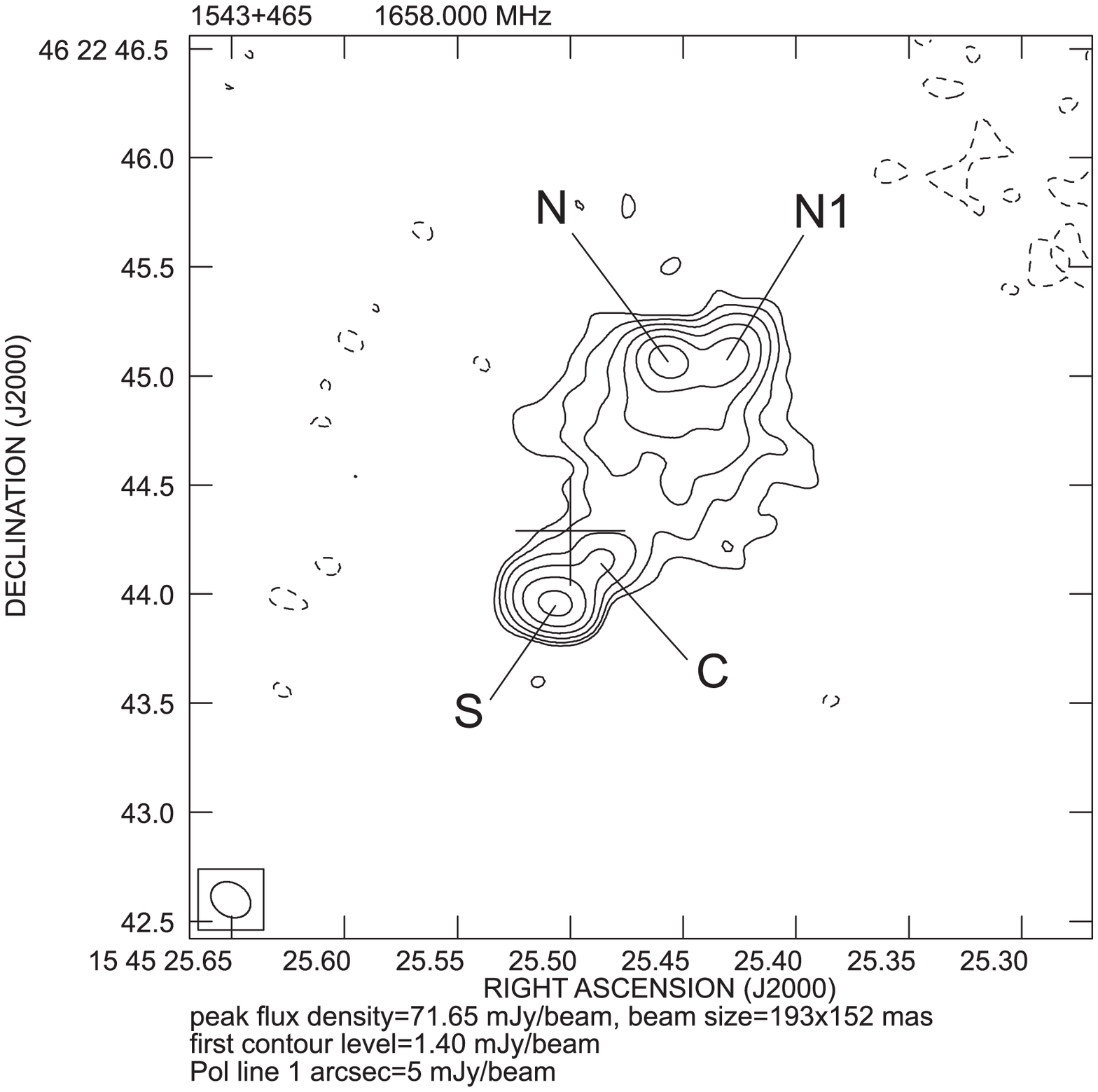}
\includegraphics[width=7cm,height=7cm]{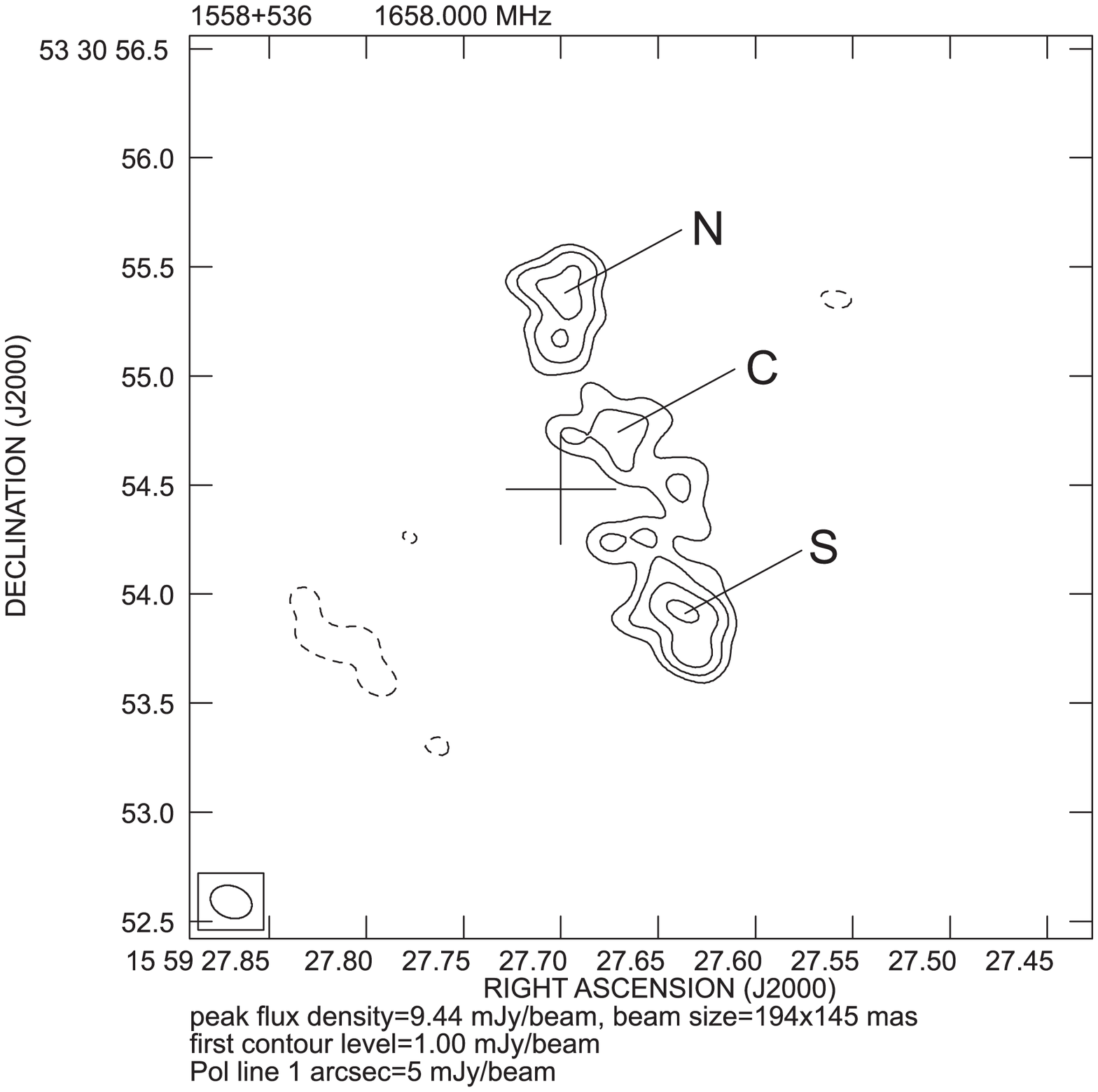}

\caption{MERLIN L-band (cont.)}
\end{figure*}

\setcounter{figure}{0}
\begin{figure*}
\centering
\includegraphics[width=7cm,height=7cm]{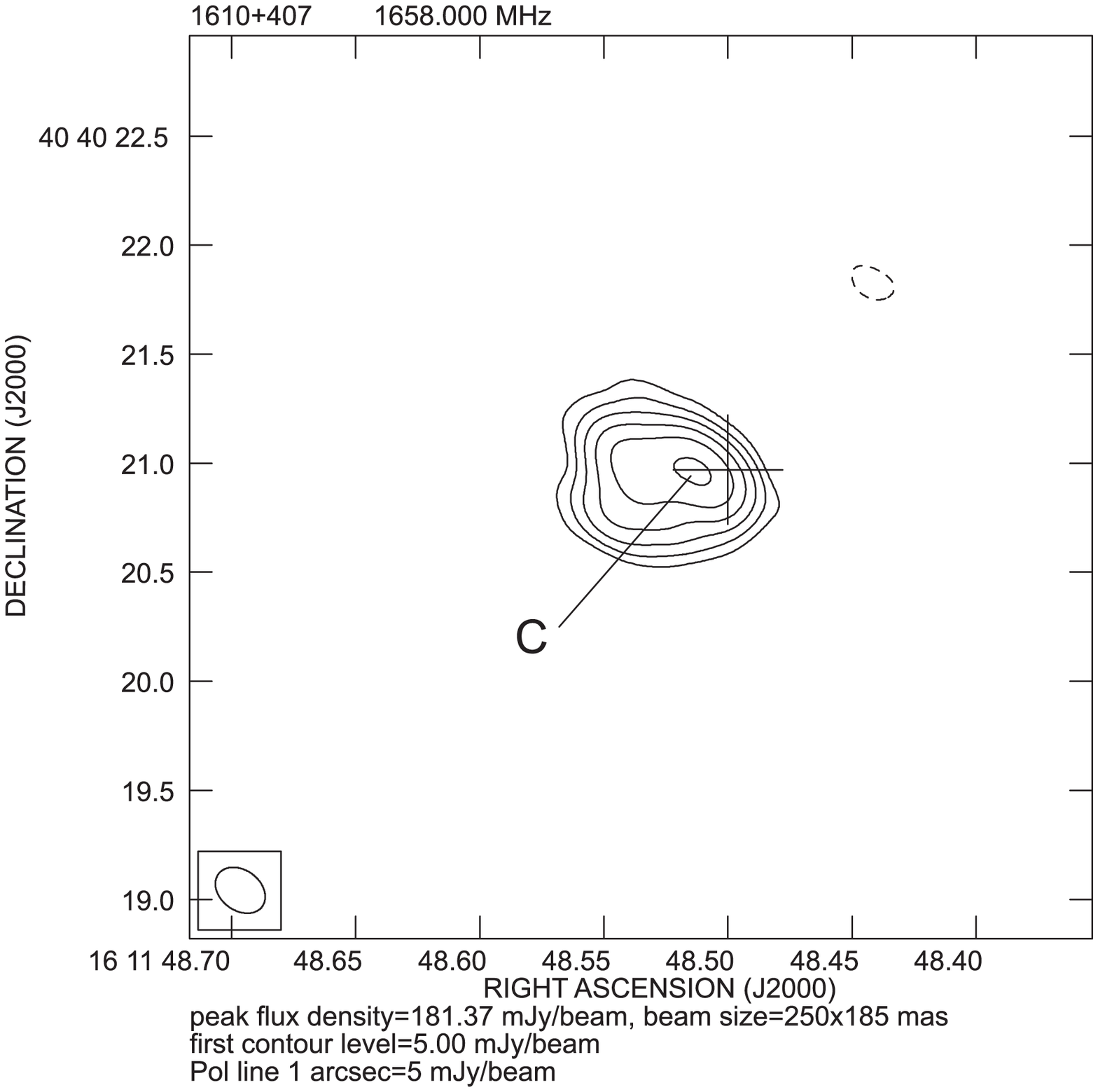}
\includegraphics[width=7cm,height=7cm]{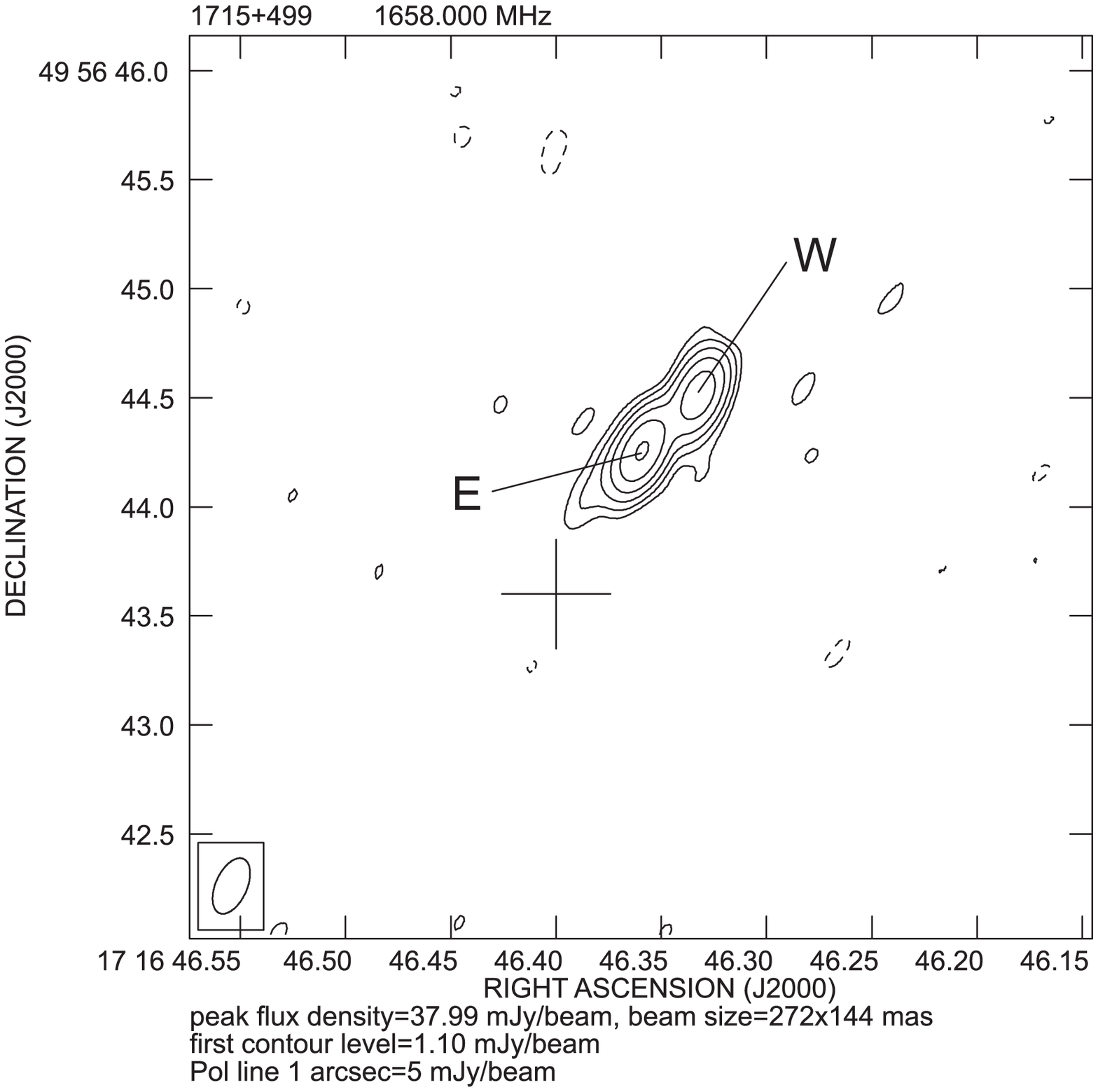}
\caption{MERLIN L-band (cont.)}
\end{figure*}

\begin{figure*}
\centering
\includegraphics[width=7cm,height=7cm]{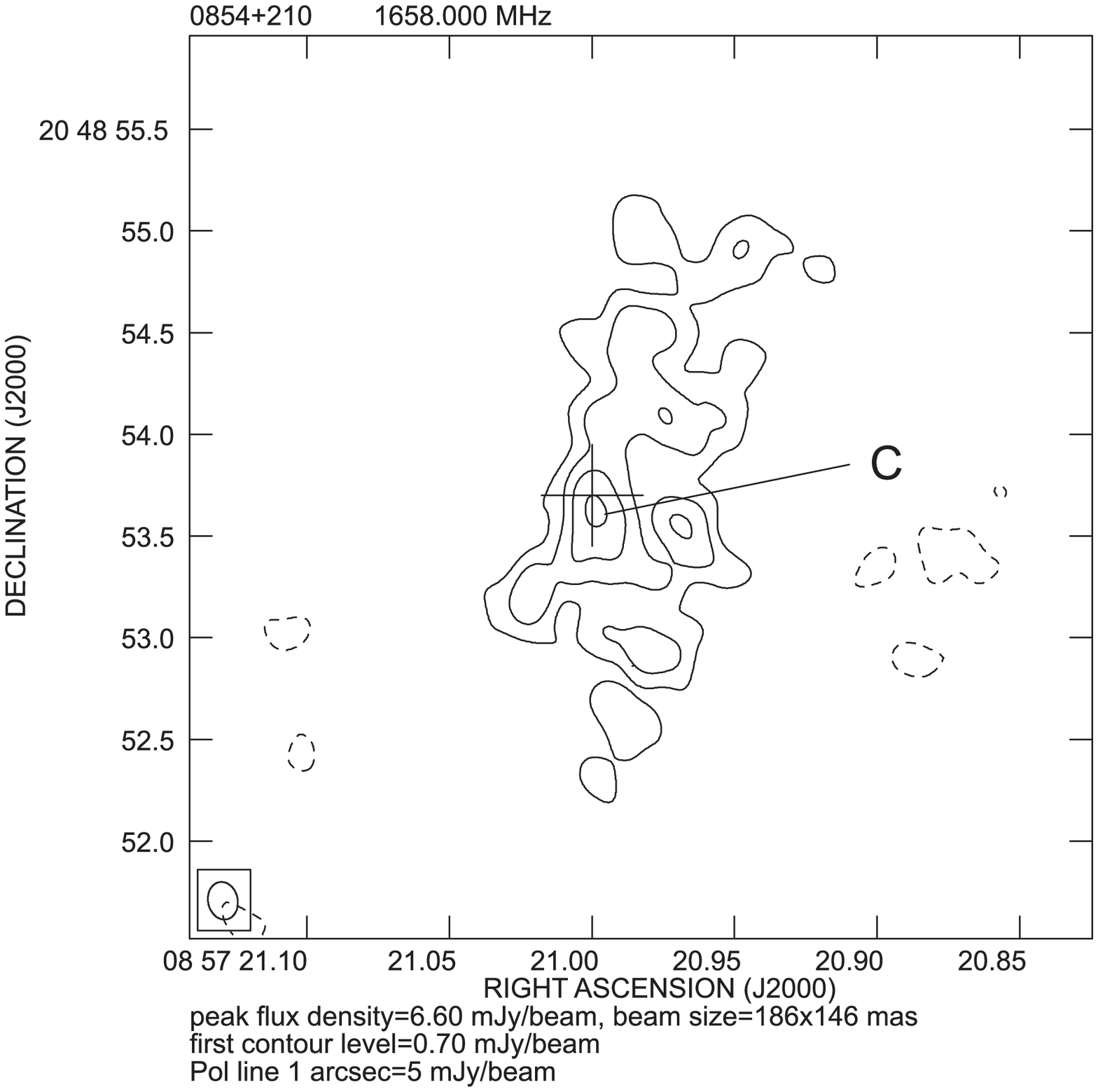}
\includegraphics[width=7cm,height=7cm]{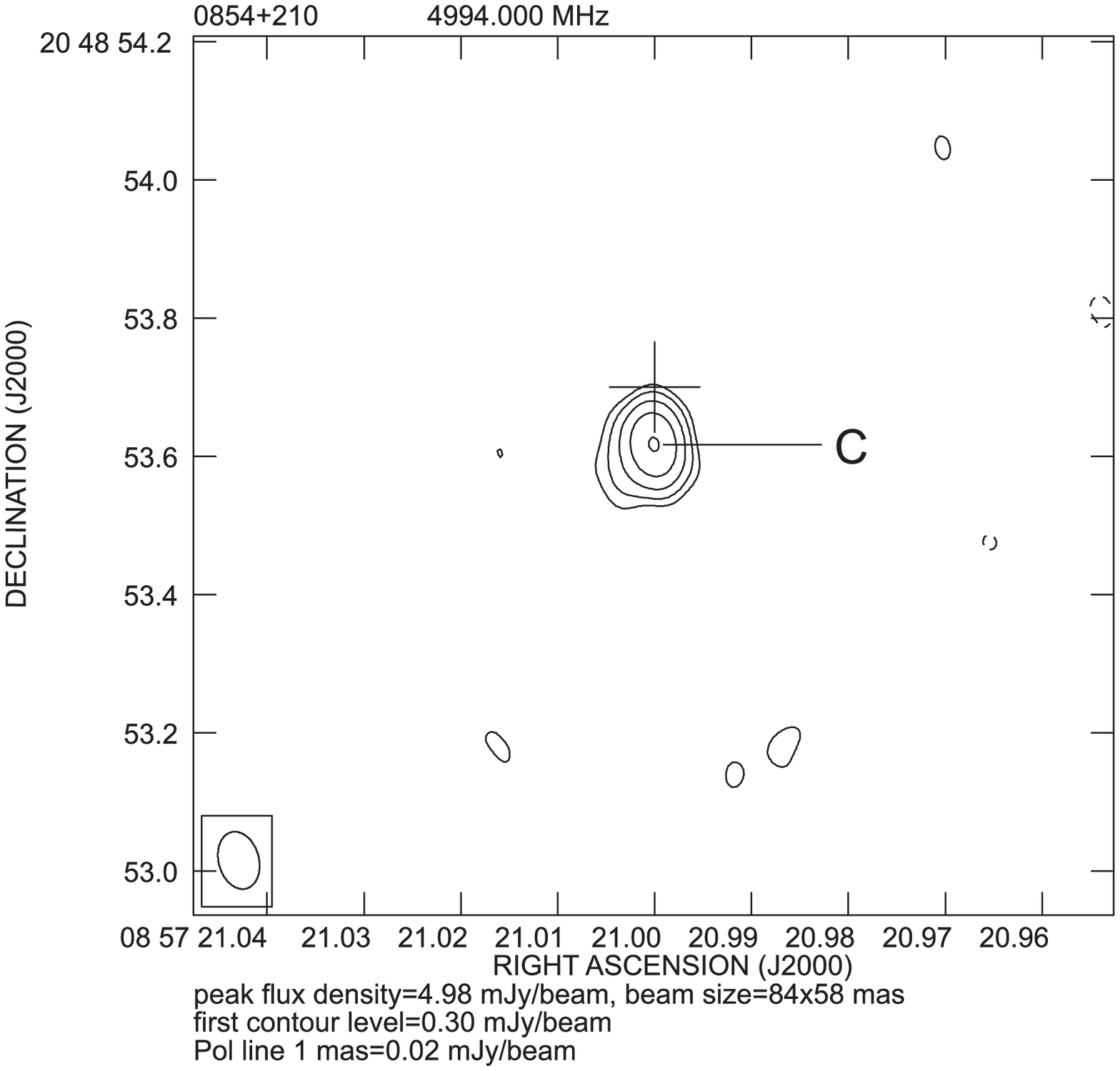}
\includegraphics[width=7cm,height=7cm]{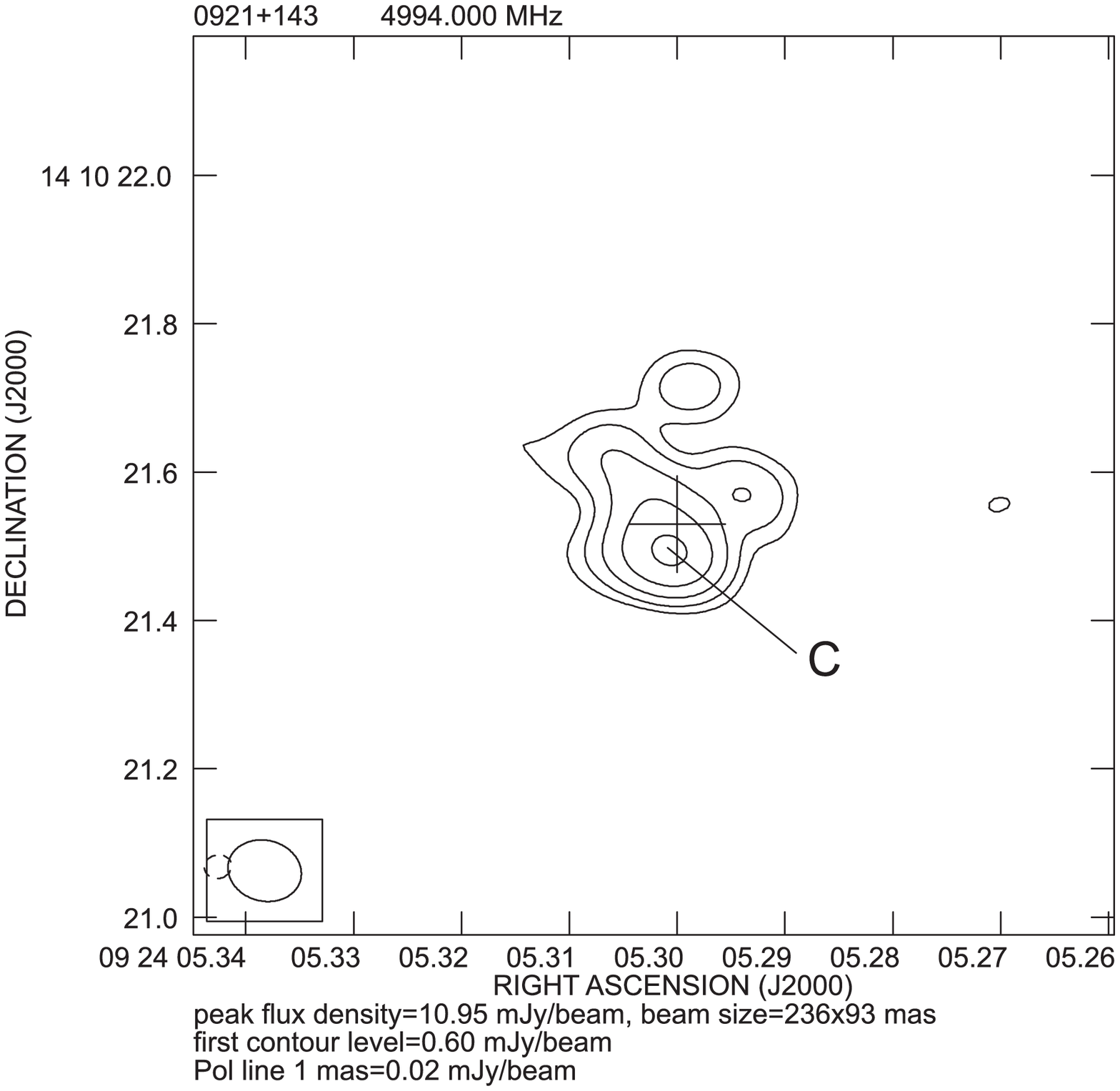}
\includegraphics[width=7cm,height=7cm]{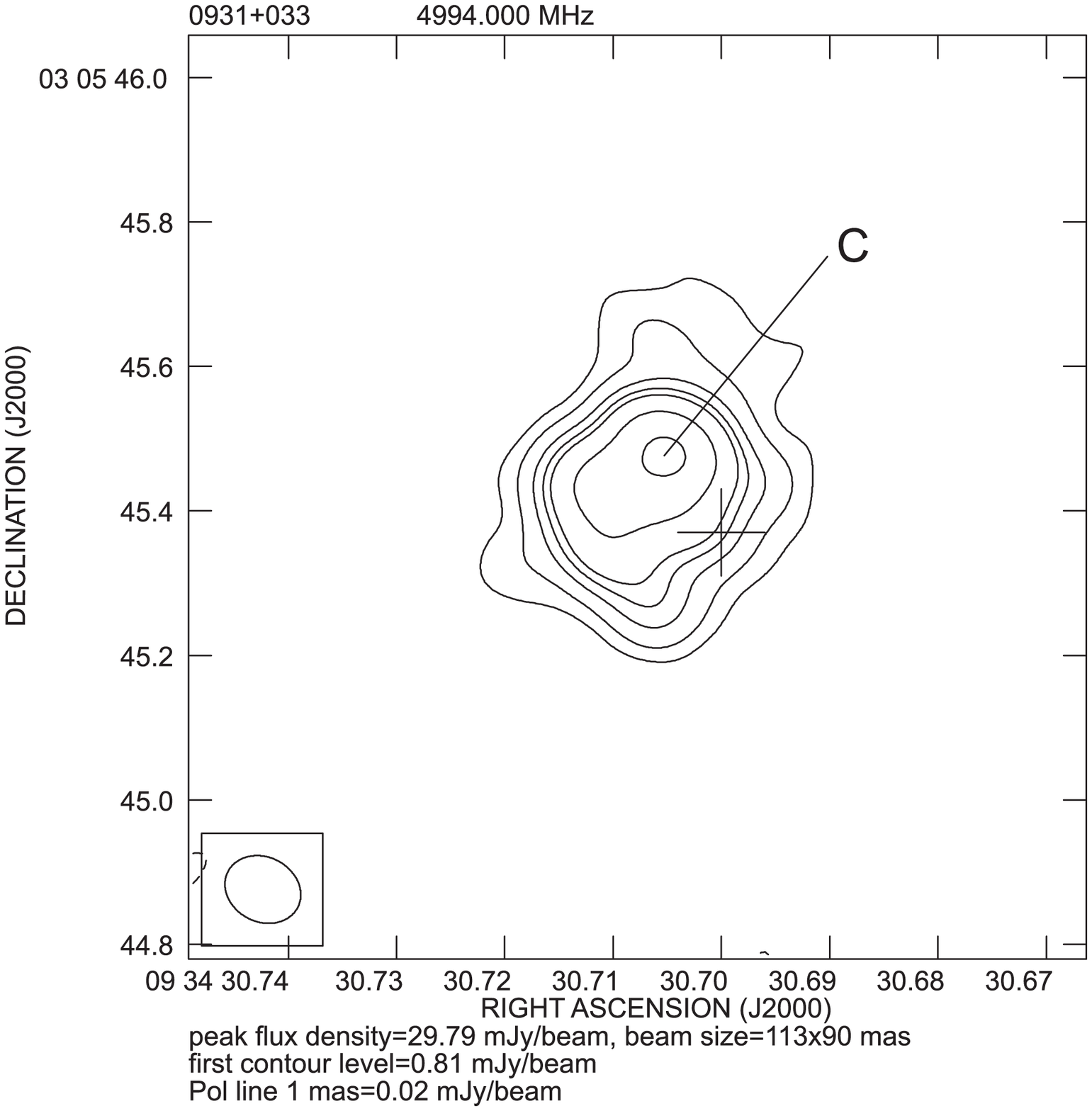}
\includegraphics[width=7cm,height=7cm]{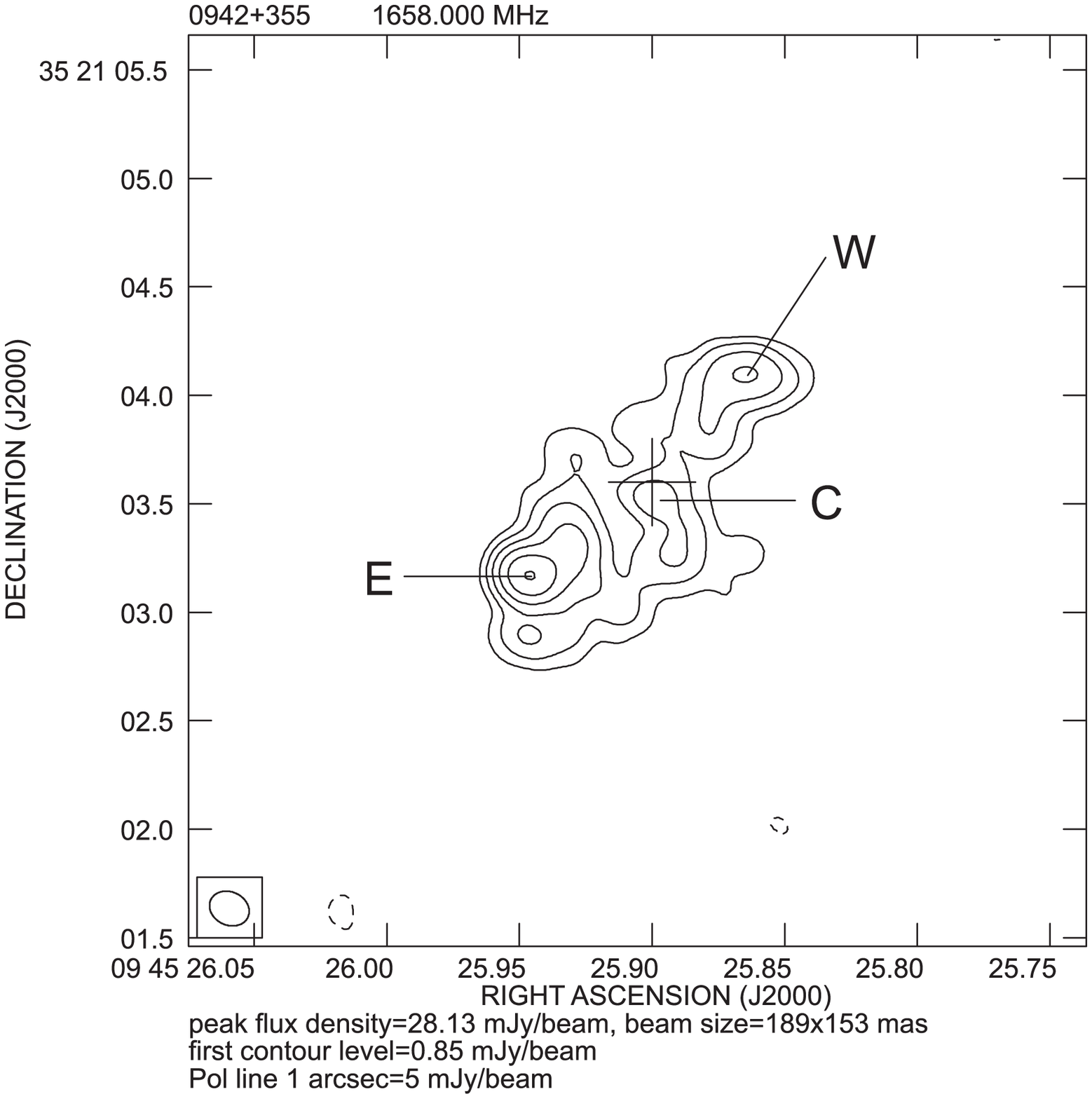}
\includegraphics[width=7cm,height=7cm]{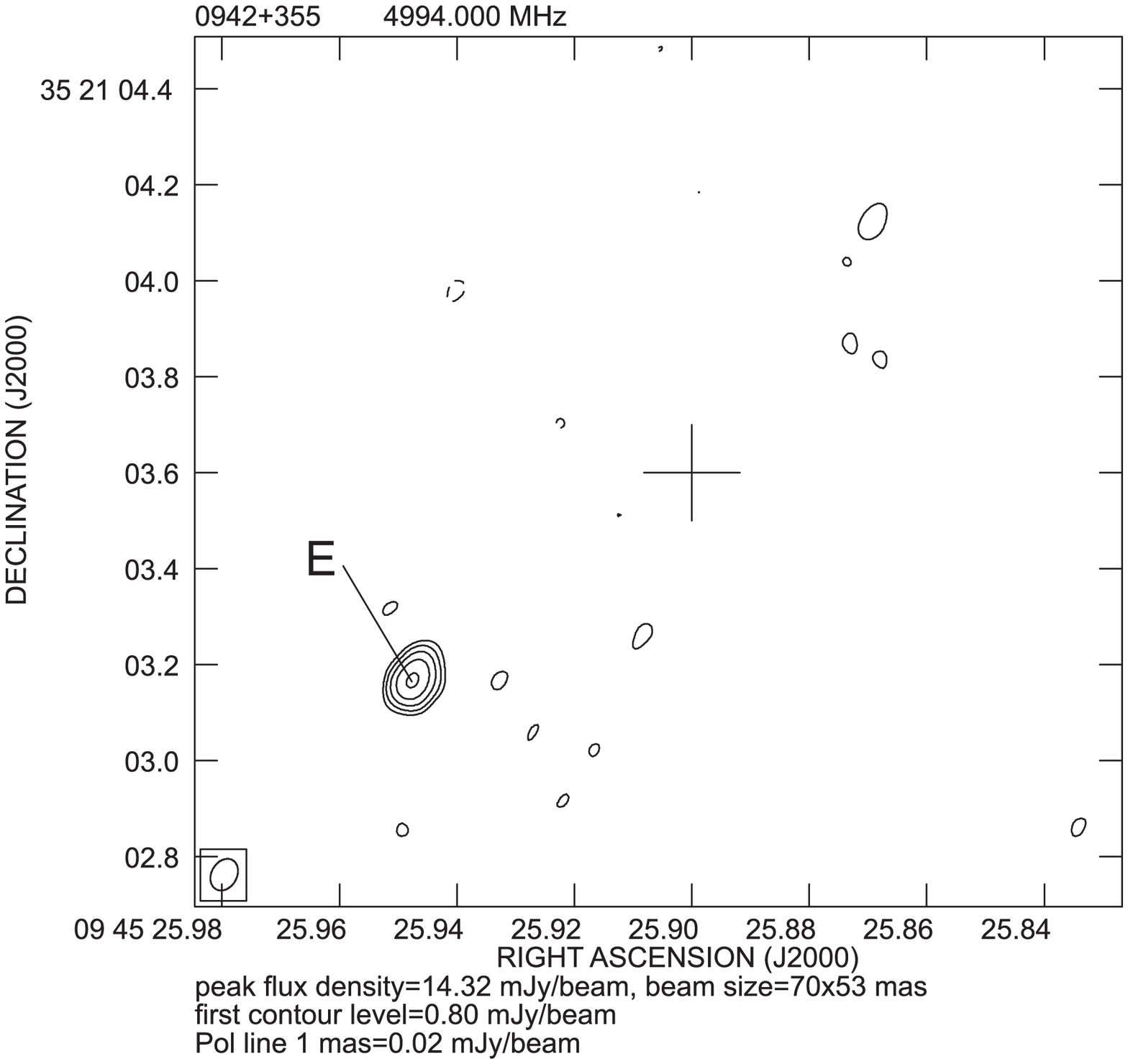}
\caption{MERLIN L-band and C-band images. Contours increase
by a factor 2, and the first contour level corresponds to $\approx
3\sigma$, vectors represent the polarized flux density. 
A cross indicates the position of an optical object found using
the most actual version of SDSS/DR7.}
\label{18i6cm_images}
\end{figure*}

\setcounter{figure}{1}
\begin{figure*}
\centering
\includegraphics[width=7cm,height=7cm]{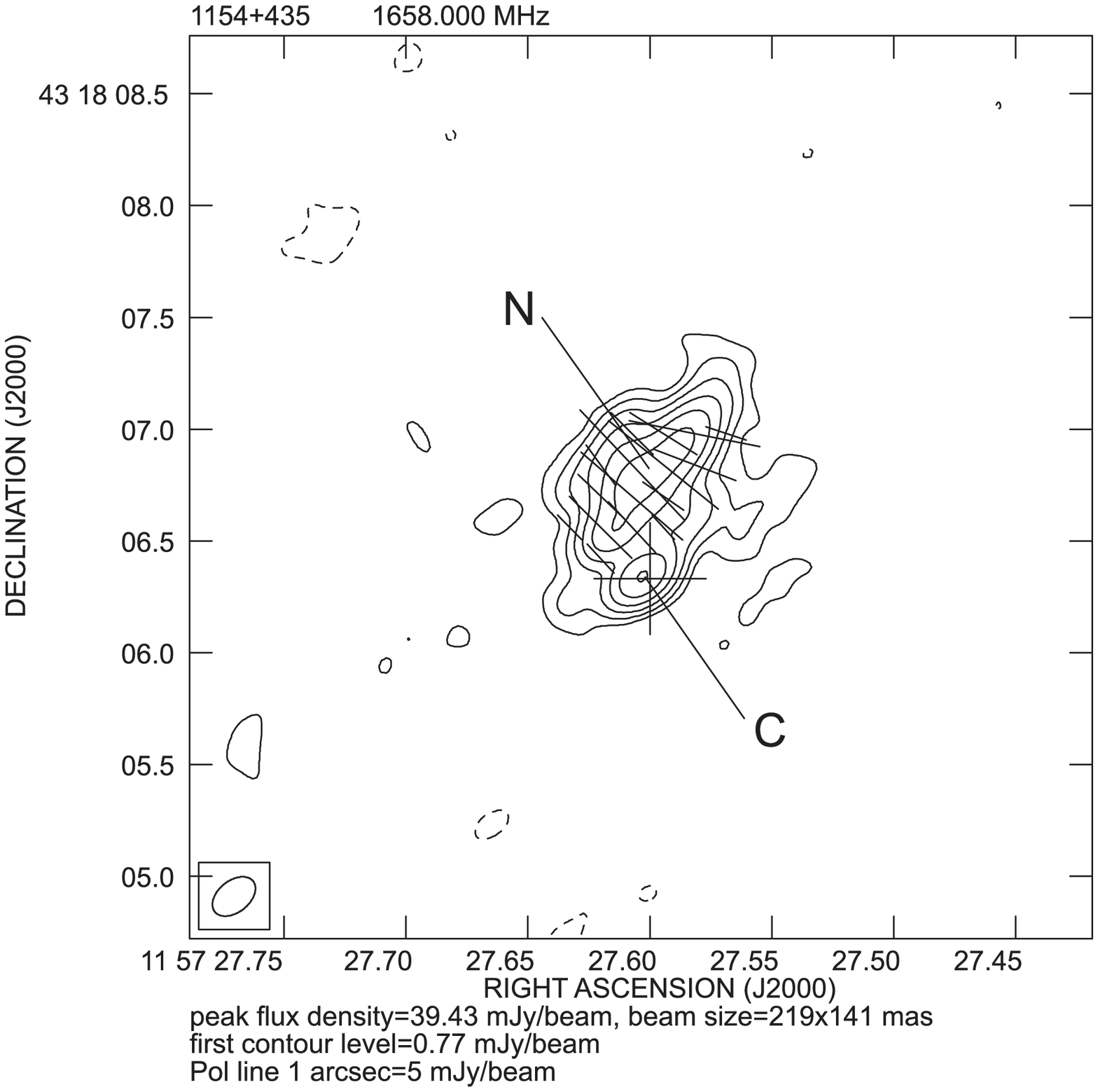}
\includegraphics[width=7cm,height=7cm]{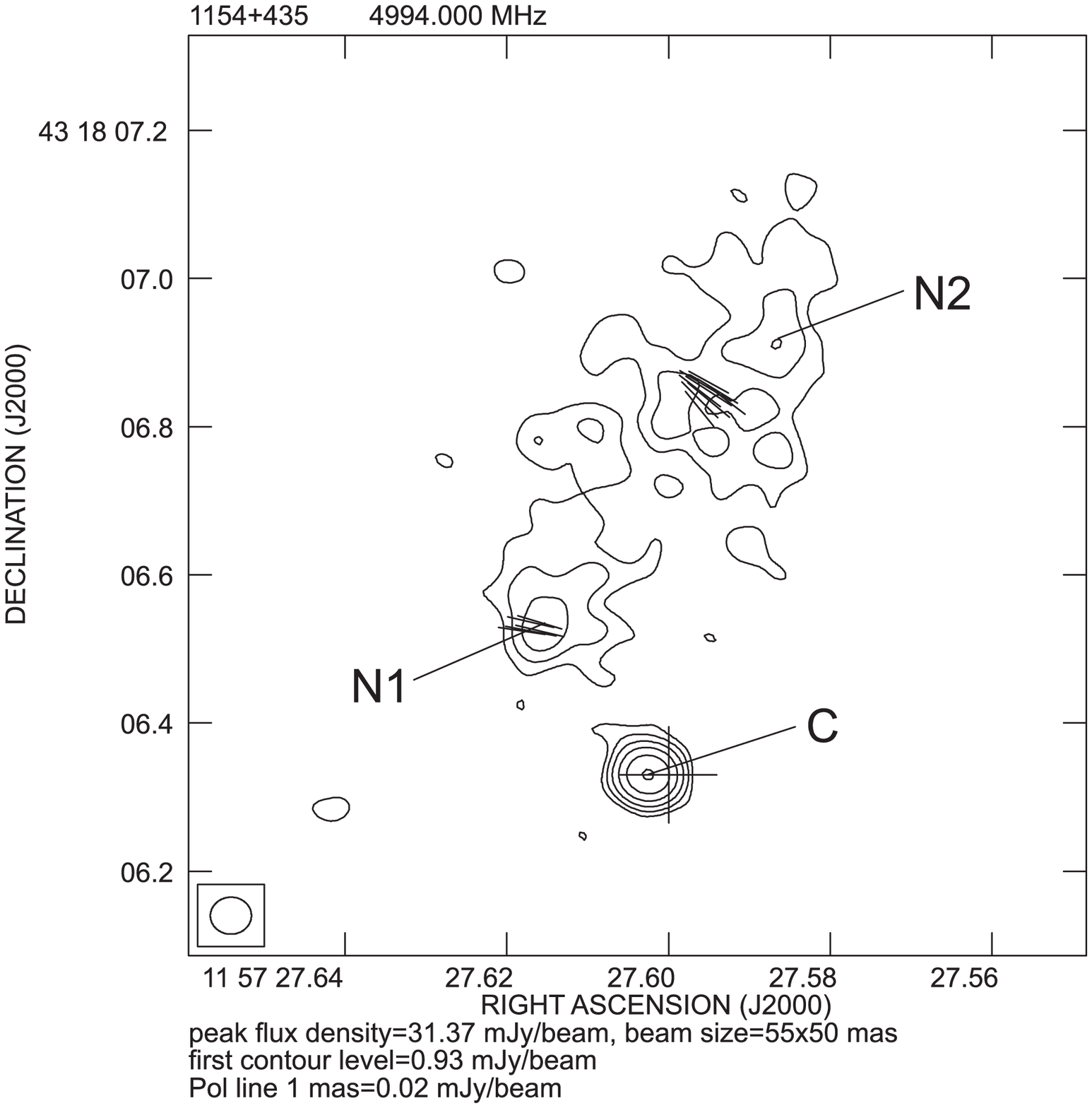}
\includegraphics[width=7cm,height=7cm]{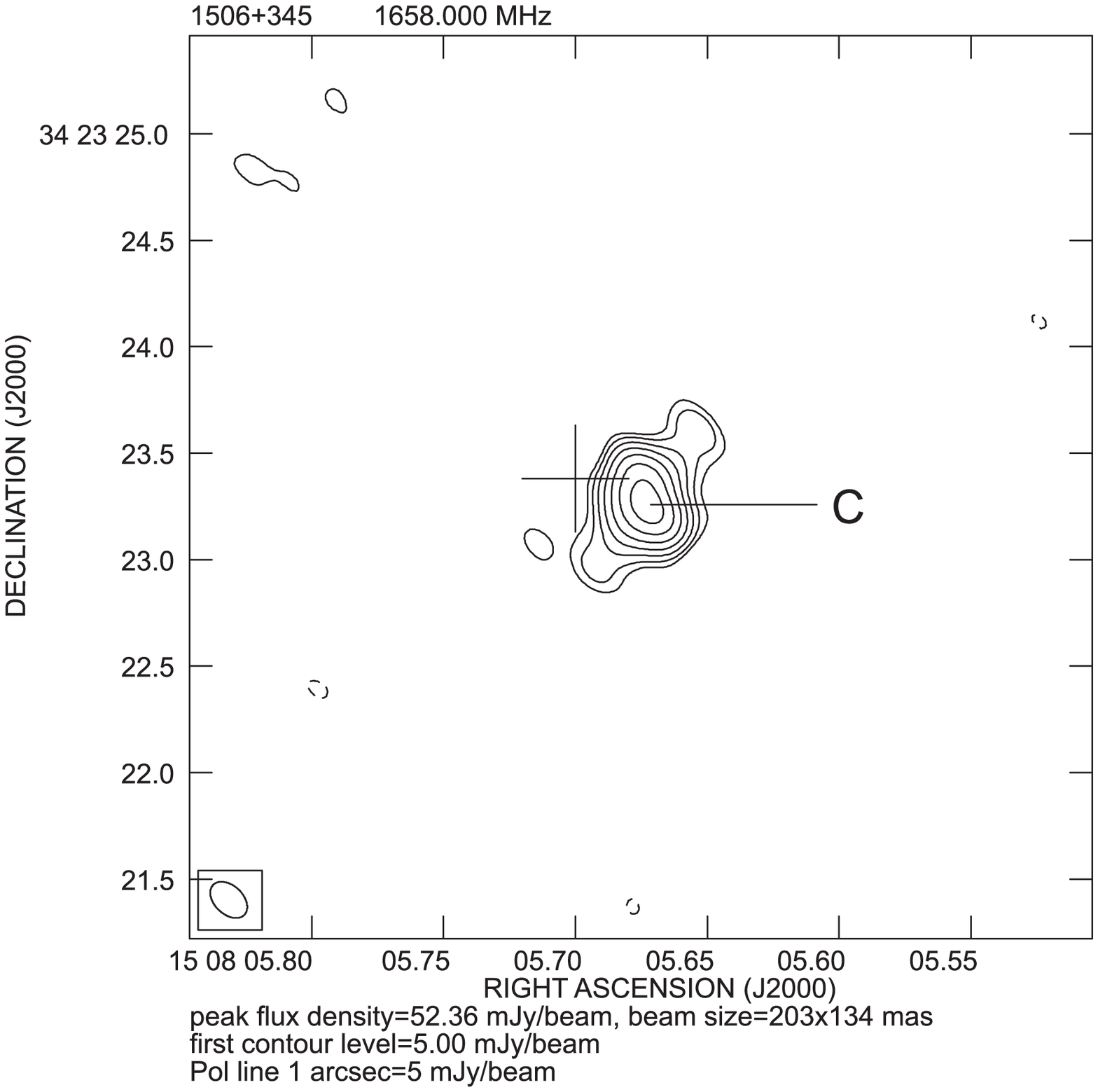}
\includegraphics[width=7cm,height=7cm]{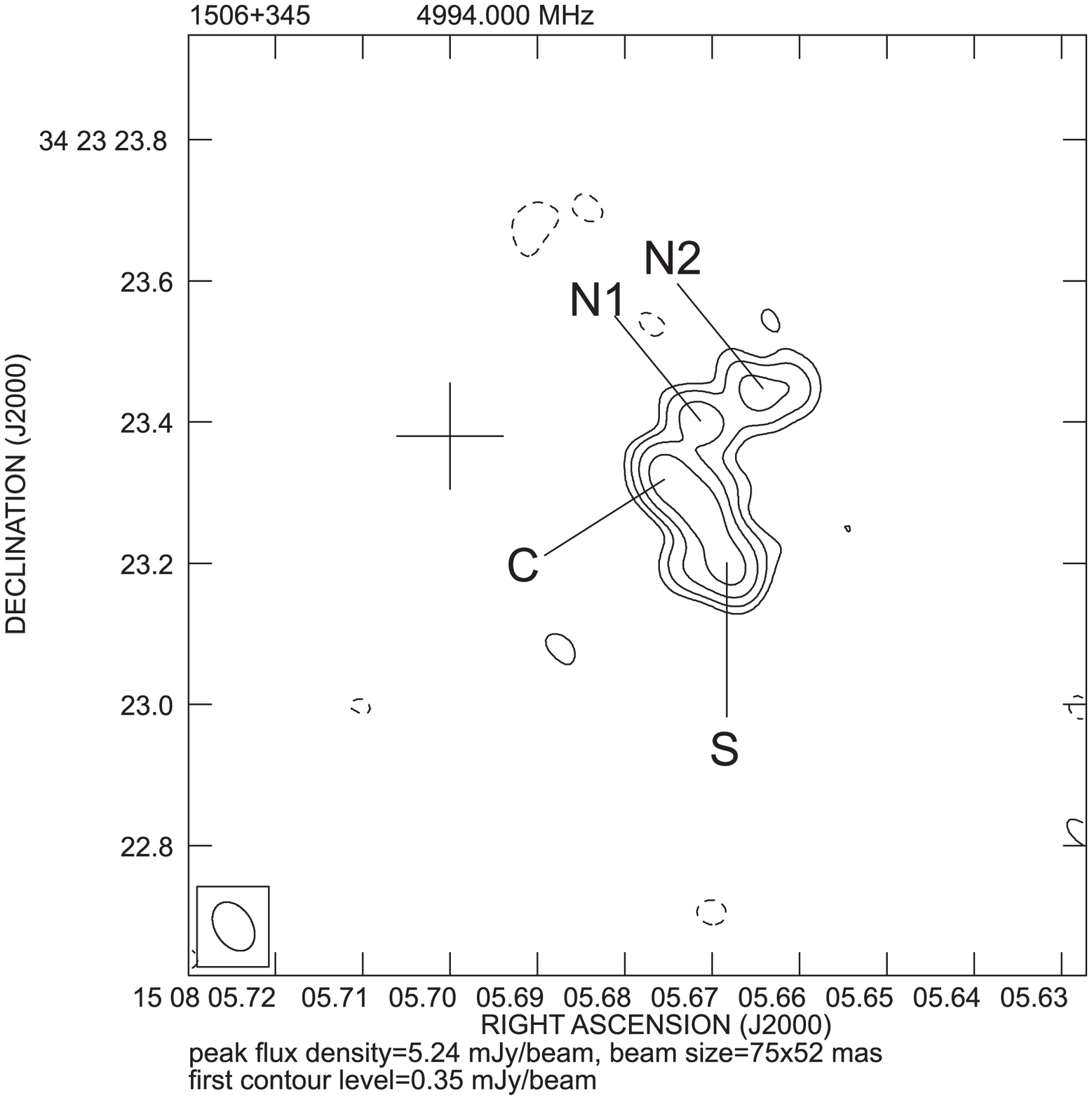}
\includegraphics[width=7cm,height=7cm]{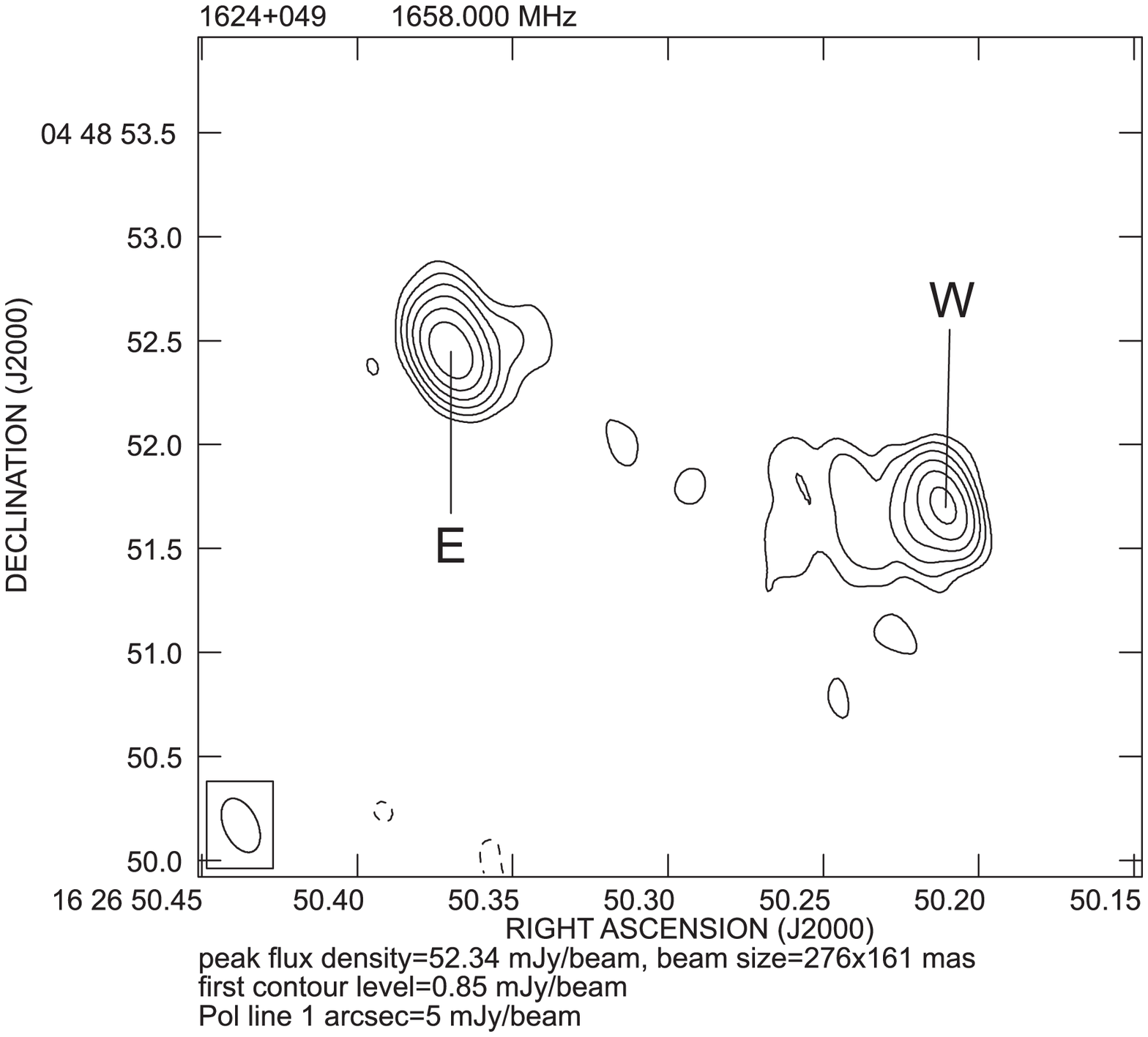}
\includegraphics[width=7.5cm,height=5.5cm]{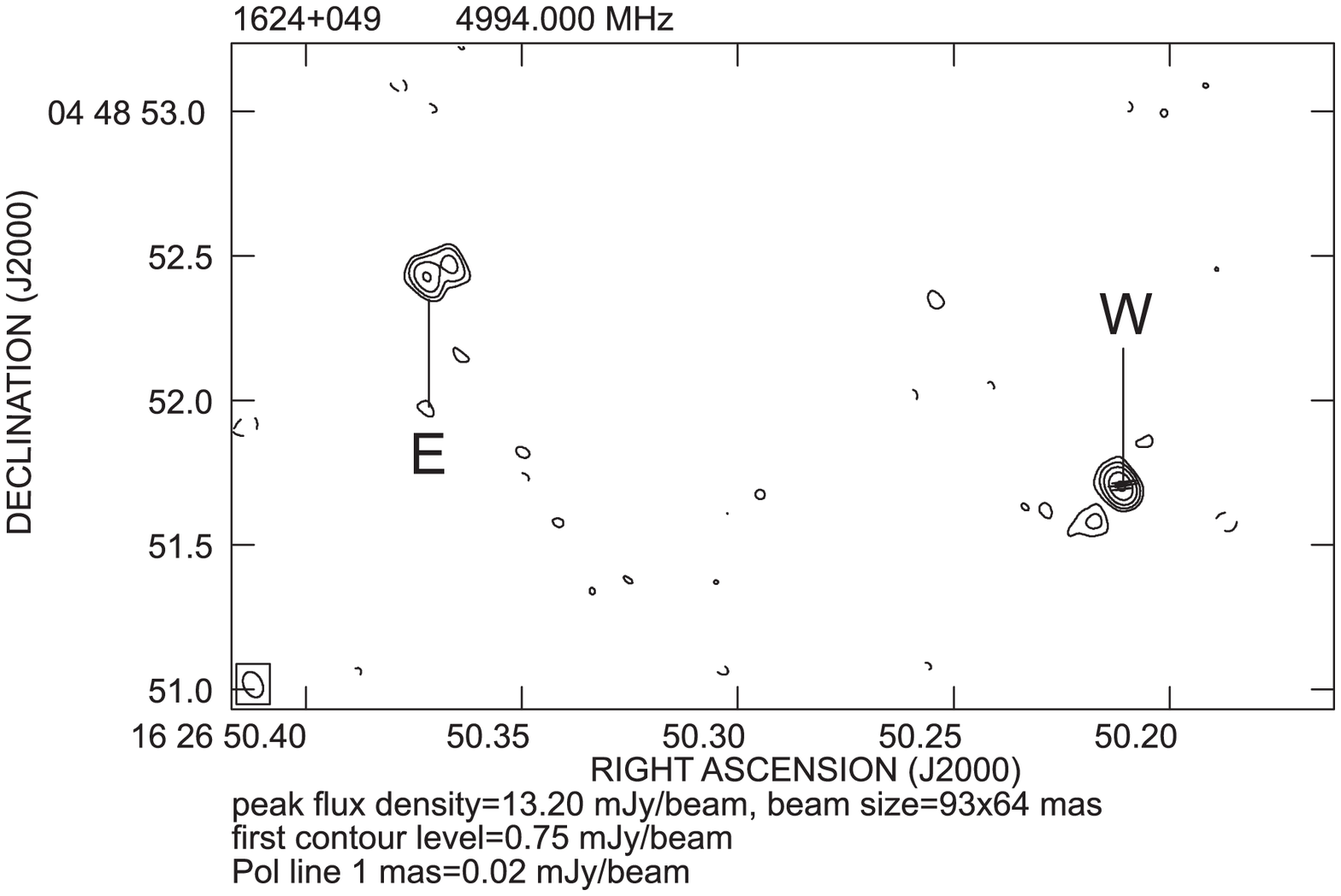}
\caption{MERLIN L-band and C-band images (cont.)}
\end{figure*}

\begin{figure*}
\centering
\includegraphics[width=7cm,height=7cm]{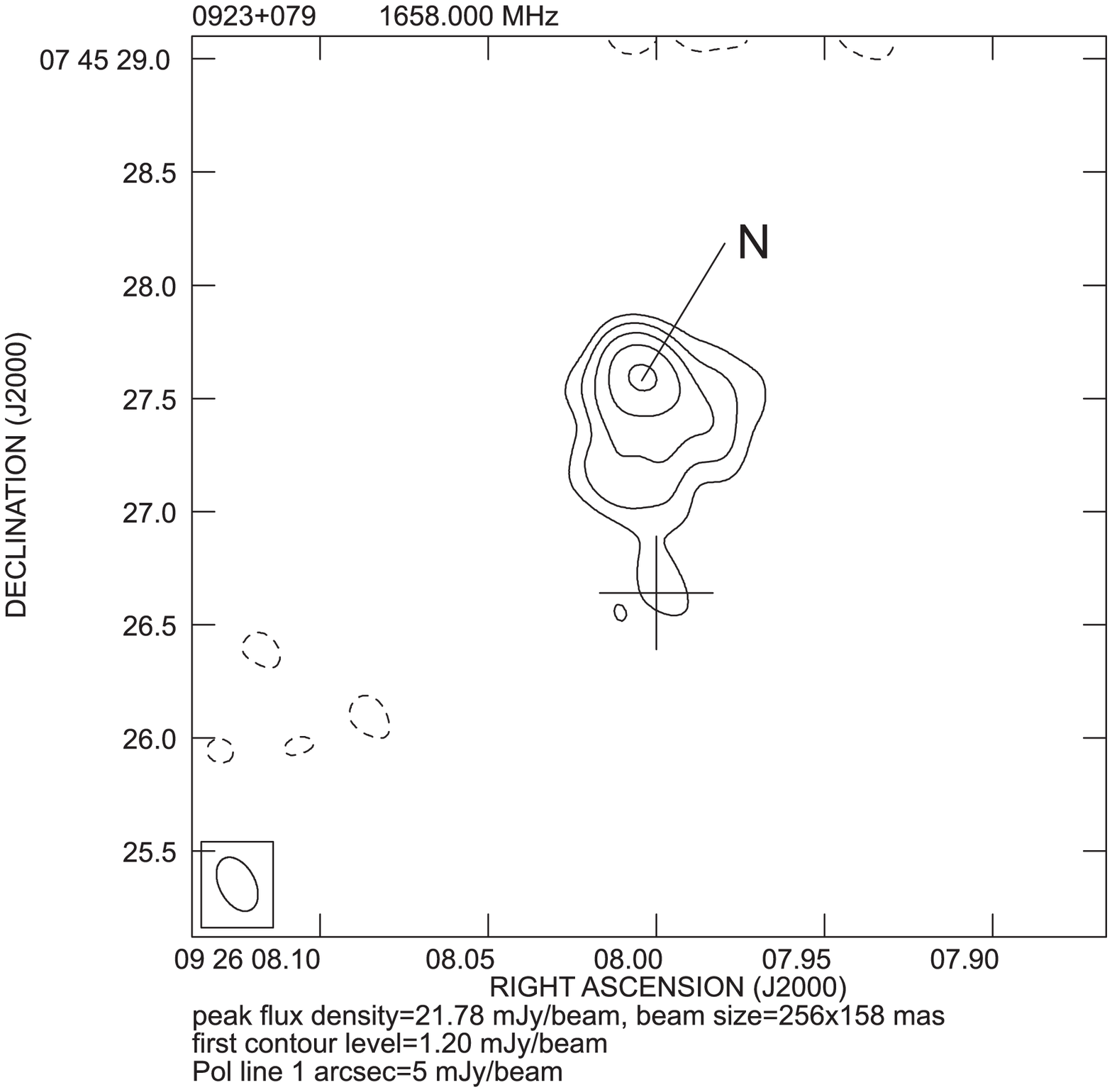}
\includegraphics[width=7cm,height=7cm]{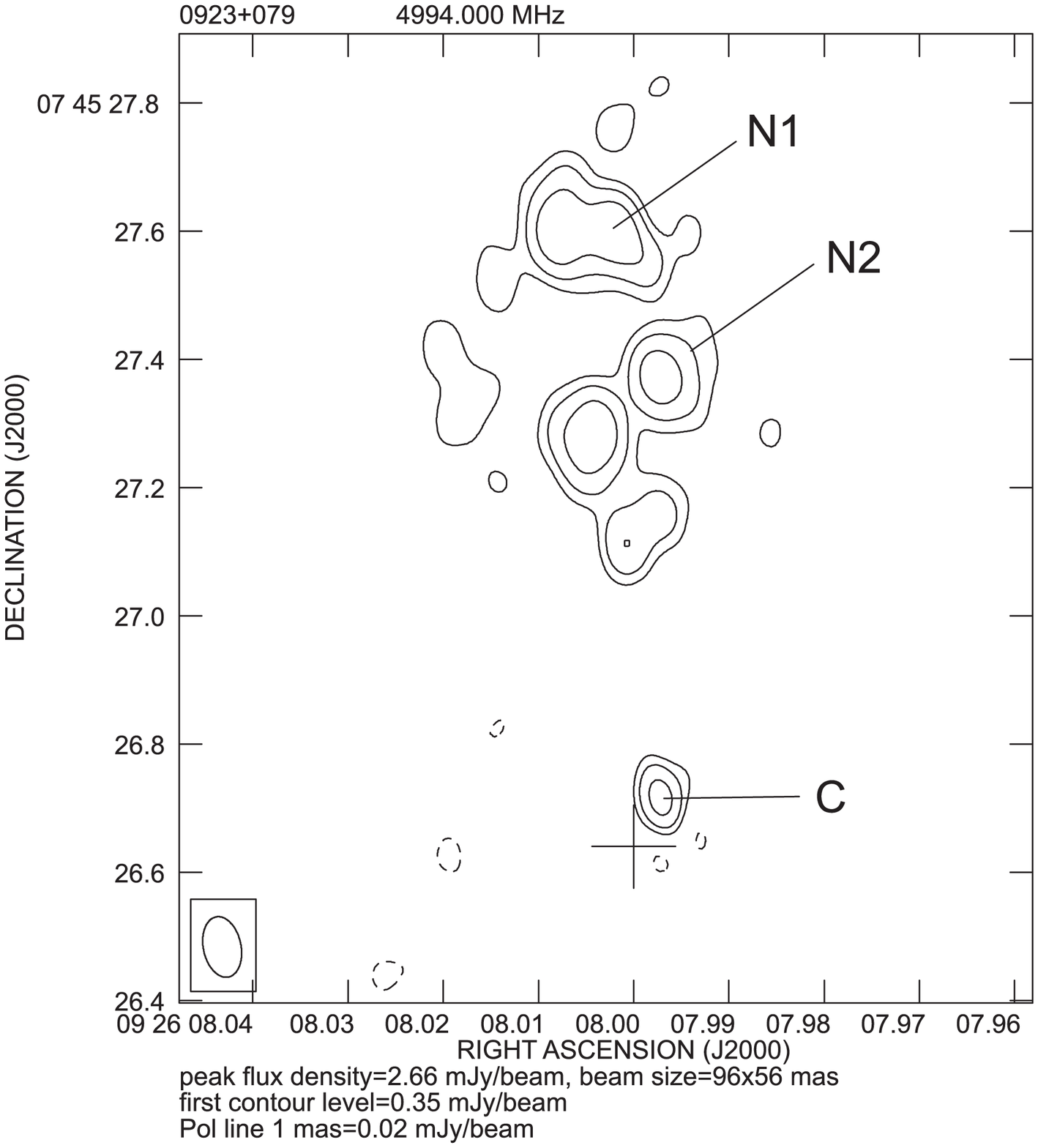}
\includegraphics[width=7cm,height=7cm]{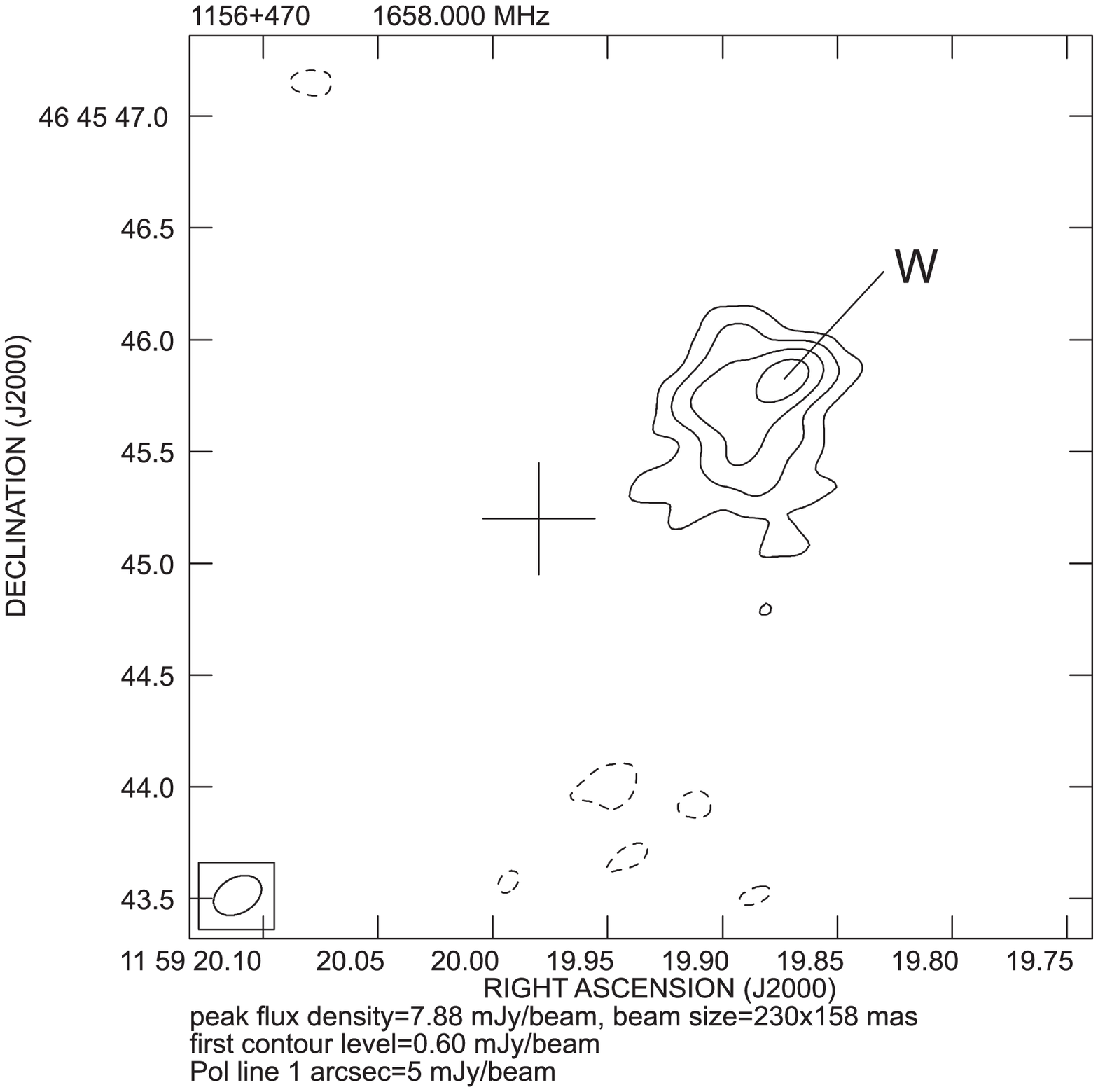}
\includegraphics[width=7cm,height=7cm]{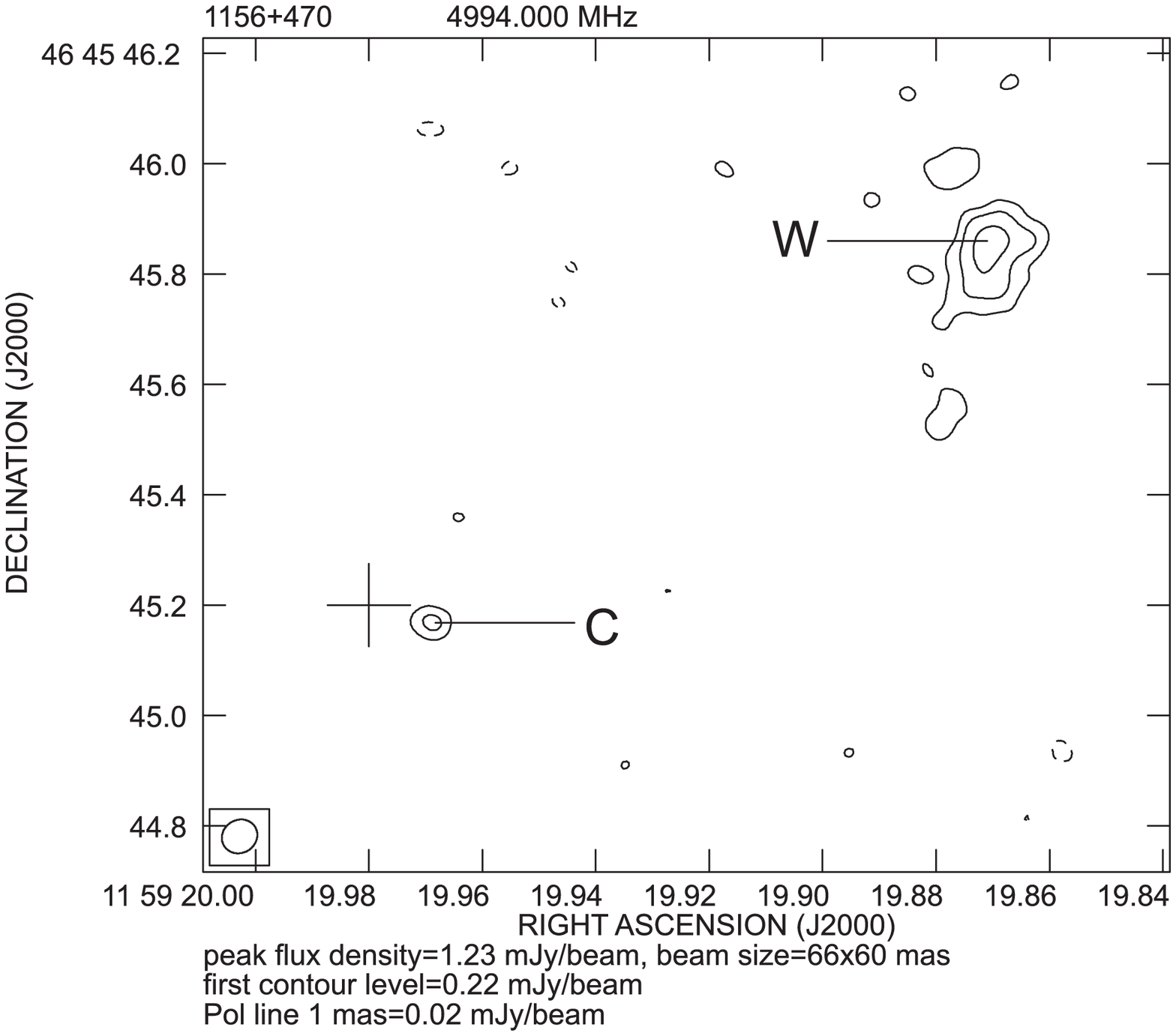}
\includegraphics[width=7cm,height=7cm]{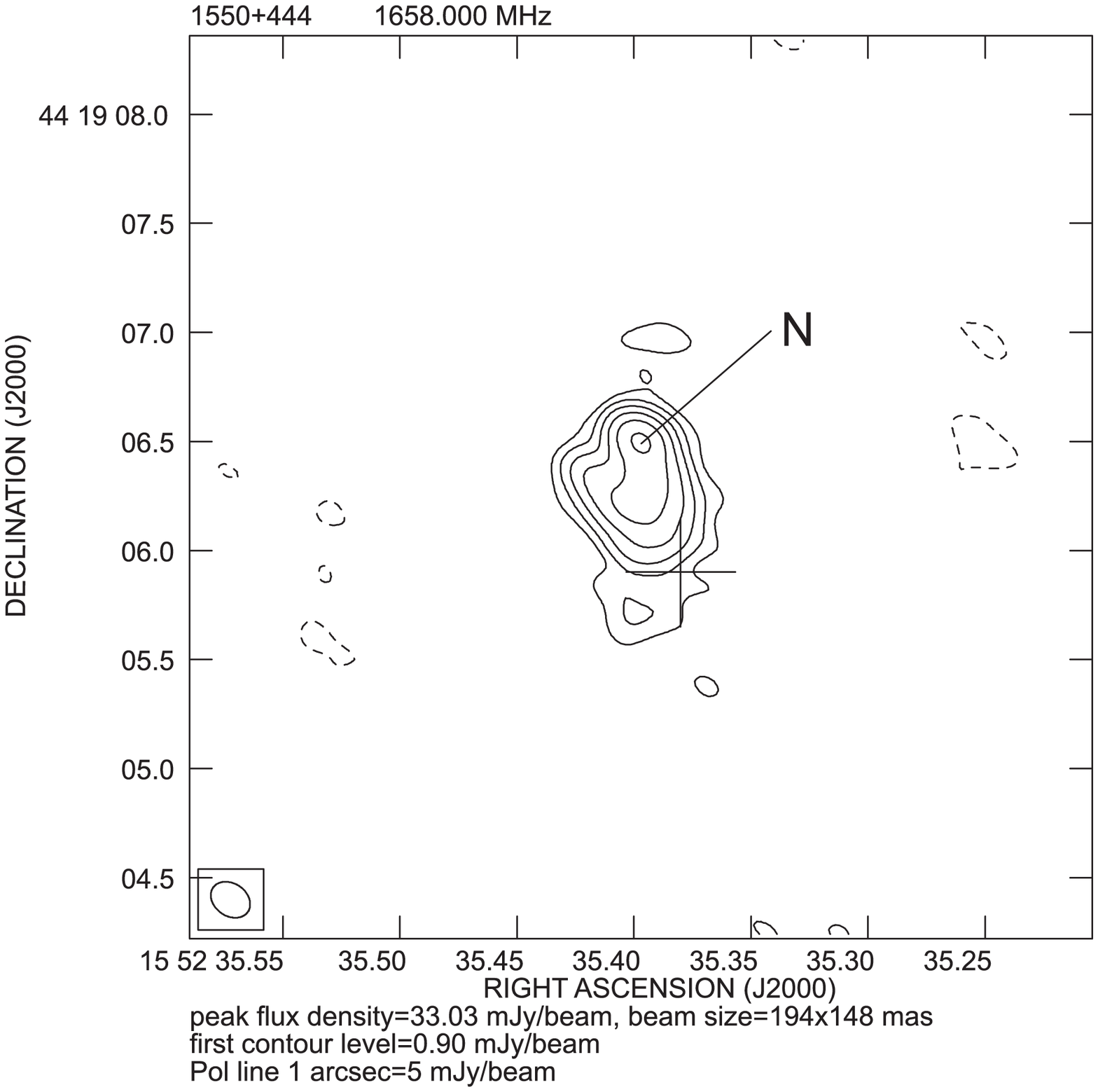}
\includegraphics[width=7cm,height=7cm]{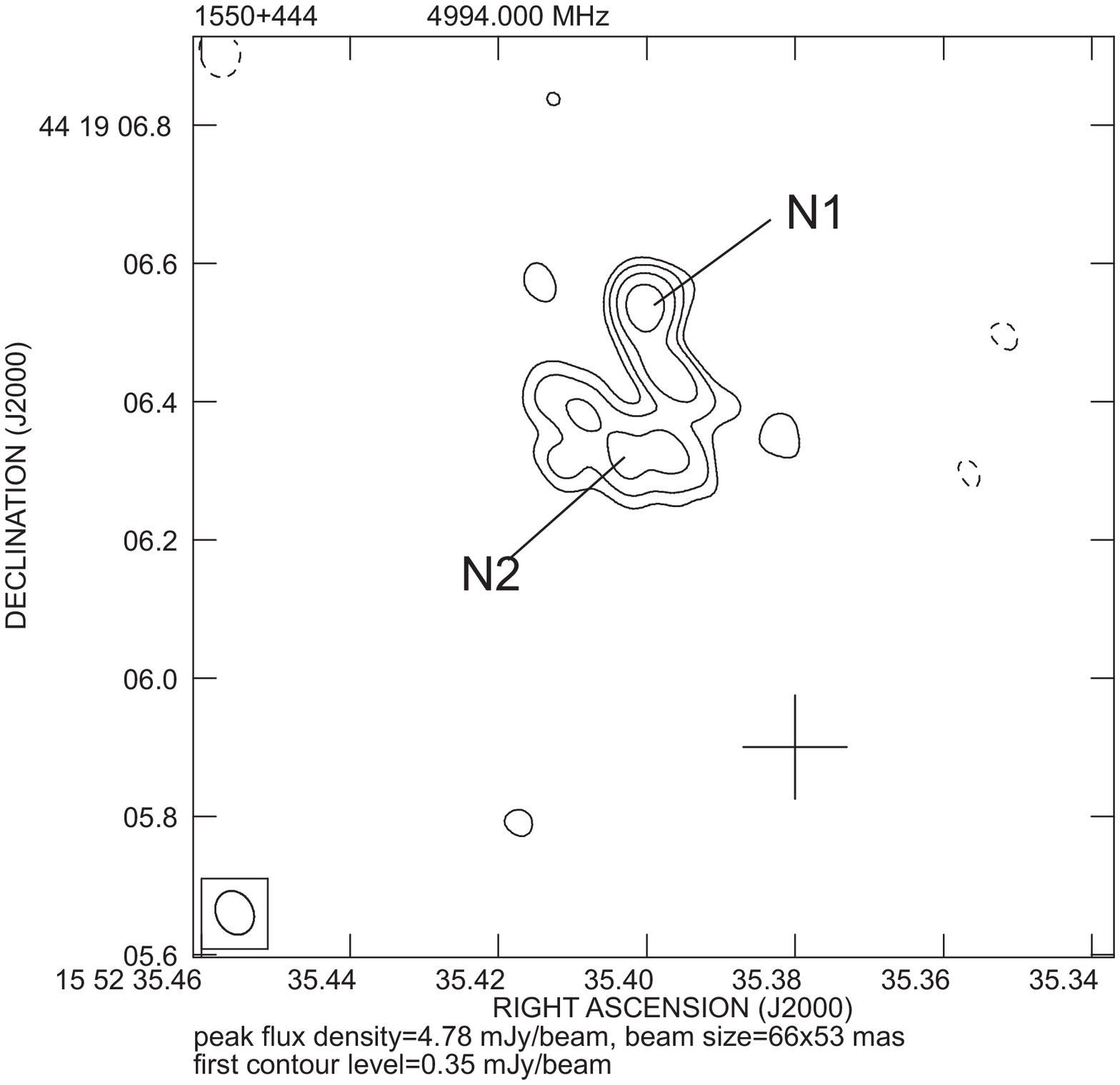}
\caption{MERLIN L-band and C-band images of 'single-lobed' objects. 
Contours increase
by a factor 2, and the first contour level corresponds to $\approx
3\sigma$, vectors represent the polarized flux density.
A cross indicates the position of an optical object found using
the most actual version of SDSS/DR7.}
\label{single-lobed}
\end{figure*}

\noindent
Finally, the new sample consists of 44 sources (Table~\ref{table1}).

The initial survey was undertaken using MERLIN at L-band. Then some of the
sources were also observed with MERLIN at C-band. 
All the L-band snapshot observations were made between December 2006 and May 2007.
Additional C-band snapshot observations were made in 2008 and 2009.
Each target source together with its associated phase reference
sources was
observed for $\sim$60\,min including telescope drive times.
The preliminary data reduction
was made using an AIPS-based PIPELINE procedure developed
at JBO. The resulting phase-calibrated images created with PI\-PE\-LI\-NE
were further improved using several cycles of self-ca\-li\-bra\-tion and
imaging, the final total intensity (I) and polarization intensity (P) images 
being produced using the task IMAGR.
We considered only emission above $3\sigma$ noise level, and we blanked out
all the points for which observed $P<3\sigma$, to avoid possible mis-use of low
signal-to-noise polarization data. The flux densities of the main components of
the target sources were then measured, by fitting Guassian models to the
components in the image, using task JMFIT
(Table~\ref{component}). For more extended features the flux
densities were evaluated by means of IMSTAT. Sub--components are referred to
as North (N), South (S), East (E), West (W) and Central (C). 
When a component is split into more
pieces, a digit (1,2, etc.) is added (e.g. N1, N2). 
The position of the optical counterparts of target sources are marked 
with a cross in the images and are taken from the SDSS
(found within the radius of 10$\arcsec$ from the radio source) or 
from the literature. The feature in the radio image closest to the cross is
indicated as 'C' and treated as probable radio core.   

Based upon available radio images (our MERLIN L-band and C-band observations
and other found in the literature), the new sources have been
divided into four categories (Table~\ref{table1}). However, for most of
the sources described here we have observation only at one frequency and
their classification should be treated as the preliminary
one ('?' means the most uncertain classification). 'Single' means the source is a
pointlike object unresolved or slightly resolved. 'Core-jet' is a
source with a bright central component and a one- or two-sided emission
(jets).
'Double-lobed' objects show at least two kinds of morphologies. Some of them
consist of two point-like components without a visible radio core 
in available radio images, whereas the
others show
a central weak component and distorted extended structure on both sides of
it. Many of the
double objects also show brightness asymmetry. For some of them (those with
probable core detection) we were able to evaluate the flux
density ratio, $r_{s}$, defined to be $>$1, of the oppositely directed
components. The last category, 'other',
comprises sources with peculiar morphologies: three objects have distorted 
radio morphologies, other three have only a
single visible lobe, and one object is resolved into four components
with high-resolution imaging.
Fig.~\ref{18cm_images}, Fig.~\ref{18i6cm_images} and Fig.~\ref{single-lobed} 
show images of the sources with resolved structures with the exception of 
one object (1641+320) which will be described in a separate paper.

In this paper, we present the analysis of MERLIN radio observations of a new sample of low
luminosity CSS sources and the study the correlation between radio power and 
linear size, and
redshift, making use of a larger sample that included also published samples of CSS
sources by \citet{f2001} and \citet{mar03}. The combined sample
of CSS sources we used in our study is gathered in Table A1 (see Appendix).
The optical data of the sample of low luminosity CSS sources will be
discussed in a forthcoming paper (Kunert-Bajraszewska \& Labiano,2010;
hereafter Paper\,II).

Throughout the paper, we assume a cosmology with
${\rm
H_0}$=71${\rm\,km\,s^{-1}\,Mpc^{-1}}$, $\Omega_{M}$=0.27,
$\Omega_{\Lambda}$=0.73.

\begin{table*}
\begin{center}
\caption[]{Flux densities of sources principal components from the 1.6\,GHz 
and 5\,GHz MERLIN images}
\begin{tabular}{cccc|cccc}
\hline
Source & Compo- & ${\rm S_{1.6\,GHz}}$&${\rm S_{5\,GHz}}$&
Source & Compo- & ${\rm S_{1.6\,GHz}}$&${\rm S_{5\,GHz}}$\\
name   & nents  &   mJy              & mJy & 
name   & nents  &   mJy              & mJy\\
(1)    & (2)    &   (3)              & (4)& 
(1)    & (2)    &   (3)              & (4)\\
\hline
0025+006& C & 66 & $-$ & 1140+058& C & 125& $-$\\
        & E & 41 & $-$ &         & W & 22 & $-$\\
0810+077& C & 53 & $-$ & 1154+435& C & 36 & 32 \\
        & E & 103& $-$ &         & N &186 & $-$\\
        & W & 175& $-$ &         & N1&$-$ & 17 \\
0821+321& E & 35 & $-$ &         & N2&$-$ & 9  \\
        & W & 33 & $-$ & 1156+470& C & $-$& 0.5 \\         
0846+017& C & 36 & $-$ &         & W & 30 & 4 \\
        & S1& 15 & $-$ & 1308+451& C & 32 & $-$\\
        & S2& 6  & $-$ &         & W & 59 & $-$\\
0850+024& E & 44 & $-$ & 1321+045& C & 13 & $-$\\
        & W & 38 & $-$ &         & E & 38 & $-$\\
0851+024& E & 50 & $-$ &         & W & 49 & $-$\\
        & W & 53 & $-$ & 1506+345& C & 99$\dagger$  & 14 \\         
0854+210& C*& 57 & 6   &         & N1& $-$& 4 \\
0907+049& C & 123& $-$ &         & N2& $-$& 3 \\
        & N & 21 & $-$ &         & S & $-$& 8 \\
        & S & 8  & $-$ & 1542+390& C & 4  & $-$\\         
0914+504& N & 51 & $-$ &         & E & 53 & $-$\\
        & S & 32 & $-$ &         & W & 119& $-$\\
0921+143& C & $-$& 18  & 1543+465& C & 9  & $-$\\         
0923+079& N & 62 & $-$ &         & N & 127& $-$\\
        & C & $-$& 1.5 &         & N1& 90 & $-$\\
        & N1& $-$& 8   &         & S & 99 & $-$\\
        & N2*& $-$& 7  & 1550+444& N & 98 & $-$\\         
0931+033& C & $-$& 83  &         & N1& $-$& 11 \\
0942+355& C & 11 & $-$ &         & N2*& $-$& 7  \\
        & E & 61 & 18  & 1558+536& C & 6  & $-$\\         
        & W & 26 & $-$ &         & N & 21 & $-$\\
1007+142& N & 482& $-$ &         & S & 28 & $-$\\ 
        & S & 291& $-$ & 1610+407& C & 514& $-$\\         
1009+053& E & 150& $-$ & 1624+049& E & 69 & 13 \\       
        & W & 9  & $-$ &         & W & 47 & 15 \\
1037+302& E & 26 & $-$ & 1715+499& E & 46 & $-$\\          
        & E1& 16 & $-$ &         & W & 33 & $-$\\
        & W & 171& $-$ &         &   &    &    \\
1053+505& E & 40 & $-$ &         &   &    &    \\
        & E1& 6  & $-$ &         &   &    &    \\ 
        & W & 32 & $-$ &         &   &    &    \\

\hline
\end{tabular}
\end{center}
$\ast$ - sum of the visible components, $\dagger$ - the whole source
\label{component}
\end{table*}

\section{Notes on individual sources}
\label{notes}

\noindent {\bf 0025+006}.
The 1.6\,GHz MERLIN map (Fig.~\ref{18cm_images}) shows an asymmetric
morphology:
the brightest central component is probably a radio core and the extended
south-eastern emission is a radio jet (E).

\noindent {\bf 0801+437}.
This source is unresolved in 1.6\,GHz MERLIN observations and only 1.6\,GHz
EVN observations show structure classified as possible double by \citet{de_Vries}.

\noindent {\bf 0810+077}.
This source has been classified as a possible double object (Fig.~\ref{18cm_images}).
The weak component (C) whose position is correlated
with the position of the optical counterpart could be a radio core, 
on opposite
sides of which there is emission from the two radio jets/lobes. The flux
density ratio, $r_{s}$ of the oppositely directed components E and W amounts to 1.7.

\noindent {\bf 0821+321}.
This is a double object with two weak radio lobes without evidence of bright 
hotspots and radio core (Fig.~\ref{18cm_images}).

\noindent {\bf 0846+017}.
This is probably a core-jet morphology (Fig.~\ref{18cm_images}) with the
brightest
component to be a radio core (C) and a bent radio jet (S1, S2).

\noindent {\bf 0850+024}.
Compact double object (Fig.~\ref{18cm_images}). The position of the optical
counterpart probably indicates the position of the slightly resolved radio
core.

\noindent {\bf 0851+024}.
Compact double object with a symmetric morphology (Fig.~\ref{18cm_images}).

\noindent {\bf 0854+210}.
Weak radio galaxy with diffuse morphology (Fig.~\ref{18i6cm_images}) also
detected in infrared \citep{drake03}. The peak of emission in 1.6\,GHz
MERLIN image is a radio core also visible in 5\,GHz MERLIN image (C).

\noindent {\bf 0907+049}.
This is a double object with the brightest central
component (C) probably containing a radio core, on opposite
sides of which there is emission from the two radio jets/lobes
(Fig.~\ref{18cm_images}).However the polarization is not detected in the southern
component, which can be an artifact. The polarized flux density of
components C and N amounts to 3\,mJy and 6\,mJy, respectively.   

\noindent {\bf 0914+114}.
Both 1.6\,GHz and 5\,GHz MERLIN images show single component.
This source has been resolved with VLBI at 2.3\,GHz into complex structure
consisting of four well separated components \citep{xiang}.

\noindent {\bf 0914+504}.
Quasar with a double morphology and asymmetric polarization. The
luminosity of the polarized southern component is lower than the northern
one, which can contain a radio core (Fig.~\ref{18cm_images}). The polarized
flux density of component S amounts to 4\,mJy.

\noindent {\bf 0921+143}.
The 1.6\,GHz MERLIN observation shows a single component which is resolved
in
5\,GHz MERLIN image into probably core-jet morphology (Fig.~\ref{18i6cm_images}).

\noindent {\bf 0923+079}.
The 1.6\,GHz morphology and the position of the optical counterpart suggest
there is only a single lobe visible in this source (Fig.~\ref{single-lobed}).
This interpretation is in agreement with 5\,GHz MERLIN image. The weak point
emission is probably a radio core and the lobe indicated as 'N' in 1.6\,GHz
image breaks into weak and probably fading components in 5\,GHz MERLIN
image.

\noindent {\bf 0931+033}.
The 1.6\,GHz MERLIN observation shows a single component which is resolved
in 5\,GHz MERLIN image into probably core-jet morphology (Fig.~\ref{18i6cm_images}).

\noindent {\bf 0942+355}.
This is a double object with asymmetric emission from two radio lobes
(Fig.~\ref{18i6cm_images}). The
position of the optical counterpart is correlated with a weak central
emission (C). The 5\,GHz MERLIN image
shows only an emission from the south-eastern radio lobe (E) with a steep
spectrum $\alpha_{1.6}^{5.0}=1.1$.
The flux density ratio, $r_{s}$ of the oppositely directed components E and W amounts
to 2.3.

\noindent {\bf 1007+142}.
Radio galaxy with a double morphology (Fig.~\ref{18cm_images}).
Large variations of polarization angle are seen in the southern lobe.
The luminosity of the polarized southern component is lower than the northern
one. The polarized flux density of component S amounts to 8.5\,mJy.

\noindent {\bf 1009+053}.
Two components with different brightness are visible
(Fig.~\ref{18cm_images}). Since the position of the optical counterpart is
outside the image range it is difficult to interpret the structure.

\noindent {\bf 1037+302}.
The 1.6\,GHz MERLIN image shows a highly asymmetric double object
(Fig.~\ref{18cm_images}) also detected with
the VLA at 8.4\,GHz and 22.5\,GHz by \citet{gir05}. The position of the optical
counterpart is well
correlated with radio core observed in 1.6\,GHz VLBA image \citep{gir05}.
Our 5\,GHz MERLIN image shows only an emission from the north-western radio
lobe (W) and is not published here.
Although the brightness asymmetry is visible in this source, the diffused
extended emission is difficult to measure and consequently the flux density ratio
is difficult to evaluate.

\noindent {\bf 1053+505}.
Three components are visible in 1.6\,GHz MERLIN image
(Fig.~\ref{18cm_images}).
The polarization is detected only in the north-western component (W), whose
luminosity is lower than the eastern one (E). 
The polarized flux density of component W amounts to 3\,mJy.
It is difficult to classify
this source without a 5\,GHz image. This is probably a double object with
emission from two radio lobes/jets.

\noindent {\bf 1140+058}.
This is a core-jet quasar (Fig.~\ref{18cm_images}).

\noindent {\bf 1154+435}.
The 1.6\,GHz MERLIN image of this source shows a complicated (double?)
morphology. The magnetic field appears to be ordered and follows the
direction of the jet visible at higher frequency. The polarized flux density of
component N amounts to 4\,mJy.
The 5\,GHz MERLIN image shows
one bright component, a radio core (C) with a flat
spectrum $\alpha_{1.6}^{5.0}=0.1$, and diffuse northern emission
which could be a bent radio jet (Fig.~\ref{18i6cm_images}).
The polarization is visible in two places, where we have peaks of
emission and where the jet interacts with the surrounding medium.

\noindent {\bf 1156+470}.
The 1.6\,GHz morphology and the position of the optical counterpart suggest
there is only a single lobe visible in this source (Fig.~\ref{single-lobed}).
The 5\,GHz MERLIN image shows very weak component 'C' which is probably
a radio core and fading lobe 'W'. 

\noindent {\bf 1308+451}.
This could be a double or a core-jet object
(Fig.~\ref{18cm_images}).

\noindent {\bf 1321+045}.
The position of the central bright component visible in 1.6\,GHz MERLIN
image (Fig.~\ref{18cm_images}) is well correlated with the position of the
optical
counterpart suggesting it is a radio core (C), on opposite sides of
which there is diffuse emission from the two radio jets/lobes.
The flux density ratio, $r_{s}$ of the oppositely directed components E and W amounts
to 1.3.

\noindent {\bf 1407+363}.
This source is unresolved in 1.6\,GHz MERLIN observations. It is
classified as a core-jet based on the 1.6\,GHz EVN+VLBA image
\citep{de_Vries}.

\noindent {\bf 1506+345}.
A point-like object visible in 1.6\,GHz MERLIN
image (Fig.~\ref{18i6cm_images}) 
is resolved into complicated structure in 5\,GHz
MERLIN image. High resolution 1.6\,GHz EVN
observations of this sources \citep{de_Vries} show two components: the
compact one (indicated as 'C' in our image) and very weak extended component
(probably 'S' in our 5\,GHz MERLIN image). 
According to the optical observations
\citep{maza93} it is an interacting galaxy pair separated by 24$\arcsec$.
The radio emission is probably associated with the north-eastern member of
this galaxy pair and its disturbed structure is probably a consequence of
the galaxy-galaxy interactions.

\noindent {\bf 1542+390}.
This is a double object (Fig.~\ref{18cm_images}) with a weak central
emission which could be a radio core (C).
There is an asymmetry in polarization between components E and W, and their
polarized flux density amounts to 4\,mJy and 1.5\,mJy, respectively.
The flux
density ratio, $r_{s}$ of the oppositely directed components E and W amounts
to 2.2.

\noindent {\bf 1543+465}.
A double galaxy with an asymmetric structure (Fig.~\ref{18cm_images}). 
A weak component indicated as
C could be a radio core. Assuming that C is indeed a radio core the flux
density ratio, $r_{s}$ of the oppositely directed components N+N1 and S amounts
to 2.2. However, without an observation at higher
frequency the identification of the components is uncertain.

\noindent {\bf 1550+444}.
The 1.6\,GHz morphology and the position of the optical counterpart suggest
there is only a single lobe visible in this source (Fig.~\ref{single-lobed}).
The 5\,GHz MERLIN image shows weak structure with a peak of emission
indicated as 'N1'.  

\noindent {\bf 1558+536}.
This is a galaxy with a double-like radio morphology. Probably one of the
weak central components can contain a radio core (Fig.~\ref{18cm_images}).
Assuming that C is indeed a radio core the flux
density ratio, $r_{s}$ of the oppositely directed components N and S
amounts to 1.3.

\noindent {\bf 1601+528}.
This source is unresolved in 1.6\,GHz MERLIN observations. It is
classified as double based on the 1.6\,GHz and 5\,GHz EVN and VLBA
observations \citep{de_Vries}.

\noindent {\bf 1610+407}.
This is probably a core-jet object (Fig.~\ref{18cm_images}), but this
classification is uncertain since this structure show more details at
higher frequency \citep{de_Vries}.

\noindent {\bf 1624+049}.
This is a galaxy with a double radio morphology visible in both 1.6\,GHz and
5\,GHz MERLIN observations (Fig.~\ref{18i6cm_images}). Both components have
steep spectrum: $\alpha_{1.6}^{5.0}=1.5$ (E) and $\alpha_{1.6}^{5.0}=1.0$
(W) and they are probably radio lobes. The flux
density ratio, $r_{s}$ of the components E and W amounts
to 1.5 and 1.15 at 1.6 and 5\,GHz, respectively.

\noindent {\bf 1715+499}.
This is a compact double object (Fig.~\ref{18cm_images}).

\noindent {\bf 1717+547}.
This source is unresolved in 1.6\,GHz MERLIN observations. It is
classified as double based on the 1.6\,GHz and 5\,GHz EVN and VLBA
observations \citep{de_Vries}.

\begin{figure*}
\includegraphics[width=8.8cm,height=8.8cm]{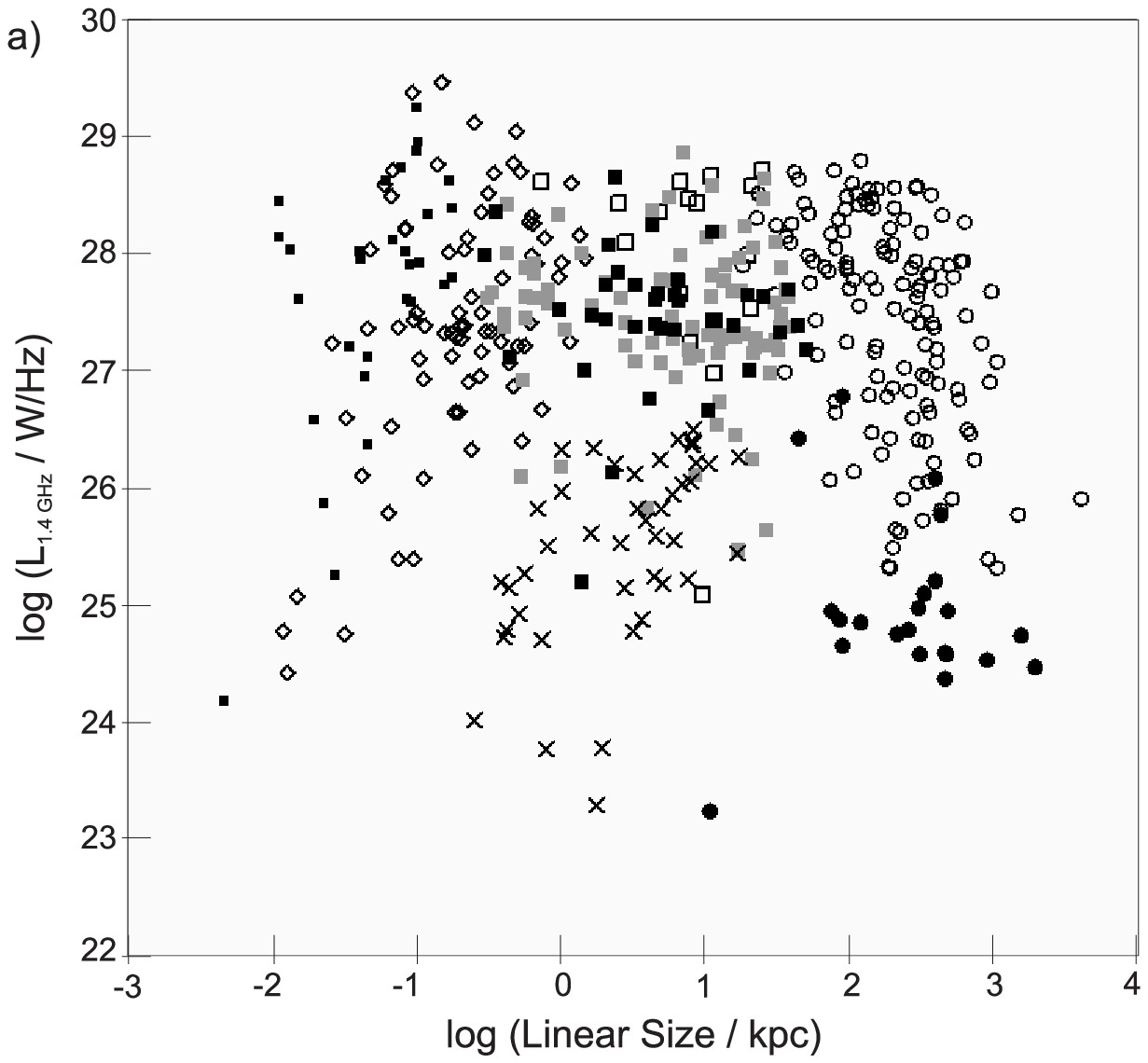}
\includegraphics[width=8.8cm,height=8.8cm]{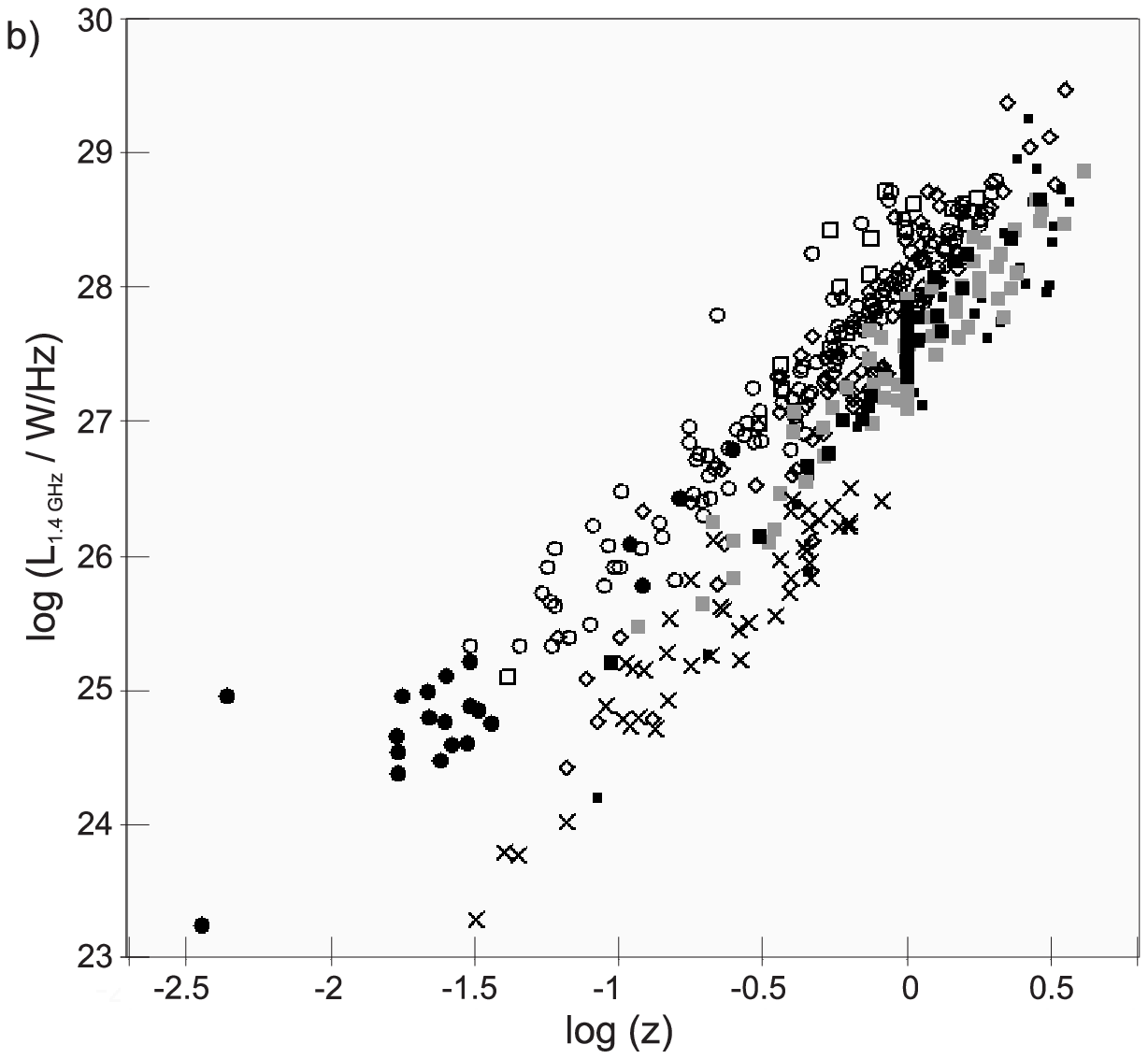}
\caption{a) Luminosity-size diagram for AGNs, 
b) Luminosity-redshift diagram for AGNs.
Squares indicate CSS sources from samples:
\citet{f2001} (grey squares), \citet{mar03} (black squares), \citet{laing83}
(empty squares). The diamonds indicate GPS objects and small black
squares indicate HFP objects from sample \citet{labiano07}.   
The filled circles indicate FR\,I objects and open circles
indicate FR\,IIs from sample \citet{laing83}. 
The crosses indicate the current sample of LLC sources
except source with the redshift indicated as '?'.}
\label{sample_plot}
\end{figure*}

\section{Discussion}
\label{dis}

The new sample of Low Luminosity Compact (LLC) sources that we present in
this paper
consists of 44 objects. We calculated their basic physical parameters
and gathered them in Table~\ref{table1}. 
In Fig.~\ref{sample_plot} we compare their luminosity and linear radio size 
with the sources from other samples.
We used combined sample of CSS sources (Table A1, Appendix),
GPS/high-frequency peaked (HFP) sources from master list created by \citet{labiano07} and 
the revised 3C sample of FR\,I and FR\,II sources (LRL sample,
\citet{laing83}) for comparison and
statistical study. The Luminosity-Size diagram (Fig.~\ref{sample_plot}a)
shows an evolutionary scheme of radio-loud AGNs.    
The selection criteria used for the new sample
resulted in approximately one third of the LLC sources
having a value of the 1.4\,GHz radio luminosity comparable to FR\,Is. 
Their luminosities are definitely lower than CSS sources
from last existing samples (crosses in Fig.~\ref{sample_plot}a). 
About 70$\%$ of the sources from the sample are galaxies and 
all of them are nearby objects with redshifts in the range $0.04<{\it
z}<0.9$, although the value of the redshift
was not a selection criterion (Fig.~\ref{sample_plot}b). 

\subsection{Characteristic of the groups of objects}
\label{double}

About 80$\%$ of the sources in our sample are resolved in available radio
maps (this paper and the literature), and at least (Table~\ref{table1})  
39$\%$ of the resolved objects are 
classified as double-lobed, although in this category we do put compact doubles 
without visible core and those with weak central component detected.  
A large precentage ($\sim$86$\%$) of these objects show asymmetry in components brightness,
but we have measured the flux density ratio only for those with probable core
detection (see Section 3). The observed asymmetry among LLC sources    
is consistent with the previous finding for the CSS source 
population \citep{saikia01} namely that, as a class, the CSSs are more asymmetric 
than larger radio sources of similar power.   
For one source, 0942+355, the brightness asymmetry of the components (probably
radio lobes) is also visible in 5\,GHz image (Fig.~\ref{18i6cm_images}). 
The 1037+302 has already been observed and investigated at higher frequencies 
with the VLA \citep{gir05}. The very weak peak of emission
visible in our 1.6\,GHz MERLIN image (Fig.~\ref{18cm_images}), well correlated with the 
position of
the optical counterpart, appeared to be a very weak radio core 
at 22\,GHz with VLA. According to \citet{gir05} this source 
will evolve to FR\,I. Two of the sources classified as double (0821+321,
1558+536) have breaking up structures without obvious signs of activity
(Fig.~\ref{18cm_images}).
About 70\,\% of the total flux density is missing in the MERLIN L-band image 
of weak 1558+536, suggesting that the diffuse emission is resolved in 
our observations. 

We have detected polarization for some of the double sources
(Fig.~\ref{18cm_images}). 
These are doubles without a visible radio core or a very weak
one (0914+504, 1007+142, 1053+505, 1542+390).
The flux density and luminosity of the polarized component are lower than
those of the second component (Table~\ref{component}). Again there is visible asymmetry
here, which can be explained in the terms
of interaction of the source with an asymmetric me\-dium in the central
regions of the host elliptical galaxy. On the contrary to large radio
objects, in GPS/CSS
sources, the jet is still crossing the ISM and the interaction with the
environment can be very strong.
Observations of the ionized gas in GPS and CSS sources show the presence of
such interactions \citep{holt, labiano}. 
The interactions with the ISM can be confirmed also by complicated
structure of 1543+465 and twisted jets in 0846+017 and 1154+435
(Fig.~\ref{18cm_images} and Fig.~\ref{18i6cm_images}). 
 
Among the LLC sources classified as 'others' we have identified so 
called 'single-lobed' objects:
0923+079, 1156+470 and 1550+444, based on their radio morphology and
position of the optical counterpart (Fig.~\ref{single-lobed}).
In the case of two sources (0923+079, 1156+470) about 60\,\% of the total
flux density is missing in our MERLIN observations,
which may suggest that they indeed possess
another component, but too weak and extended to be detected in a snapshot
mode with MERLIN. The 5\,GHz MERLIN observations made for all three sources
revealed, in two cases (0923+079, 1156+470), very weak central emission which is probably radio
core, and breaking up, fading radio lobe in all cases.  
This kind of objects have been already
observed in one of the previous samples of CSS sources \citep{mar03}, 
and amongst large-scale sources \citep{sub03}.

The last three objects belong to the category 'others' we labelled 
as the 'interacting systems', and these are: the two galaxy pairs (0854+210,
1506+345), and a very compact 
quasar binary (1641+320) which will be discussed in detail in a separate
paper (Fig.~\ref{18i6cm_images}). It has to be noted here that interpretation 
of 0854+210 as a galaxy
pair has not been confirmed yet by detailed optical study. 
Two of these sources - 0854+210 and 1506+345 - belong to group of
radio-excess {\it IRAS} galaxies \citep{drake03}. The radio powers of 
compact radio-excess galaxies are
generally lower than the well known CSS objects, with a median of $\sim
10^{24.5}~{\rm W~Hz^{-1}}$ at 5\,GHz, and the lower jet energy fluxes are
responsible for that \citep{drake03}. Also their morphologies do not
resemble typical CSS/GPS sources with symmetric double
lobes. They rather appear to be highly disrupted which is consistent in all
three objects observed by us. 0854+210 has an extended weak
emission indicating a trace of previous activity, and 1506+345 has a
complex 5\,GHz morphology which is probably a consequence of galaxy-galaxy
interactions. The galaxy pair 1506+345 has been also observed in a CO line survey made by
\citet{maza93}, which revealed that both components of this interacting pair
contain $\sim 10^{10}~M_{\odot}$ of molecular gas each. According to the
authors what we observe here is a colliding disc galaxies which will evolve
into gas-rich galaxy. A young phase of radio activity has been ignited here.

Summing up we have observed four main categories of radio morphology
(Table~\ref{table1}) among
LLC sources and the most interesting cases among them have been described
above.
It is difficult to compare radio morphologies of LLC sources with those
observed in previous samples \citep{f2001, mar03} because of different
observing frequencies and resolutions. However, what we can confirm is that
the fraction of sources with complex morphologies (breaking up structures
and with single lobe) is higher than that reported in \citet{f2001} and
\citet{mar03}.   
We suggest in section ~\ref{evolution} that many of them can be interpreted
as short-lived objects.

\begin{figure*}
\centering
\includegraphics[width=12cm,height=7cm]{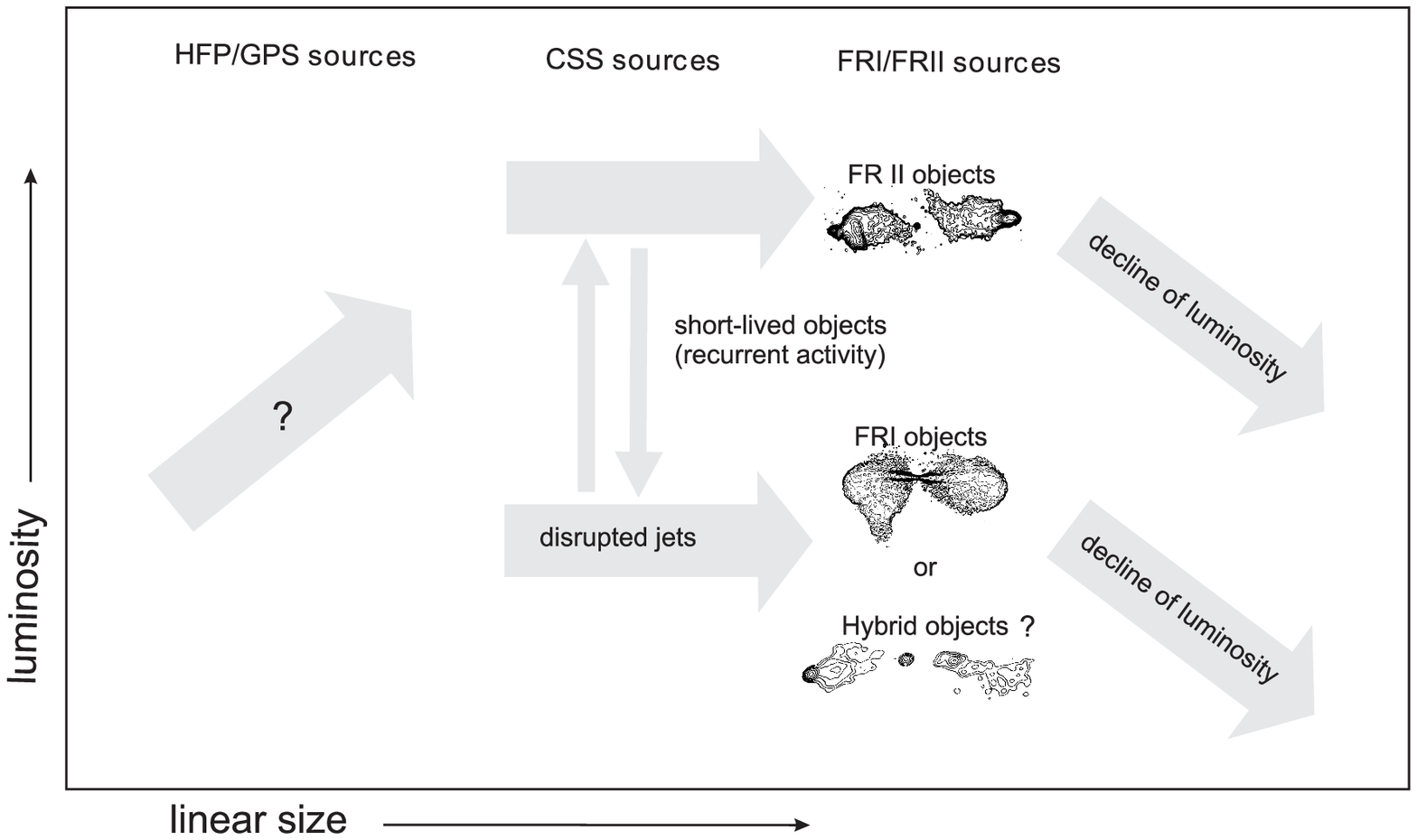}
\caption{Evolutionary scheme of radio-loud AGNs.}
\label{evolution_plot}
\end{figure*}

\subsection{Evolution of CSS sources}
\label{evolution}

Fig.~\ref{sample_plot}a shows a full range of luminosity and size covered by
radio sources and can be used to study the radio source evolution.
We propose an evolutionary scheme that is based on these observed parameters in
Fig.~\ref{evolution_plot}. The flattening of the slope at small size visible
on the diagram is real and has been first reported by \cite{odea97}
suggesting different evolution and excess of compact objects comparing to the
large scale radio sources. \citet{sn20} discussed the evolutionary
scheme of radio sources and concluded that the radio
luminosities of GPS sources increase as they evolve, reach a maximum in 
the CSS phase and then gra\-dually decrease as these objects grow further 
to become FR\,II/FR\,I sources.  
However, to match observations, the radio jets of the
young sources must slow down as they cross the host ISM and
dim faster than predicted. Recently \citet{de_Vries} showed that indeed
the expansion speed of GPS radio sources is correlated with their radio 
luminosity and is lower for less powerful young radio sources. 
Most of the GPS and CSS sources known so far are
powerful objects and the clue to solve the evolution puzzle can be in the analysis of less
powerful compact sources.  

According to the latest theoretical
predictions \citep{kb07,czerny} and observational facts \citep{gug05, kun06} 
the situation in the
group of young radio objects can be complicated. The episodic nature of the
radio activity is a fact among large scale objects where we observe two or
more of active periods in the radio images. This is still debated
for young objects, as is the nature of the mechanism causing it.    
Based on our latest observations of LLC sample we suggest we should treat the
'young source phase' of the evolution of radio object as the one which
determines its further evolution. Different mechanisms and interactions may
play an important role here. The most likely mechanism to
produce these effects is interaction of the radio source with
the host environment. \citet{labiano09} found that the GPS and CSS
sources (galaxies and quasars) show a strong correlation between 
[O\,III]$\lambda$5007 line luminosity and size of the radio source suggesting 
a possible deceleration in the jet as it crosses the host ISM.
However this correlation may not be present in fainter radio sources where
the radio and [O\,III]$\lambda$5007 can be produced in a different way (Paper\,II).
There are indications \citep{tasse}, based on the environmental studies of
radio sources
at moderate redshifts (z$\sim$~0.5-1), that low-luminosity radio AGNs  lie in
denser environments than the powerful objects, have lower stellar masses and
show excess mid-IR emission consistent with a
hidden radiatively efficient active nucleus. \citet{tasse} also argue that
the low luminosity radio activity is associated with the re-fuelling of
massive black holes.
Recently \citet{czerny} discussed mechanism of instabilities in the accretion
flow suggesting that the duration of the active phase should be longer than
$10^{4}$ years to enable the young source to escape from the host galaxy. For the
shorter active phase the radio source will be confined within the host
galaxy and such short outbursts may occur many times. 
We have presented this scenario on the scheme of the radio source evolution
as the two-sided arrows (Fig.~\ref{evolution_plot}) and labeled these sources as
short-lived objects. To this group of objects belong observed by us CSSs with
weak breaking up structures, without compact hotspots or jets and core visible,
which means compact but already fading objects that will not evolve into large
scale sources in this phase of activity. We interpret some double sources
(0821+321, 1558+536)
and 'single-lobed' objects (0923+079, 1156+470, 1550+444) observed in LLC 
sample as compact faders.
We suggest that these double objects from our sample with brightness and
polarization asymmetry may undergo disturbed evolution and starting to fade
away as single-lobed objects or to become FR\,Is or hybrids. 
According to \citet{alex00} sources with
disrupted jets are not expected to evolve into FR\,I objects, but fade
so strongly that they become virtually undetectable.
Also sources we called 'interacting systems' (galaxy pairs and quasar
binaries) can be interpreted as short-lived objects. 
The galaxy-ISM interactions observed in these double systems can change the direction of
the jets and disturb the environment fuelling the black hole and
consequently prevent the growing of the radio structure. 
As an effect of such interactions we observe complicated
radio structures consisting of weak fading emission together with current phase
and direction of the activity. Two of the weak 'interacting systems'
(0854+210 and 1506+345) are
radio-excess {\it IRAS} galaxies. As has been discussed by \citet{drake03}
the radio-excess sources, with the radio powers in general lower than the
well known CSS objects, are not expected to evolve in the same manner as 
CSS/GPS sources, but rather fade
even further to return to their place among the radio-quiet
FIR-luminous AGNs.

The evolutionary path for powerful compact objects is the one established so
far namely that the CSS sources evolve to become FR\,IIs.

According to \citet{sn20} and also visible in our
plot (Fig.~\ref{sample_plot}a), the smallest radio objects (GPS/HFP sources)
should increase their luminosity and size during the evolution. However in
the evolutionary scheme (Fig.~\ref{evolution_plot}) we put a question mark
while illustrating this path. We have noticed that most of the GPS/HFP
sources are already as bright as CSSs and FR\,IIs, and these small objects
with lower luminosity can either follow the predicted path or fade earlier
becoming weak CSS object. 

Another question mark in the evolutionary scheme (Fig.~\ref{evolution_plot})
concerns the rare class of radio sources - hybrids, which evolutionary path
is suggested by \citet{kb07}. However, recent results based on Chandra
observations show that hybrid sources which show broad-line optical
spectra have X-ray properties comparable with those of FR\,IIs radio
galaxies \citep{miller}. It is possible that in the case of these rare 
objects the transition from FR\,II-type jet to FR\,I-type plum structure takes 
place at the later stages of evolution due to interaction with a dense
and non-homogeneous intergalactic medium. Numerical works
\citep[e.g.][]{meliani} 
suggest that a powerful FR\,II-type jet can be decelerated by a high 
density medium and transform to FR\,I-type.
 
\section{Summary}
In this paper, we have presented results from MERLIN L-band and C-band observations of
a new sample of low luminosity compact sources. Most of the sources have
been resolved showing complex, asymmetric radio morphologies suggesting
presence of dense environment and jet-ISM interactions. We suggest that some 
of the sources with the breaking up structures or one-sided morphology are 
candidates for compact faders. In this case the fraction of
faders is higher ($\sim$15\%) than that reported in CSS samples 
from B3-VLA and FIRST catalogues ($\sim$4\%). 
We suspect we can increase the number of these sources by a factor of 2 or
even 3, by extending the studies to lower frequencies and luminosities.

We studied correlation between radio power and linear size, and
redshift with a larger sample that included also published samples of
compact objects and large scale FR\,IIs and FR\,Is.
The low luminosity compact objects occupy the space in radio power versus
linear size diagram below the main evolutionary path of radio objects. We
conclude that many of them can be short-lived objects, at least in the
current phase of evolution. These undergo disrupted evolution many times as
they are be able to escape the host galaxy and evolve to FR\,IIs. 
The presented here analysis of the radio properties of LLC sources and also
analysis of their spectroscopic features which will be discussed in a
second paper of the series indicate that most of them will evolve finally to
FR\,II. 
However, we
suggest that there exists a much larger population of short-lived low luminosity
compact objects unexplored so far and among them we can find precursors of
large scale FR\,Is.

\section*{Acknowledgments}

We thank Peter Thomasson and Anita Richards for their help with 
MERLIN observations and data reduction.
MERLIN is a UK National Facility operated by
the University of Manchester on behalf of STFC.

The research has made use of the NASA/IPAC Extragalactic Database (NED)
which is operated by
the Jet Propulsion Laboratory, California Institute of Technology, under
contract with the
National Aeronautics and Space Administration.

This research has made use of Sloan Digital Sky Survey.
Funding for the SDSS and SDSS-II has been provided by the Alfred P. Sloan
Foundation, the Participating Institutions, the National Science Foundation,
the U.S. Department of Energy, the National Aeronautics and Space
Administration,
the Japanese Monbukagakusho, the Max Planck Society, and the Higher
Education
Funding Council for England. The SDSS Web Site is
http://www.sdss.org/.  The SDSS is managed by the Astrophysical Research
Consortium for the Participating Institutions. The Participating
Institutions
are the American Museum of Natural History, Astrophysical Institute Potsdam,
University of Basel, University of Cambridge, Case Western Reserve
University,
University of Chicago, Drexel University, Fermilab, the Institute
for Advanced Study, the Japan Participation Group, Johns Hopkins University,
the Joint Institute for Nuclear Astrophysics, the Kavli Institute for
Particle
Astrophysics and Cosmology, the Korean Scientist Group, the Chinese Academy
of
Sciences (LAMOST), Los Alamos National Laboratory, the Max-Planck-Institute
for
Astronomy (MPIA), the Max-Planck-Institute for Astrophysics (MPA), New
Mexico
State University, Ohio State University, University of
Pittsburgh, University of Portsmouth, Princeton University, the United
States Naval Observatory, and the University of Washington.

\appendix

\section[]{Combined sample of CSS sources}

Table A1 consists of a large sample of CSS sources which includes last
published CSS samples by \citet{f2001} and \citet{mar03}.  The data were
updated by using the most actual version of the SDSS/DR7 and were used for
evolution study presented in Fig.4 and Paper II.

\begin{table*}
\begin{center}
\caption[]{Combined sample of CSS sources}
\begin{tabular}{@{}c c c c c c c c c c@{}}
\hline
~~~Source & RA & Dec &
\multicolumn{1}{c}{\it z}&
\multicolumn{1}{c}{${\rm S_{1.4}}$}&
\multicolumn{1}{c}{log${\rm L_{1.4}}$}&
\multicolumn{1}{c}{${\rm S_{4.85}}$}&  
\multicolumn{1}{c}{log${\rm L_{4.85}}$}&
\multicolumn{1}{c}{LAS}&
\multicolumn{1}{c}{LLS}\\
~~~Name   & h~m~s & $\degr$~$\arcmin$~$\arcsec$ & &
\multicolumn{1}{c}{Jy}&
\multicolumn{1}{c}{$\rm W~Hz^{-1}$} &
\multicolumn{1}{c}{Jy}&
\multicolumn{1}{c}{$\rm W~Hz^{-1}$} &
\multicolumn{1}{c}{$\arcsec$} &
\multicolumn{1}{c}{$h^{-1}~{\rm kpc}$} \\
~~~(1)& (2)& (3) &(4)&
\multicolumn{1}{c}{(5)}&
\multicolumn{1}{c}{(6)}&
\multicolumn{1}{c}{(7)}&
\multicolumn{1}{c}{(8)}&
\multicolumn{1}{c}{(9)}&
\multicolumn{1}{c}{(10)}\\
\hline
2358+406&  00 00 53.1 & 40 54 02    & 1.000  & 1.300 & 27.84 & 0.688 &
27.56 & 0.08 & 0.64\\
0003+387&  00 06 20.6 & 39 00 28    & 1.470  & 0.482 & 27.82 & 0.112 &
27.19 & 1.30 & 11.09\\
0034+444&  00 36 53.5 & 44 43 21    & 2.790  & 0.675 & 28.65 & 0.173 &
28.06 & 3.20 & 25.54\\
0039+391&  00 41 54.9 & 39 25 21    & 1.006  & 0.315 & 27.23 & 0.069 &
26.57 & 0.34 & 2.74\\
0039+373&  00 42 07.1 & 37 39 36    & 1.006  & 0.939 & 27.70 & 0.228 &
27.09 & 0.10 & 0.81\\
0039+398&  00 42 17.4 & 40 09 48    & 1.000  & 0.748 & 27.60 & 0.200 &
27.02 & 3.70 & 29.79\\
0039+412&  00 42 18.5 & 41 29 26    & 1.000  & 0.371 & 27.29 & 0.129 &
26.83 & 2.00 & 16.10\\
0041+425&  00 44 39.2 & 42 48 02    & 1.000  & 0.463 & 27.39 & 0.146 &
26.89 & 1.00 & 8.05\\
0049+379&  00 52 16.8 & 38 15 28    & 1.700* & 0.793 & 28.20 & 0.192 &
27.58 & 1.45 & 12.40\\
0110+401&  01 13 17.8 & 40 26 13    & 1.479  & 0.546 & 27.88 & 0.283 &
27.60 & 3.90 & 33.28\\
0120+405&  01 23 26.2 & 40 46 59    & 0.840  & 0.608 & 27.32 & 0.199 &
26.83 & 2.40 & 18.38\\
0123+402&  01 25 59.2 & 40 28 37    & 1.000  & 0.256 & 27.13 & 0.066 &
26.54 & 1.10 & 8.86\\
0128+394&  01 31 29.5 & 39 42 58    & 0.929  & 0.327 & 27.16 & 0.062 &
26.43 & 2.70 & 21.32\\
0137+401&  01 40 33.7 & 40 24 14    & 1.620  & 0.287 & 27.70 & 0.113 &
27.30 & 4.20 & 35.94\\
0140+387&  01 43 33.0 & 39 02 11    & 2.900* & 0.426 & 28.50 & 0.063 &
27.67 & 0.70 & 5.53\\
0144+432&  01 47 55.6 & 43 32 15    & 1.260  & 0.334 & 27.50 & 0.085 &
26.90 & 4.00 & 33.64\\
0147+400&  01 50 19.6 & 40 17 30    & 1.000  & 0.716 & 27.58 & 0.259 &
27.14 & 0.10 & 0.80\\
0213+412&  02 16 30.3 & 41 31 51    & 0.515  & 0.541 & 26.75 & 0.259 &
26.43 & 2.00 & 12.39\\
0222+422&  02 25 32.0 & 42 29 41    & 3.500* & 0.260 & 28.48 & 0.066 &
27.88 & 3.40 & 25.33\\
0228+409&  02 31 38.8 & 41 09 54    & 1.000  & 0.351 & 27.27 & 0.115 &
26.78 & 3.80 & 30.59\\
0254+406&  02 57 50.2 & 40 50 32    & 1.224  & 0.497 & 27.64 & 0.154 &
27.13 & 4.50 & 37.69\\
0255+460&  02 58 30.2 & 46 16 06    & 1.210  & 0.713 & 27.78 & 0.256 &
27.34 & 0.66 & 5.52\\
0701+392&  07 04 31.3 & 39 11 23    & 1.283  & 0.503 & 27.69 & 0.170 &
27.22 & 1.80 & 15.17\\
0703+468&  07 06 48.0 & 46 47 56    & 1.000  & 1.585 & 27.92 & 0.625 &
27.52 & 0.08 & 0.64\\
0722+393&  07 25 50.0 & 39 17 25    & 1.000  & 1.104 & 27.76 & 0.236 &
27.09 & 0.25 & 2.01\\
0729+437&  07 32 43.7 & 43 35 41    & 1.000  & 0.385 & 27.31 & 0.132 &
26.84 & 1.30 & 10.47\\
0744+464&  07 47 43.6 & 46 18 58    & 2.926  & 0.514 & 28.59 & 0.134 &
28.00 & 1.40 & 11.03\\
0748+413&  07 52 19.9 & 41 15 53    & 1.000  & 0.234 & 27.09 & 0.067 &
26.55 & 0.40 & 3.22\\
0754+396&  07 58 08.8 & 39 29 28    & 2.119  & 0.523 & 28.25 & 0.134 &
27.66 & 2.20 & 18.52\\
0800+472&  08 04 13.9 & 47 04 43    & 0.510  & 0.888 & 26.95 & 0.333 &
26.52 & 1.00 & 6.16\\
0805+406&  08 09 03.1 & 40 32 57    & 1.775* & 0.530 & 28.07 & 0.183 &
27.60 & 2.50 & 21.36\\
0809+404&  08 12 53.1 & 40 19 00    & 0.551  & 1.062 & 27.11 & 0.446 &
26.73 & 1.20 & 7.70\\
0810+460&  08 14 30.3 & 45 56 40    & 0.620  & 1.106 & 27.25 & 0.276 &
26.65 & 0.63 & 4.28\\
0814+441&  08 17 36.0 & 43 59 35    & 1.000  & 0.289 & 27.18 & 0.079 &
26.62 & 4.00 & 32.20\\
0822+394&  08 25 23.7 & 39 19 46    & 1.210  & 1.198 & 28.01 & 0.341 &
27.46 & 0.05 & 0.42\\
0840+424&  08 43 31.6 & 42 15 30    & 1.000  & 1.409 & 27.87 & 0.583 &
27.49 & 0.08 & 0.64\\
0856+406&  08 59 59.6 & 40 24 35    & 2.290  & 0.240 & 28.00 & 0.042 &
27.24 & 0.80 & 6.66\\
0902+416&  09 05 22.2 & 41 28 40    & 1.000  & 0.496 & 27.42 & 0.159 &
26.92 & 0.34 & 2.74\\
0930+389&  09 33 06.9 & 38 41 50    & 2.395  & 0.284 & 28.12 & 0.088 &
27.61 & 3.70 & 30.55\\
0935+428&  09 38 16.5 & 42 38 35    & 1.291  & 0.439 & 27.64 & 0.136 &
27.13 & 1.30 & 10.97\\
0951+422&  09 54 12.6 & 42 01 09    & 1.783  & 0.428 & 27.98 & 0.167 &
27.57 & 2.00 & 17.08\\
0955+390&  09 58 44.3 & 38 48 22    & 1.000  & 0.460 & 27.38 & 0.135 &
26.85 & 4.80 & 38.64\\
1007+422&  10 10 24.8 & 41 59 33    & 1.000  & 0.430 & 27.35 & 0.141 &
26.87 & 0.13 & 1.05\\
1008+423&  10 11 54.2 & 42 04 33    & 1.000  & 0.593 & 27.50 & 0.199 &
27.02 & 0.05 & 0.40\\
1014+392&  10 17 14.2 & 39 01 23    & 0.210  & 1.392 & 26.25 & 0.485 &
25.79 & 6.10 & 20.77\\
1016+443&  10 19 48.2 & 44 08 25    & 0.330* & 0.354 & 26.10 & 0.071 &
25.40 & 0.11 & 0.52\\
1025+389&  10 28 44.3 & 38 44 37    & 0.361  & 0.658 & 26.46 & 0.336 &
26.17 & 3.20 & 16.07\\
1027+392&  10 30 16.5 & 38 57 54    & 1.000  & 0.385 & 27.31 & 0.128 &
26.83 & 1.60 & 12.88\\
1039+424&  10 42 06.3 & 42 10 31    & 1.000  & 0.277 & 27.16 & 0.061 &
26.51 & 1.50 & 12.07\\
1044+454&  10 47 33.7 & 45 08 52    & 4.100* & 0.440 & 28.87 & 0.088 &
28.17 & 1.00 & 7.01\\
1049+384&  10 52 11.8 & 38 11 44    & 1.018  & 0.682 & 27.58 & 0.202 &
27.05 & 0.10 & 0.81\\
1055+404&  10 58 00.4 & 40 10 19    & 1.000  & 0.369 & 27.29 & 0.107 &
26.75 & 2.80 & 22.54\\
1128+455&  11 31 38.9 & 45 14 51    & 0.404  & 2.049 & 27.07 & 0.656 &
26.58 & 0.90 & 4.85\\
1133+432&  11 35 56.0 & 42 58 45    & 1.000  & 1.449 & 27.88 & 0.442 &
27.37 & 0.07 & 0.56\\
1136+420&  11 38 59.1 & 41 48 39    & 0.829  & 0.459 & 27.18 & 0.187 &
26.79 & 1.00 & 7.62\\
1136+383&  11 39 34.0 & 38 03 42    & 1.000  & 0.460 & 27.38 & 0.195 &
27.01 & 0.05 & 0.40\\
1141+466&  11 43 39.6 & 46 21 20    & 0.116  & 0.863 & 25.47 & 0.132 &
24.65 & 8.10 & 16.86\\
1143+456&  11 46 15.2 & 45 20 38    & 0.762  & 0.711 & 27.28 & 0.149 &
26.60 & 0.80 & 5.92\\
1157+460&  12 00 31.2 & 45 48 42    & 0.743  & 1.156 & 27.46 & 0.311 &
26.89 & 0.80 & 5.86\\
1159+395&  12 01 49.9 & 39 19 11    & 2.370  & 0.606 & 28.43 & 0.261 &
28.07 & 0.05 & 0.41\\
\hline
\end{tabular}
\end{center} 
\label{combined}
\end{table*}

\setcounter{table}{0}
\begin{table*}
\begin{center}
\caption[]{Combined sample of CSS sources}
\begin{tabular}{@{}c c c c c c c c c c@{}}
\hline
~~~Source & RA & Dec &
\multicolumn{1}{c}{\it z}&
\multicolumn{1}{c}{${\rm S_{1.4}}$}&
\multicolumn{1}{c}{log${\rm L_{1.4}}$}&
\multicolumn{1}{c}{${\rm S_{4.85}}$}&  
\multicolumn{1}{c}{log${\rm L_{4.85}}$}&
\multicolumn{1}{c}{LAS}&
\multicolumn{1}{c}{LLS}\\
~~~Name   & h~m~s & $\degr$~$\arcmin$~$\arcsec$ & &
\multicolumn{1}{c}{Jy}&
\multicolumn{1}{c}{$\rm W~Hz^{-1}$} &
\multicolumn{1}{c}{Jy}&
\multicolumn{1}{c}{$\rm W~Hz^{-1}$} &
\multicolumn{1}{c}{$\arcsec$} &
\multicolumn{1}{c}{$h^{-1}~{\rm kpc}$} \\
~~~(1)& (2)& (3) &(4)&
\multicolumn{1}{c}{(5)}&
\multicolumn{1}{c}{(6)}&
\multicolumn{1}{c}{(7)}&
\multicolumn{1}{c}{(8)}&
\multicolumn{1}{c}{(9)}&
\multicolumn{1}{c}{(10)}\\
\hline
1201+394&  12 04 06.8 & 39 12 18    & 0.445  & 0.483 & 26.54 & 0.168 &
26.08 & 2.10 & 12.00\\
1204+401&  12 07 06.2 & 39 54 39    & 2.066  & 0.259 & 27.92 & 0.058 &
27.27 & 1.60 & 13.51\\
1212+380&  12 14 56.7 & 37 48 51    & 1.500* & 0.293 & 27.63 & 0.067 &
26.99 & 0.30 & 2.56\\
1216+402&  12 18 37.0 & 40 00 45    & 0.756  & 0.366 & 26.98 & 0.101 &
26.42 & 3.80 & 28.04\\
1217+427&  12 19 53.8 & 42 29 51    & 1.000  & 0.303 & 27.20 & 0.092 &
26.69 & 2.70 & 21.74\\
1220+408&  12 22 35.2 & 40 36 21    & 1.000  & 0.463 & 27.39 & 0.121 &
26.80 & 4.10 & 33.01\\
1225+442&  12 27 42.0 & 44 00 42    & 0.348  & 0.384 & 26.19 & 0.114 &
25.66 & 0.20 & 0.98\\
1233+418&  12 35 35.7 & 41 37 07    & 0.250* & 0.690 & 26.11 & 0.285 &
25.73 & 2.20 & 8.55\\
1241+411&  12 44 20.0 & 40 51 37    & 0.249  & 0.361 & 25.83 & 0.171 &
25.50 & 1.00 & 3.88\\
1242+410&  12 44 49.2 & 40 48 06    & 0.813  & 1.342 & 27.63 & 0.707 &
27.35 & 0.04 & 0.30\\
1313+453&  13 16 11.9 & 45 04 29    & 1.544  & 0.667 & 28.02 & 0.307 &
27.68 & 0.16 & 1.37\\
1340+439&  13 43 05.9 & 43 43 24    & 1.000  & 0.547 & 27.46 & 0.133 &
26.85 & 0.07 & 0.56\\
1343+386&  13 45 36.9 & 38 23 13    & 1.853  & 0.895 & 28.34 & 0.399 &
27.99 & 0.11 & 0.94\\
1350+432&  13 52 28.5 & 42 59 22    & 2.149  & 0.170 & 27.78 & 0.029 &
27.01 & 1.60 & 13.44\\
1432+428&  14 34 27.8 & 42 36 20    & 1.000  & 0.912 & 27.68 & 0.312 &
27.22 & 0.04 & 0.32\\
1441+409&  14 42 59.3 & 40 44 29    & 1.000  & 0.963 & 27.71 & 0.331 &
27.24 & 0.10 & 0.80\\
1445+410&  14 47 12.7 & 40 47 45    & 0.195  & 0.409 & 25.64 & 0.156 &
25.22 & 8.10 & 26.02\\
1449+421&  14 51 07.3 & 41 54 41    & 1.000  & 0.813 & 27.63 & 0.131 &
26.84 & 0.08 & 0.64\\
1458+433&  15 00 30.0 & 43 09 52    & 0.927  & 0.438 & 27.28 & 0.121 &
26.72 & 1.60 & 12.63\\
2301+443&  23 03 45.3 & 44 39 06    & 1.700* & 1.229 & 28.39 & 0.274 &
27.73 & 0.50 & 4.28\\
2302+402&  23 04 54.6 & 40 28 54    & 1.000  & 1.163 & 27.79 & 0.386 &
27.31 & 0.60 & 4.83\\
2304+377&  23 07 01.0 & 38 02 42    & 0.400* & 1.487 & 26.92 & 0.642 &
26.56 & 0.10 & 0.54\\
2311+469&  23 13 48.1 & 47 12 15    & 0.742  & 1.927 & 27.69 & 0.856 &
27.33 & 2.40 & 17.59\\
2322+403&  23 24 48.5 & 40 40 23    & 1.000  & 0.318 & 27.22 & 0.089 &
26.67 & 3.20 & 25.76\\
2330+402&  23 32 52.9 & 40 30 37    & 1.000  & 0.831 & 27.64 & 0.302 &
27.20 & 0.07 & 0.56\\
2348+450&  23 51 27.8 & 45 18 30    & 0.978  & 0.742 & 27.57 & 0.230 &
27.06 & 0.20 & 1.60\\
2349+410&  23 51 52.8 & 41 21 15    & 2.046  & 0.452 & 28.15 & 0.106 &
27.52 & 1.20 & 10.14\\
\hline
0744+291& 07 48 05.3& 29 03 23 & 1.000  & 0.480 & 27.40 & 0.170 &
26.95 & 0.55 & 4.43\\
0747+314& 07 50 12.3& 31 19 48 & 1.000  & 1.050 & 27.74 & 0.350 &
27.27 & 0.25 & 2.01\\
0801+303& 08 04 42.2& 30 12 37 & 1.452  & 1.189 & 28.20 & 0.404 &
27.73 & 1.30 & 11.08\\
0811+360& 08 14 49.0& 35 53 50 & 1.000  & 0.580 & 27.49 & 0.170 &
26.95 & 0.20 & 1.61\\
0850+331& 08 53 21.1& 32 55 01 & 1.000  & 0.465 & 27.39 & 0.208 &
27.04 & 1.95 & 15.70\\
0853+291& 08 56 01.2& 28 58 35 & 1.085  & 0.630 & 27.61 & 0.190 &
27.09 & 0.80 & 6.56\\
0922+322& 09 25 32.7& 31 59 53 & 1.000  & 0.530 & 27.45 & 0.200 &
27.02 & 1.40 & 11.27\\
0949+287& 09 52 06.1& 28 28 32 & 1.000  & 1.364 & 27.86 & 0.529 &
27.45 & 0.31 & 2.49\\
1009+408& 10 12 51.8& 40 39 04 & 1.000  & 0.411 & 27.34 & 0.194 &
27.01 & 4.06 & 32.69\\
1045+352& 10 48 34.2& 34 57 25 & 1.604  & 1.051 & 28.25 & 0.439 &
27.88 & 0.50 & 4.28\\
1056+316& 10 59 43.2& 31 24 21 & 0.307* & 0.459 & 26.14 & 0.209 &
25.80 & 0.50 & 2.25\\
1059+351& 11 02 08.7& 34 55 11 & 0.594* & 0.702 & 27.01 & 0.252 &
26.56 & 3.03 & 20.17\\
1123+340& 11 26 23.7& 33 45 27 & 1.247  & 1.320 & 28.08 & 0.380 &
27.54 & 0.25 & 2.10\\
1125+327& 11 28 02.4& 32 30 47 & 0.750* & 0.600 & 27.19 & 0.205 &
26.72 & 6.81 & 50.10\\
1126+293& 11 29 21.7& 29 05 06 & 1.000  & 0.729 & 25.20 & 0.213 &
24.67 & 0.79 & 1.37\\
1132+374& 11 35 05.9& 37 08 41 & 2.880  & 0.638 & 28.66 & 0.218 &
28.20 & 0.30 & 2.37\\
1232+295& 12 34 54.4& 29 17 44 & 1.000  & 0.460 & 27.38 & 0.160 &
26.93 & 0.40 & 3.22\\
1236+327& 12 39 09.1& 32 30 27 & 1.000  & 0.832 & 27.64 & 0.256 &
27.13 & 3.12 & 25.12\\
1242+364& 12 44 49.7& 36 09 26 & 1.000  & 0.780 & 27.61 & 0.210 &
27.04 & 0.55 & 4.43\\
1251+308& 12 53 25.7& 30 36 35 & 1.311  & 0.450 & 27.67 & 0.200 &
27.32 & 0.55 & 4.65\\
1302+356& 13 04 34.5& 35 23 34 & 1.000  & 0.483 & 27.40 & 0.185 &
26.99 & 0.20 & 1.61\\
1315+396& 13 17 18.6& 39 25 28 & 1.562  & 0.615 & 27.99 & 0.227 &
27.56 & 0.03 & 0.29\\
1334+417& 13 36 26.4& 41 31 13 & 1.000  & 0.470 & 27.39 & 0.154 &
26.91 & 5.47 & 44.04\\
1401+353& 14 03 19.2& 35 08 12 & 0.450* & 0.630 & 26.67 & 0.180 &
26.13 & 1.80 & 10.35\\
1407+369& 14 09 09.5& 36 42 08 & 0.996* & 0.538 & 27.45 & 0.216 &
27.05 & 0.25 & 2.01\\
1425+287& 14 27 40.3& 28 33 26 & 1.000  & 0.859 & 27.65 & 0.198 &
27.02 & 0.75 & 6.03\\
1502+291& 15 04 26.7& 28 54 31 & 2.283$\dagger$  & 0.567 & 28.37 & 0.261 &
28.03 & 0.04 & 0.35\\
1542+323& 15 44 48.4& 32 08 45 & 1.000  & 0.854 & 27.65 & 0.325 &
27.23 & 2.41 & 19.40\\
1601+382& 16 03 35.1& 38 06 43 & 1.000  & 0.430 & 27.35 & 0.200 &
27.02 & 0.75 & 6.04\\
1616+366& 16 18 23.6& 36 32 02 & 0.734  & 0.536 & 27.12 & 0.268 &
26.82 & 0.06 & 0.44\\
1619+378& 16 21 11.3& 37 46 05 & 1.273  & 0.640 & 27.79 & 0.200 &
27.28 & 0.75 & 6.32\\
1627+289& 16 29 12.3& 28 51 34 & 1.000 & 0.526 & 27.44 & 0.162 &
26.93 & 0.65 & 5.23\\
\hline
\end{tabular}
\end{center} 
\end{table*} 

\setcounter{table}{0}
\begin{table*}
\begin{center}
\caption[]{Combined sample of CSS sources}
\begin{tabular}{@{}c c c c c c c c c c@{}}
\hline
~~~Source & RA & Dec &
\multicolumn{1}{c}{\it z}&
\multicolumn{1}{c}{${\rm S_{1.4}}$}&
\multicolumn{1}{c}{log${\rm L_{1.4}}$}&
\multicolumn{1}{c}{${\rm S_{4.85}}$}&  
\multicolumn{1}{c}{log${\rm L_{4.85}}$}&
\multicolumn{1}{c}{LAS}&
\multicolumn{1}{c}{LLS}\\
~~~Name   & h~m~s & $\degr$~$\arcmin$~$\arcsec$ & &
\multicolumn{1}{c}{Jy}&
\multicolumn{1}{c}{$\rm W~Hz^{-1}$} &
\multicolumn{1}{c}{Jy}&
\multicolumn{1}{c}{$\rm W~Hz^{-1}$} &
\multicolumn{1}{c}{$\arcsec$} &
\multicolumn{1}{c}{$h^{-1}~{\rm kpc}$} \\
~~~(1)& (2)& (3) &(4)&
\multicolumn{1}{c}{(5)}&
\multicolumn{1}{c}{(6)}&
\multicolumn{1}{c}{(7)}&
\multicolumn{1}{c}{(8)}&
\multicolumn{1}{c}{(9)}&
\multicolumn{1}{c}{(10)}\\
\hline
1632+391& 16 34 02.9& 39 00 00 & 1.085  & 0.930 & 27.78 & 0.370 &
27.38 & 0.80 & 6.56\\
1656+391& 16 58 22.2& 39 06 26 & 1.000  & 0.650 & 27.53 & 0.240 &
27.10 & 0.12 & 0.97\\
1709+303& 17 11 19.9& 30 19 18 & 1.000  & 1.030 & 27.73 & 0.370 &
27.29 & 0.40 & 3.22\\
1717+315& 17 19 30.0& 31 28 48 & 1.000  & 0.450 & 27.38 & 0.150 &
26.89 & 0.60 & 4.83\\
1723+406& 17 25 16.3& 40 36 41 & 1.000  & 0.962 & 27.71 & 0.221 &
27.07 & 4.65 & 37.44\\
\hline
\end{tabular}
\end{center} 

\begin{minipage}{165mm}
Description of the columns:
(1) source name;
(2) and (3) source coordinates (J2000);
(4) redshift; $\ast$ - photometric redshift taken form the literature or
from the SDSS; $\dagger$ - this is a new value from the SDSS; we adopt  
${\it z}=1.0$ for sources with unknown redshift;
(5) total flux density at 1.4\,GHz extracted from FIRST;
(6) log of the radio luminosity at 1.4\,GHz;
(7) total flux density at 4.85\,GHz extracted from GB6;
(8) log of the radio luminosity at 4.85\,GHz;
(9) largest angular size (LAS);
(10) largest linear size (LLS) calculated based on the largest angular size
measurements on resolved structure taken from the literature. 
Data are taken from \citet{f2001} sample (first part of the table) and
\citet{mar03}.
sample. There is a small overlap of this two samples that is why the
overlaping sources are included only in a first part of the table.  
\end{minipage}
\end{table*}

\end{document}